\documentclass[aps,prb,twocolumn,showpacs,amsmath,amssymb,superscriptaddress,floatfix]{revtex4-2}
\usepackage[english]{babel}
\usepackage{blindtext}
\usepackage[utf8]{inputenc}
\usepackage{graphicx}
\usepackage{bm}
\usepackage{latexsym}
\usepackage{physics}
\usepackage{amsfonts}
\usepackage{appendix}
\usepackage[normalem]{ulem}
\usepackage[colorlinks = true,
            linkcolor = blue,
            urlcolor  = blue,
            citecolor = blue,
            anchorcolor = blue]{hyperref}
\newcommand{\eps}{\epsilon}
\newcommand{\up}{\uparrow}
\newcommand{\down}{\downarrow}
\newcommand{\bs}[1]{\mathbf{#1}}

\newcommand{\bk}{\mathbf{k}}
\newcommand{\bq}{\mathbf{q}}
\newcommand{\bQ}{\mathbf{Q}}

\newcommand{\old}[1]{}
\newcommand{\new}[1]{#1}

\graphicspath{{figures/}}
\begin{document}

\author{Pietro M. Bonetti}
\affiliation{Max Planck Institute for Solid State Research, Heisenbergstrasse 1, D-70569 Stuttgart, Germany}

\author{Alessandro Toschi}
\affiliation{Institute of Solid State Physics, Vienna University of Technology, 1040 Vienna, Austria}

\author{Cornelia Hille}
\affiliation{Institut f\"ur Theoretische Physik and Center for Quantum Science, Universit\"at T\"ubingen, Auf der Morgenstelle 14, 72076 T\"ubingen, Germany}

\author{Sabine Andergassen}
\affiliation{Institut f\"ur Theoretische Physik and Center for Quantum Science, Universit\"at T\"ubingen, Auf der Morgenstelle 14, 72076 T\"ubingen, Germany}

\author{Demetrio Vilardi}
\email{d.vilardi@fkf.mpg.de}
\affiliation{Max Planck Institute for Solid State Research, Heisenbergstrasse 1, D-70569 Stuttgart, Germany}

\title{Single~boson~exchange~representation of~the~functional~renormalization~group for~strongly~interacting~many-electron~systems}%
\date{\today}
\begin{abstract}
We present a reformulation of the functional renormalization group (fRG) for many-electron systems, which relies on the recently introduced single boson exchange (SBE) representation of the parquet equations [\href{https://link.aps.org/doi/10.1103/PhysRevB.100.155149}{Phys.~Rev.~B \textbf{100}, 155149 (2019)}]. The latter exploits a diagrammatic decomposition, which classifies the contributions to the full scattering amplitude in terms of their reducibility with respect to cutting one interaction line, naturally distinguishing the processes mediated by the exchange of a single boson in the different channels. We apply this idea to the fRG by splitting the one-loop fRG flow equations for the vertex function into SBE contributions and a residual four-point fermionic vertex. 
Similarly as in the case of parquet solvers, recasting the fRG algorithm in the SBE representation offers both computational and interpretative advantages: the SBE decomposition not only significantly reduces the numerical effort of treating the high-frequency asymptotics of the flowing vertices, but it also allows for a clear physical identification of the collective degrees of freedom at play. We illustrate the advantages of an SBE formulation of fRG-based schemes, by computing through the merger of dynamical mean-field theory and fRG the 
susceptibilities and the Yukawa couplings of the two-dimensional Hubbard model from weak to strong coupling, for which we also present an intuitive physical explanation 
of the results. The SBE formulation of the one-loop flow equations paves a promising route for future multiboson and multiloop extensions of fRG-based algorithms. 
\end{abstract}
\maketitle
\section{Introduction}

A 
major challenge for the theory of many-electron systems is to correctly describe the competing microscopical processes occurring on very different length and time scales.
This task becomes particularly hard in the nonperturbative regime of intermediate to strong interactions, where several fascinating phenomena of condensed matter physics take place.

The recently developed  \cite{Taranto2014,Wentzell2015,Vilardi2019}   merger of the dynamical mean field theory (DMFT)~\cite{Metzner1989,Georges1996} and the functional renormalization group (fRG) \cite{Salmhofer1999,Berges2002,Kopietz2010,Metzner12,Dupuis2021}, coined as DMF\textsuperscript2RG \cite{Taranto2014}, can be regarded as one of the diagrammatic extensions \cite{Rohringer2018} of DMFT designed to be applied in the most challenging parameter regimes of quantum many-body Hamiltonians. 
By combining the unbiased \cite{Metzner12} diagrammatic structure of the fRG to the intrinsic nonperturbative content \cite{Chalupa2021} of DMFT, the DMF\textsuperscript2RG offers a particularly promising route to tackle crucial nonperturbative features of the many-electron physics.
However, the impact of DMF\textsuperscript2RG calculations will eventually depend on the numerical performance of its actual implementations, where the evident bottleneck is posed by the treatment of the two-particle vertex functions \cite{Rohringer2012,Wentzell2016}, due to the large number of momentum and frequency variables needed for their definition. 

The single boson exchange (SBE) decomposition, recently introduced \cite{Krien2019_I} 
to rationalize the treatment of parquet-type diagrams  \cite{Dedominicis1964,Dedominicis1964A,Bickers2004,Yang2009,Tam2013,Rohringer2012,Valli2015,Kauch2019,Li2019}, leads to a significant reduction of the computational effort \cite{Krien2020,Krien2021}. Due to the qualitative similarity of the diagrammatic structure in parquet- 
and fRG-based approximations, it seems quite natural to exploit similar ideas also in an fRG context, recasting the fRG flow equations within the SBE formalism.

On a general level, we recall that applying the SBE formalism corresponds to recasting the two-particle diagrams in terms of their reducibility/irreducibility with respect to the cutting of an interaction line. Due to the two-particle nature of the electronic (Coulomb) interaction, the SBE classification shares important qualitative features with the parquet formalism as, for example, the high-frequency asymptotic properties \cite{Rohringer2012,Tagliavini2018,Wentzell2016} of the corresponding irreducibile diagrams. 
At the same time, the SBE classification of diagrams circumvents important problems (such as the multiple divergences of the irreducible vertices \cite{Schaefer2013,Gunnarsson2016,Schaefer2016,Ribic2016,Gunnarsson2017,Vucicevic2018,Chalupa2018,Springer2020,Chalupa2021}) which affect parquet-based approaches in the nonperturbative regime.


The specific application of the SBE formalism to the fRG presented here relies on the partial bosonization of the vertex function~\cite{Krahl2007,Friederich2010,Denz2020}, similar to the channel decomposition~\cite{Karrasch2008,Husemann2009,Wang2012,Vilardi2017,Tagliavini2019,Vilardi2019} already adopted in the context of fRG 
and parquet solvers~\cite{Eckhardt2020,Krien2020} (for recent developments in this direction see also Refs.~\cite{Hille2020,Astretsov2020,Astleitner2020,Harkov2021}).
In addition to the screened interaction, 
a fermion-boson Yukawa coupling~ \cite{Schmalian1999,Sadovskii2019}
(or Hedin vertex~\cite{Hedin1965}) is  determined 
from the vertex asymptotics, 
similarly to the construction of the kernel functions describing the high-frequency asymptotics, see Ref.~\cite{Wentzell2016}. Their relation 
allows to recover 
the flow equations of the screened interaction and Yukawa coupling in the SBE representation. 
We note that the obtained 
structure is apparently the same as the one reported in
Ref.~\cite{Husemann2012}, where instead of the high-frequency limit the zero frequency value has been used. 

Several bosonization procedures have been already developed for the weak coupling fRG, by applying, for instance, the Hubbard-Stratonovich transformation on the bare action~\cite{Diehl2007,Diehl2007_II,Strack2008,Bartosch2009,Obert2013}, or by means of the dynamical bosonization~\cite{Baier2004,Friederich2010,Friederich2011}. Instead, the description of the DMF\textsuperscript2RG vertex in terms of exchanged bosons is a highly nontrivial task. 
Indeed, the complex frequency structure of the initial DMFT vertex function \cite{Rohringer2012} prevents a straightforward application of the Gaussian integration in the Hubbard-Stratonovic procedure or of the dynamical bosonization. In this perspective, the SBE decomposition offers a relatively simple way to circumvent this problem, expressing the DMFT vertex in terms of bosonic propagators and Yukawa couplings, which can then be used as initial conditions of a mixed boson-fermion flow. In this respect, it is worth stressing that the fermion-boson approach \cite{Stepanov2018,Stepanov2019A} has 
proven to be useful also in the context of diagrammatic extensions of DMFT \cite{Stepanov2019} {\sl not} based on the fRG formalisms.

\new{The purpose of the present manuscript is twofold. 
On the one hand, we show how to apply the  recently introduced SBE decomposition to both the fRG and the DMF\textsuperscript2RG treatment of the 2D Hubbard model \footnote{See Ref.~\cite{Qin21} for a recent overview of computational results for the 2D Hubbard model.}. On the other hand, we analyze the quality of the approximation resulting  from the SBE decomposition on the fRG/DMF$^2$RG flows from weak to strong coupling, at half filling as well as away from it. In this respect, we also exploit the transparent nature of the SBE formulation to investigate the physical mechanisms responsible for the enhanced $d$-wave pairing fluctuations found in the DMF\textsuperscript2RG calculations.}

The paper is organized as follows: In Sec.~\ref{sec:method} we introduce the Hubbard model and present the SBE decomposition together with its implementation in both the fRG and DMF\textsuperscript2RG flow. 
In Secs.~\ref{sec:results half filling} and \ref{sec:results doped regime} we discuss the results for the susceptibilities and Yukawa couplings at half filling and out of it, providing also a comparison with the fermionic formalism. In order to identify the mechanisms responsible for strong $d$-wave pairing correlations, we perform a diagnostics \cite{Gunnarsson2015,Gunnarsson2016,Rohringer2020,Schaefer2021,Delre2021} of the corresponding fluctuations. 
We finally conclude with a summary and an outlook in Sec.~\ref{sec:concl}.

%
\section{Method}
\label{sec:method}
%
\subsection{The Hubbard model}
We consider the single-band Hubbard model in two dimensions,
\begin{equation}
    \begin{split}
        \mathcal{H}=\sum_{i\neq j,\sigma}t_{ij}c^\dagger_{i\sigma}c_{j\sigma}+U\sum_i n_{i\up}n_{i\down}-\mu\sum_{i,\sigma} n_{i\sigma},
    \end{split}
    \label{eq: Hubbard model}
\end{equation}
where $c_{i\sigma}$ ($c^{\dagger}_{i\sigma}$) annihilates (creates) an electron with spin $\sigma$ at the lattice site $i$ ($n_{i\sigma}=c^{\dagger}_{i\sigma}c_{i\sigma}$), $t_{ij}=-t$ is the hopping between nearest-neighbor sites, $t_{ij}=-t'$ the hopping between next-nearest-neighbor sites (the Fourier transform of $t_{ij}$ gives the bare dispersion $\eps_\bk$), and $U$ the on-site Coulomb interaction.  
The filling is fixed by adjusting the chemical potential $\mu$.

In the following we use $t\equiv 1$ as the energy unit.
\subsection{SBE decomposition}
\begin{figure}
    \centering
    \includegraphics[width = 0.45 \textwidth]{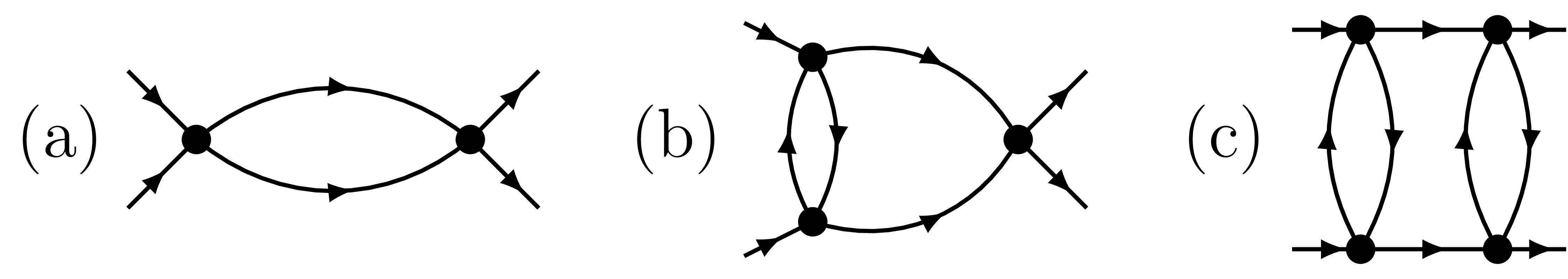}
    \caption{Representative diagrams of the $U$ irreducibility. Here we choose the \emph{particle-particle} ($pp$) channel as an example. (a) two-particle- and $U$-reducible diagram in the $pp$ channel. Since this diagram has two interaction vertices at its both ends, it contributes to the screened interaction. (b) Two-particle- and $U$-reducible diagram, contributing to the Yukawa coupling, as there is only one bare interaction vertex which can be removed to disconnect the diagram. (c) Diagram which is two-particle reducible but $U$-irreducible, therefore contributing to the SBE rest function.}
    \label{fig: Ured & irred diagrams}
\end{figure}

In this section we present the SBE decomposition as introduced in Ref.~\cite{Krien2019_I}. All diagrams contributing to the two-particle vertex can be divided into $U$-reducible and $U$-irreducible, depending on whether the removal of a bare interaction vertex cuts the diagram into two disconnected parts or not, respectively. Moreover, among the $U$-reducible diagrams, we can identify three different channels, depending on how the fermionic legs are connected to the removed interaction $U$. In particular, we find $U$-particle-particle ($U$-$pp$), $U$-particle-hole ($U$-$ph$) and $U$-particle-hole-crossed ($U$-$\overline{ph}$) reducible diagrams~\cite{Krien2019_I}. This classification of diagrams is alternative, and not equivalent, to the more common one, based on the notion of two-particle reducibility (for the relation to the latter we refer to the Appendix~\ref{app: fermionic weak coupling}). 
Indeed, there are some diagrams which are two-particle reducible in a given channel, but $U$-irreducible. These diagrams are the so-called box diagrams (see Fig.~\ref{fig: Ured & irred diagrams}).
In general, a diagram which is $U$-reducible in a given channel is also two-particle reducible in the same channel, but not vice versa, the only exception being the diagram composed by a single interaction vertex. 
Within the SBE decomposition, it is quite natural to interpret the collection of all $U$-reducible
diagrams in a given channel 
as an effective interaction 
between two fermions, 
mediated by the exchange of a boson, representing a collective fluctuation. 
As mentioned in Refs.~\cite{Krien2019_I,Krien2019_II,Krien2020} and illustrated explicitly in the following,
the SBE decomposition not only significantly reduces the numerical complexity of the vertex functions, but it also allows for a clear physical identification of the collective degrees of freedom arising in a strongly correlated electron system. 

In a system with the U(1)-charge and SU(2)-spin symmetries, along with translational invariance, the two-particle vertex can be expressed as~\cite{Rohringer2012}:
\begin{equation}
    \begin{split}
        V_{\sigma_1\sigma_2\sigma_3\sigma_4}(k_1,k_2,k_3,k_4) &=
        V(k_1,k_2,k_3)\delta_{\sigma_1\sigma_3}\delta_{\sigma_2\sigma_4}\\
        &-V(k_2,k_1,k_3)\delta_{\sigma_1\sigma_4}\delta_{\sigma_2\sigma_3},
    \end{split}
    \label{eq: SU(2) U(1) vertex}
\end{equation}
where $\sigma_i$ represents the spin quantum number, and $k_i=(\bk_i,\nu_i)$ is a collective variable including the crystal momentum and a fermionic Matsubara frequency. The fourth fermionic variable, $k_4$, on which the vertex depends, is fixed by momentum and energy conservation. According to the SBE decomposition, we represent the function $V=V_{\up\down\up\down}$ as
\begin{equation}
    \begin{split}
        V(k_1,k_2,k_3) &= \Lambda_\text{$U$irr}(k_1,k_2,k_3) - 2U+ \mathcal{S}_{k_1 k_3}(k_1+k_2)\\
                        &+ \mathcal{M}_{k_1 k_3}(k_2-k_3) +\frac{1}{2} \mathcal{M}_{k_1 k_4}(k_3-k_1)\\
                       &  +\frac{1}{2} \mathcal{C}_{k_1 k_4}(k_3-k_1),
    \end{split}
    \label{eq: vertex parametrization}
\end{equation}
where $k_4=k_1+k_2-k_3$, $\mathcal{S}$, $(\mathcal{M}+\mathcal{C})/2$, and $\mathcal{M}$ represent the sum of all $U$-$pp$, $U$-$ph$, and $U$-$\overline{ph}$ reducible diagrams, respectively, the function $\Lambda_\text{$U$irr}$ accounts for all fully $U$-irreducible diagrams, and a term $2U$ has been subtracted in order to avoid double counting of the bare interaction, which is already included in the $U$-reducible channels. The functions $\mathcal{S}$, $\mathcal{M}$, and $\mathcal{C}$, corresponding to the pairing, charge and magnetic channel respectively, depend on two fermionic variables, and a bosonic one, indicated in brackets in Eq.~\eqref{eq: vertex parametrization}. Since each of them is $U$-reducible in a given channel, their dependencies on the fermionic arguments can be factorized. We can therefore express them as 
\begin{subequations}
    \label{eq: bosonic decomposition}
    \begin{align}
        \mathcal{M}_{kk'}(q)&\equiv h^m_{k}(q)\,D^m(q)\,\Bar{h}^m_{k'}(q),\label{eq: magn bosonic decomposition}\\
        \mathcal{C}_{kk'}(q)&\equiv h^c_{k}(q)\,D^c(q)\,\Bar{h}^c_{k'}(q),\label{eq: char bosonic decomposition}\\
        \mathcal{S}_{kk'}(q)&\equiv h^s_{k}(q)\,D^s(q)\,\Bar{h}^s_{k'}(q), \label{eq: Swave bosonic decomposition}
    \end{align}
\end{subequations}
where we name $h$ ($\Bar{h}$) as left-sided (right-sided) Yukawa coupling and $D$ as screened interaction, which plays the role of an effective bosonic propagator. The right-sided Yukawa couplings $\Bar{h}$ can be related to their respective left-sided ones, $h$ through the relations
\begin{subequations}
    \begin{align}
        \Bar{h}^X_k(q)& = h^X_{k+q}(-q), \hskip 0.7cm X=m,c, \\
        \Bar{h}^s_k(q)& = h^s_k(q).
    \end{align}
\end{subequations}

By considering the symmetries
\begin{subequations}
    \begin{align}
        &V(k_1,k_2,k_3) = V(k_2,k_1,k_1+k_2-k_3), \label{eq: ras vertex}\\
        &V(k_1,k_2,k_3) = V(k_3,k_1+k_2-k_3,k_1), \label{eq: trs vertex}
    \end{align}
\end{subequations}
where Eq.~\eqref{eq: ras vertex} corresponds to the simultaneous exchange of the ingoing and outgoing variables, and Eq.~\eqref{eq: trs vertex} to the simultaneous exchange of the two ingoing variables with the two outcoming ones, one can easily prove that $h^X_{k+q}(-q)=h^X_k(q)$, with $X=m,c$.  Therefore the relation
\begin{equation}
    h_k^X(q)=\bar{h}_k^X(q)    
\end{equation}
holds for all channels. For this reason, from now on we label by $h^X$ both the left- and the right-sided Yukawa couplings. It is 
worthwhile to stress that 
the equivalence $h=\bar{h}$ holds
because of the choice of notation~\eqref{eq: vertex parametrization},  different choices may lead to to more complicated relations between $h$ and $\bar{h}$.

The screened interactions $D$ are related to the physical susceptibilities~\cite{Krien2019_I,Krien2019_II}, that is
\begin{subequations}
\label{eq: D}
    \begin{align}
         D^m(q) &= U + U^2 \chi^m(q), \label{eq: Dm from chi}\\
         D^c(q) &= U - U^2 \chi^c(q), \label{eq: Dc from chi}\\
         D^s(q) &= U - U^2 \chi^s(q), \label{eq: Ds from chi}
    \end{align}
\end{subequations}
where $\chi^m$, $\chi^c$, and $\chi^s$ are the magnetic, charge, and pairing susceptibilities of the system, respectively. 
The Yukawa couplings are connected to the so-called three-legged correlators via
\begin{subequations}
\label{eq: h}
    \begin{align}
         h^m_k(q) &= \frac{1-\sum_{\sigma}\text{sgn}\sigma\,\langle c^\dagger_{k+q,\sigma}c_{k,\sigma} \Hat{m}_{-q} \rangle_\text{1PI}}{1+U\chi^m(q)},\label{eq: hm from chi3}\\
         h^c_k(q) &= \frac{1-\sum_{\sigma}\langle c^\dagger_{k+q,\sigma}c_{k,\sigma} \Hat{\varrho}_{-q} \rangle_\text{1PI}}{1-U\chi^c(q)},\label{eq: hc from chi3}\\
         h^s_k(q)& = \frac{1+\langle c_{q-k,\down}c_{k,\up} \Hat{\Delta}^\dagger_q \rangle_\text{1PI}}{1-U\chi^s(q)},\label{eq: hs from chi3}
    \end{align}
\end{subequations}
where the symbol $\langle \cdots \rangle_\text{1PI}$ indicates the connected average with amputated external propagators, and the fermionic bilinears are defined as
\begin{subequations}
    \begin{align}
        \Hat{m}_q &= \int_k\sum_\sigma \text{sgn}\sigma\, c^\dagger_{k+q,\sigma}c_{k,\sigma}, \\
        \Hat{\varrho}_q& = \int_k\sum_\sigma c^\dagger_{k+q,\sigma}c_{k,\sigma}, \\
        \Hat{\Delta}_q& = \int_k c_{q-k,\down}c_{k,\up}.
    \end{align}
\end{subequations}
Here, and from now on, the symbol $\int_k=T\sum_\nu\int_\mathrm{B.Z.}\frac{d^2\bk}{(2\pi)^2}$ denotes a sum over fermionic Matsubara frequencies and a momentum integration over the Brillouin zone. Notice that in Eq.~\eqref{eq: h} the division by a term $1\pm U\chi$ is necessary to avoid double counting as it removes from the three-legged correlators all those $U$-reducible diagrams which are already included in the screened interaction $D$.

It is interesting to consider 
the asymptotic high frequency behavior of the Yukawa couplings and bosonic propagators:
\begin{subequations}
\label{eq: limit}
    \begin{align}
        &\lim_{\Omega\rightarrow\infty} D^X(\bs{q},\Omega) = U, \label{eq: D limit}\\
        &\lim_{\nu\rightarrow\infty} h^X_{(\bk,\nu)}(q) = \lim_{\Omega\rightarrow\infty} h^X_{k}(\bs{q},\Omega) = 1. \label{eq: h limit}
    \end{align}
\end{subequations}
The limits of large bosonic frequency $\Omega$ are quite trivial to prove, as in this case both the susceptibilities and the three-legged correlators are zero. More interesting is the large fermionic frequency limit for the Yukawa coupling. In this limit, one can show 
by diagrammatic arguments~\cite{Wentzell2016} that the three-legged correlator approaches $\pm U \chi$ (the sign depending on the channel), which makes the Yukawa coupling approaching $1$. 


\subsection{Flow equations}
In this section we derive the one-loop ($1\ell$) fRG flow equations within the SBE decomposition. The dependencies of the various functions on the RG scale $\Lambda$ are implicit to keep the notation lighter. 

We start by focusing on the 
flow equations within the channel decomposition of Husemann and Salmhofer~\cite{Husemann2009,Husemann2012}, who divided the different contributions to the vertex according to the notion of two-particle reducibility. The flow equation for a given two-particle reducible channel $\phi$ reads
\begin{equation}
    \begin{split}
        \partial_\Lambda\phi^X_{kk'}(q) = \int_p L^X_{kp}(q) \left[\widetilde{\partial}_\Lambda \Pi^X_p(q) \right] L^X_{pk'}(q), \label{eq: channel flow equation}
    \end{split}
\end{equation}
with $X=m,c$ or $s$. The bubbles are defined as 
\begin{subequations}
    \begin{align}
        \Pi^m_k(q)& = -G(k)G(k+q),\\
        \Pi^c_k(q)& = G(k)G(k+q),\\
        \Pi^s_k(q)& = -G(k)G(q-k),
    \end{align}
\end{subequations}
with $G(k)=[(\Theta^\Lambda(k)/(i\nu-\eps_\bk+\mu))^{-1}-\Sigma(k)]^{-1}$ the propagator, and $\Sigma$ the self-energy. The symbol $\widetilde{\partial}_\Lambda$ denotes a derivative only on the explicit RG scale dependence of the propagator introduced by the cutoff function $\Theta^\Lambda(k)$.
The $L$-functions are given by
\begin{subequations}
    \begin{align}
        L^m_{kk'}(q)& = V(k',k,k+q),\label{eq: Lm}\\
        L^c_{kk'}(q)& = 2V(k,k',k+q) - V(k',k,k+q),\label{eq: Lc} \\
        L^s_{kk'}(q)& = V(k,q-k,k'),\label{eq: Ls}
    \end{align}
\end{subequations}
with $V$ defined in Eqs.~\eqref{eq: SU(2) U(1) vertex} and \eqref{eq: vertex parametrization}.

Each two-particle reducible channel 
consists of two contributions, one which is also $U$-reducible and one which is not. We can therefore express $\phi$ as
\begin{equation}
    \phi^X_{kk'}(q) = h^X_{k}(q)\,D^X(q)\,h^X_{k'}(q) + \mathcal{R}^X_{kk'}(q)-U,
\end{equation}
where we identify 
the $U$-irreducible contribution $\mathcal{R}^X$ as rest function. It obeys the asymptotic relations~\cite{Wentzell2016}
\begin{equation}
    \begin{split}
        \lim_{\nu\rightarrow\infty} \mathcal{R}^X_{(\bk,\nu),k'}(q) = 
        &\lim_{\nu'\rightarrow\infty} \mathcal{R}^X_{k,(\bk',\nu')}(q)  \\
    =   &\lim_{\Omega\rightarrow\infty} \mathcal{R}^X_{kk'}(\bs{q},\Omega) = 0. \label{eq: R limit}
    \end{split}
\end{equation}
With the help of Eqs.~\eqref{eq: vertex parametrization} and \eqref{eq: limit}, we further notice that 
\begin{equation}
    \lim_{\nu'\rightarrow\infty}L^X_{k,(\bk,\nu')}(q) = h^X_k(q) D^X(q).
\end{equation}
By inserting the limits \eqref{eq: limit} and \eqref{eq: R limit} into Eq.~\eqref{eq: channel flow equation}, 
we hence obtain the flow equations for the screened interactions, Yukawa couplings and rest functions: 
%
\begin{subequations}
    \label{eq: flow eq general}
    \begin{align}
        \partial_\Lambda D^X(q)& = \left[D^X(q)\right]^2\!\!\int_p h^X_p(q) \left[\widetilde{\partial}_\Lambda \Pi^X_p(q) \right] h^X_p(q), \label{eq: flow eq D general}\\
        \partial_\Lambda h^X_k(q)& = \int_p \mathcal{L}^X_{kp}(q) \left[\widetilde{\partial}_\Lambda \Pi^X_p(q) \right] h^X_p(q), \label{eq: flow eq h general}\\
        \partial_\Lambda \mathcal{R}^X_{kk'}(q) &= \int_p \mathcal{L}^X_{kp}(q) \left[\widetilde{\partial}_\Lambda \Pi^X_p(q) \right] \mathcal{L}^X_{pk'}(q), \label{eq: flow eq R general}
    \end{align}
\end{subequations}
where $\mathcal{L}^X_{kp}(q)=L^X_{kp}(q)-h^X_{k}(q)\,D^X(q)\,h^X_{p}(q)$. 
Alternatively, the above equations can be derived through the explicit introduction of three bosonic fields via as many Hubbard-Stratonovich transformations (see Appendix~\ref{app: Hubbard-Stratonovich}).

We note that a similar decomposition has been used in Ref.~\cite{Husemann2012}, where the flow equations for the screened interactions and Yukawa couplings
exhibit the same structure. However, instead of using the vertex asymptotics, they are 
determined by its value at the lowest Matsubara frequency. 

By applying the same line of reasoning, one can generalize the flow Eq.~\eqref{eq: flow eq general}, derived within the $1\ell$ truncation, to the multiloop extension introduced in Refs.~\cite{Kugler2018_I,Kugler2018_II,Tagliavini2019,Hille2020}.
In the following we discuss its application to the conventional fRG 
as well as its merger with the DMFT in the DMF\textsuperscript2RG.

\subsection{fRG}

Within the conventional 
fRG, the fully $U$-irreducible vertex function $\Lambda_{U_\mathrm{irr}}$ corresponds to 
the sum of the rest functions of the three channels, and hence does not include any contributions from diagrams that are fully two-particle irreducible. 

\subsubsection{Cutoff scheme}
In order to run a fRG flow, we regularize the bare Green's function by making use of the so-called $\Omega$-cutoff~\cite{Husemann2009,Husemann2012}, that is
\begin{equation}
    G_0^\Lambda(\bk,\nu) = \frac{\Theta^\Lambda(\nu)}{i\nu+\mu-\eps_\bk},
\end{equation}
where the cutoff function is given by 
\begin{equation}
    \Theta^\Lambda(\nu) = \frac{\nu^2}{\nu^2+\Lambda^2}.
    \label{eq: cutoff function}
\end{equation}
\subsubsection{Initial conditions}
The fRG initial conditions at 
$\Lambda=\Lambda_\text{ini}$ read as
\begin{subequations}
    \begin{align}
        D^X_\text{ini}(q)& = U, \\
        h^X_{\text{ini},k}(q)& = 1,\\
        \mathcal{R}^X_{\text{ini},kk'}(q)& = 0,
    \end{align}
    \label{eq: init frg}
\end{subequations}
that, by comparing with Eq.~\eqref{eq: vertex parametrization}, is equivalent to imposing $V_{\text{ini}}=U$.
%
\subsection{\texorpdfstring{DMF\textsuperscript2RG}{DMF2RG}}
\begin{figure}[t]
    \centering
    \includegraphics[width = 0.475 \textwidth]{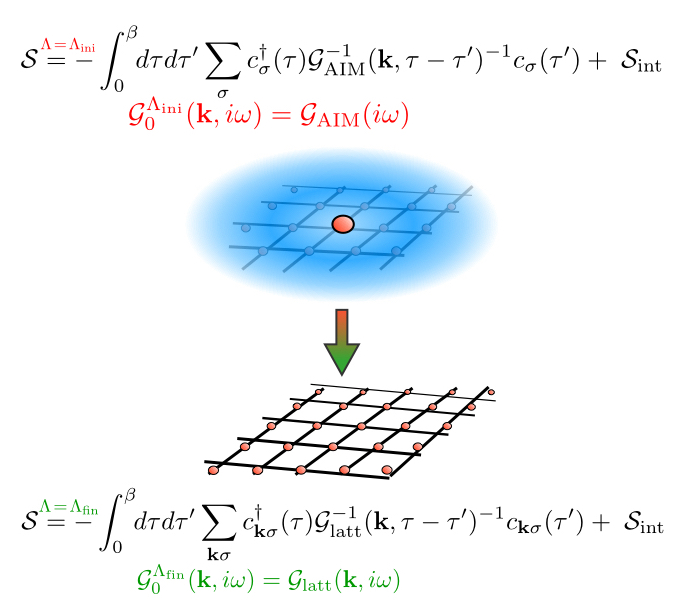}
    \caption{Central idea of DMF\textsuperscript2RG: Starting from an effective AIM, nonlocal correlation effects are gradually included by the flow.}
    \label{fig: DMF2RG easy}
\end{figure}

The DMF\textsuperscript2RG~\cite{Taranto2014,Vilardi2019} flow differs from the fRG one in its initial conditions, \new{ which are defined by the DMFT self-consistent solution of the same problem (see Fig.~\ref{fig: DMF2RG easy}), as well as in the cutoff scheme used}. {
\new{ Consistent to previous DMF$^2$RG studies~\cite{Taranto2014,Vilardi2019}, our DMFT calculations have been performed using the DOS corresponding to the same 2D square lattice dispersion used for the subsequent fRG flow. 
It should be further noticed that, in order determine the initial conditions within the SBE implementation, one needs to}}  apply the decomposition in Eq.~\eqref{eq: vertex parametrization} also to the DMFT vertex. The local bosonic propagators and Yukawa couplings can be computed directly from the Anderson impurity model (AIM), from Eqs.~\eqref{eq: D}-\eqref{eq: h}, while the fully $U$-irreducible vertex can be extracted by subtraction from Eq.~\eqref{eq: vertex parametrization}.
%

\subsubsection{Cutoff scheme}
When introducing a scale dependence on the bare fermionic propagator through a cutoff, one has to keep in mind that the regularized bare propagator has to smoothly interpolate between the bare propagator of the self-consistent AIM and the bare lattice one, specifically: 
\begin{subequations}
\label{eq: DMF2RG bound}
    \begin{align}
        & G_{0,\text{ini}}(\bk,\nu)=\mathcal{G}_{0,\text{AIM}}(\nu)=\frac{1}{i\nu+\mu-\Delta_\text{hyb}(\nu)}, \label{eq: DMF2RG bound cond1}\\
        & G_{0,\text{fin}}(\bk,\nu)= G_0(\bk,\nu)=\frac{1}{i\nu+\mu-\epsilon_\bk}, \label{eq: DMF2RG bound cond2}
    \end{align}
\end{subequations}
where $\Delta_{\text{hyb}}(\nu)$ is the hybridization function of the AIM.
Furthermore, we require the DMFT solution for the local propagator to be conserved at each step of the flow, that is \cite{Vilardi2019}
\begin{equation}
    \int_\text{B.Z.}\frac{d^2\bk}{(2\pi)^2}G(\bk,\nu)\Big\rvert_{\Sigma=\Sigma_{\rm dmft}} = \mathcal{G}_{\text{AIM}}(\nu),
    \label{eq: DMFT conservation}
\end{equation}
with $\mathcal{G}_{\text{AIM}}(\nu)=[\mathcal{G}_{0,\text{AIM}}^{-1}(\nu)-\Sigma_{\rm dmft}(\nu)]^{-1}$ the full Green's function of the AIM, and $\Sigma_\mathrm{dmft}$ the DMFT self-energy.
We therefore make the following choice for the regularized Green's function
\begin{equation}
  \begin{split}
      G^\Lambda_0(\bk,\nu) &= \Theta^\Lambda(\nu)G_0(\bk,\nu) + \Xi^\Lambda(\nu)\mathcal{G}_{0,\text{AIM}}(\nu),
      \label{eq: DMF2RG cutoff}
  \end{split}
\end{equation}
where we choose the function $\Theta^\Lambda(\nu)$ to be a smooth frequency cutoff, namely the same as in Eq.~\eqref{eq: cutoff function}, which imposes the boundary values for the RG scale $\Lambda_\mathrm{ini}=+\infty$, and $\Lambda_\mathrm{fin}=0$.
The function $\Xi^\Lambda(\nu)$ is determined by the DMFT self-consistency relation \eqref{eq: DMFT conservation}. With the above definitions it is straightforward to check that the boundary conditions \eqref{eq: DMF2RG bound} are consistently fulfilled. 

This choice for the DMF\textsuperscript2RG cutoff \eqref{eq: DMF2RG cutoff} has a direct intuitive physical meaning: at a given scale $\Lambda$, indeed, the fermionic modes with energies $|\nu|\gtrsim \Lambda $ are nonlocal and their contributions on top of the DMFT solution have been included, while the ones with energies $|\nu|\lesssim \Lambda $ still belong to the AIM and do not yet contribute to the generation of nonlocal correlations. 
%
\subsubsection{Initial conditions}
The DMF\textsuperscript2RG employs the DMFT solution as a correlated starting point for the RG flow. The initial conditions for the screened interactions and Yukawa couplings therefore read
\begin{subequations}
    \begin{align}
        D^X_{\text{ini}}(q)& = D^X_{\text{loc}}(\Omega),\\
        h^X_{\text{ini},k}(q)& = h^X_{\text{loc},\nu}(\Omega),
    \end{align}
    \label{eq: init dmf2rg}
\end{subequations}
where $D_\text{loc}$ and $h_\text{loc}$ are the bosonic propagator and Yukawa couplings of the self-consistent AIM. Concerning the rest functions, we set their initial value 
to zero, so that they will represent only the nonlocal contributions, 
the local ones being already included in the $U$-irreducible vertex function of the AIM, $\Lambda_{U\text{irr}}^{\text{loc}}$. 

The flow equations for the screened interactions, Yukawa couplings, and rest functions are the same as in the plain fRG. 
%
\subsection{Form factor decomposition}
In order to simplify the set of flow equation described above, throughout this work we project the Yukawa coupling dependence onto the secondary spatial momentum $\bk$ onto $s$-wave form factors, $f^s_\bk\equiv1$, that is we approximate
\begin{equation}
    h^X_{k}(q) \sim h^X_\nu(q) = \int_\bk f^s_\bk\, h^X_{(\bk,\nu)}(q),
\end{equation}
with $\int_\bk$ a shorthand for $\int_\mathrm{B.Z.}\frac{d^2\bk}{(2\pi)^2}$. Within this approximation, Eq.~\eqref{eq: flow eq general} becomes
\begin{subequations}
    \begin{align}
        \partial_\Lambda D^X(q) &= \left[ D^X(q)\right]^2 T\sum_\nu h^X_\nu(q)\left[\widetilde{\partial}_\Lambda \Pi^X_\nu(q)\right] h^X_\nu(q),\\
        \partial_\Lambda h^X_\nu(q) &= T\sum_{\nu'} \mathcal{L}^X_{\nu\nu'}(q)\left[\widetilde{\partial}_\Lambda \Pi^X_{\nu'}(q)\right] h^X_{\nu'}(q),\\
        \partial_\Lambda \mathcal{R}^X_{\nu\nu'}(q)& = T\sum_{\nu^{\prime\prime}} \mathcal{L}^X_{\nu\nu^{\prime\prime}}(q)\left[\widetilde{\partial}_\Lambda \Pi^X_{\nu^{\prime\prime}}(q)\right] \mathcal{L}^X_{\nu^{\prime\prime}\nu'}(q),
    \end{align}
\end{subequations}
where 
\begin{subequations}
    \begin{align}
        \Pi^X_\nu(q) &= \int_\bk \left(f^s_\bk\right)^2 \Pi^X_{(\bk,\nu)}(q),\\
         \mathcal{L}_{\nu\nu'}^X(q) &= \int_\bk\int_{\bk'} f^s_\bk f^s_{\bk'} \mathcal{L}^X_{(\bk,\nu),(\bk',\nu')}(q).
    \end{align}
\end{subequations}

We finally note that if the residual vertex develops a strong momentum dependence, considering a limited number of form factors is not justified a priori and has to be verified.

\subsection{\texorpdfstring{$d$-wave}{d-wave} pairing channel}
\label{Sec: dwave pairing channel methods}
In the doped regime we include $d$-wave pairing fluctuations into our parametrization of the two-particle vertex by
\begin{equation}
    \begin{split}
        V(k_1,k_2,k_3) &= \Lambda_\text{$U$irr}^{\mathrm{loc}}(\nu_1,\nu_2,\nu_3) - 2U\\
                        &+ \mathcal{S}_{\nu_1 \nu_3}(k_1+k_2) - \mathcal{D}_{k_1 k_3}(k_1+k_2)\\
                        &+ \mathcal{M}_{\nu_1 \nu_3}(k_2-k_3) +\frac{1}{2} \mathcal{M}_{\nu_1 \nu_4}(k_3-k_1)\\
                        &+\frac{1}{2} \mathcal{C}_{\nu_1 \nu_4}(k_3-k_1),
    \end{split}
    \label{eq: vertex decomposition Dwave}
\end{equation}
where $\mathcal{M}_{\nu\nu'}(q)$, $\mathcal{C}_{\nu\nu'}(q)$, and $\mathcal{S}_{\nu\nu'}(q)$ are the the $s$-wave projections of the couplings of Eq.~\eqref{eq: bosonic decomposition}, and, in order to deal with a positive quantity, we have chosen $\mathcal{D}$ with a minus sign in front of it.
The function $\Lambda_\text{$U$irr}^{\mathrm{loc}}(\nu_1,\nu_2,\nu_3)$ represents the sum of $U$-irreducible diagrams at the DMFT level and therefore does not flow, that is in the doped regime we neglect the flow of rest functions in the magnetic, charge, and $s$-wave pairing channels. The $d$-wave pairing channel $\mathcal{D}_{kk'}(q)$ is given by
\begin{equation}
    \mathcal{D}_{kk'}(q) = \mathcal{D}_{\nu\nu'}(q) f^d_{\bk-\bq/2} f^d_{\bk^\prime-\bq/2},
\end{equation}
with the $d$-wave form factor $f^d_\bk = \cos k_x - \cos k_y$. In essence, the function $\mathcal{D}_{\nu\nu'}(q)$, for transfer momentum $\bs{q}=0$, represents the sum of all diagrams that are reducible in the two-particle-$pp$ channel and at the same time exhibit a $d$-wave symmetry in the dependency on secondary momenta $\bk$ and $\bk^\prime$. Due to the locality of the bare Hubbard interaction $U$, \textit{all} the above mentioned diagrams are $U$-irreducible, that is the $d$-wave pairing channel will consist exclusively of its rest function. Its flow equation reads 
\begin{equation}
    \partial_\Lambda \mathcal{D}_{\nu\nu'}(q) = T\sum_{\nu^{\prime\prime}} L^d_{\nu\nu^{\prime\prime}}(q) \left[\widetilde{\partial}_\Lambda \Pi^d_{\nu^{\prime\prime}}(q)\right] L^d_{\nu^{\prime\prime}\nu'}(q),
    \label{eq: d-wave flow equation}
\end{equation}
with
\begin{equation}
    \Pi^d_\nu(q) = \int_\bk \left(f^d_\bk\right)^2 G(k)G(q-k),
\end{equation}
and
\begin{equation}
    L^d_{\nu\nu^{\prime}}(q) = \int_\bk\int_{\bk'} f^d_\bk f^d_{\bk'} V(k,q-k,k'),
    \label{eq: Ld}
\end{equation}
where $k=(\bk,\nu)$, and $k'=(\bk',\nu')$. 
This corresponds to restrict ourselves to the pure $d$-wave first harmonic contribution. 
The flow equations for the Yukawa couplings and bosonic propagators of the magnetic, charge, and $s$-wave pairing channels are the same as in the previous section, with the difference that the feedback of the $d$-wave pairing channel on them is included through Eq.~\eqref{eq: vertex decomposition Dwave}. 
%
%
%
\section{Results at half filling}
\label{sec:results half filling}
To illustrate the potential of our SBE implementation of the DMF\textsuperscript2RG scheme (see Appendix~\ref{app: technical aspects} for the details on the numerical implementation), we first consider the testbed case of half filling.  The expected physical behavior, encoded in the response functions of the different channels, is quite clear in this situation, due to the underlying particle-hole symmetry of the problem.

\new{  In the spirit of the SBE, for most of the calculations presented in this section we have {\sl neglected} the flow of the 
nonlocal 
corrections to the rest functions ($\mathcal{R}^X=0$).  Such an SBE-based simplification allows for a substantial gain in memory cost, compared to the conventional formalism, as the Yukawa couplings and the screened interactions depend on less arguments than the full vertex. This turns also into a gain in computational time, which is however not trivial to estimate.  At the same time, it is important to underline that, within the DMF$^2$RG framework, this approximation corresponds to neglect {\sl only} the flow describing the 
nonlocal 
corrections to the rest function, as its purely local part is per construction 
included nonperturbatively via the DMFT initial condition. 

The quality of our SBE-based approximation to the DMF$^2$RG flow has been eventually tested by performing specific calculations {\sl with} and {\sl without} the inclusion of the flow for the nonlocal corrections to $\mathcal{R}^X$ and by comparing the obtained results. The quite positive outcome of this test has been explicitly discussed in Sec.~\ref{subsec: rest}.
}


\subsection{Susceptibilities}
\label{subsec: susc}

\subsubsection{Weak-coupling fRG and \texorpdfstring{DMF\textsuperscript2RG results}{DMF2RG}}
\begin{figure*}[t]
    \centering
    \includegraphics[width = 1.0 \textwidth]{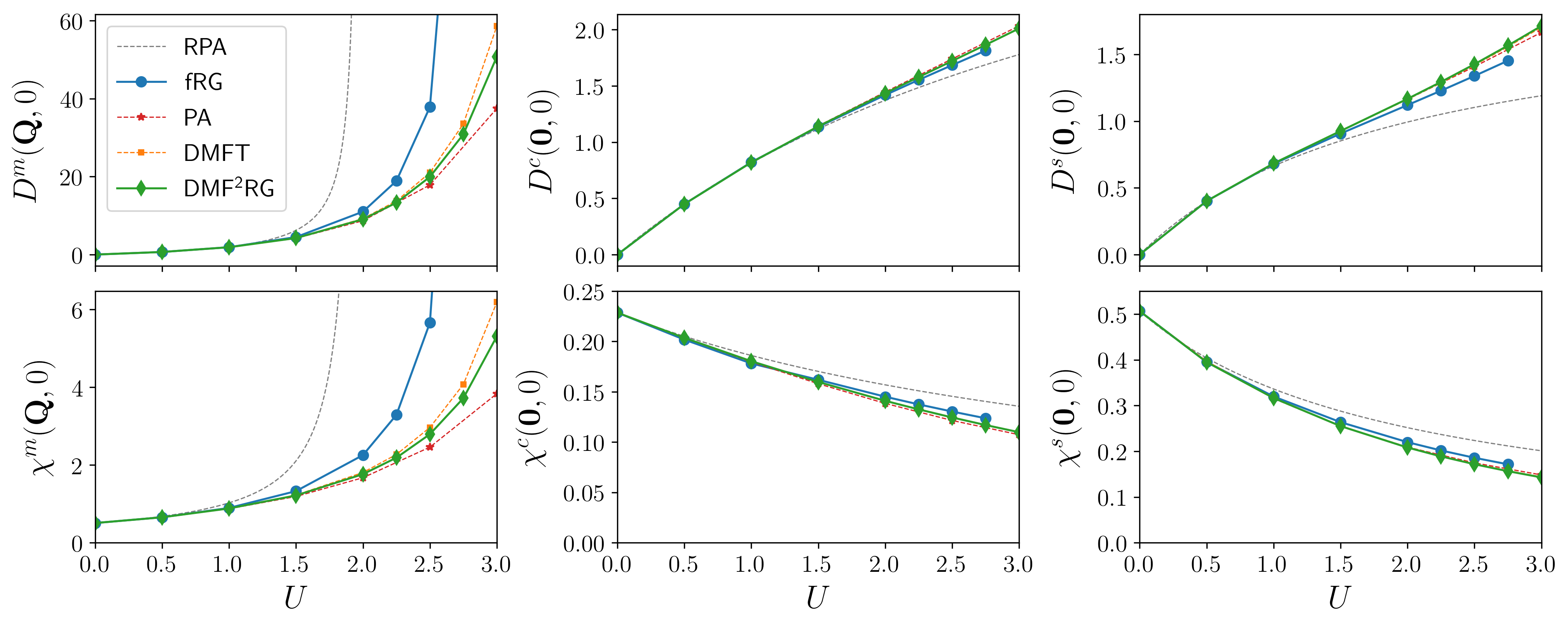}
    \caption{Comparison between RPA, fRG, PA, 
    DMFT and DMF\textsuperscript2RG for 
    at $n=1$ and for $T=0.2t$ ($t'=0$).  
    From left to right: magnetic $D^m$, charge $D^c$, and $s$-wave pairing $D^s$ screened interactions and susceptibilities, 
    respectively, at zero transfer frequency. The magnetic screened interaction is shown at $\bs{Q}=(\pi,\pi)$, while the others at $\bs{0}=(0,0)$.}
    \label{fig: fRG vs DMF2RG}
\end{figure*}
%
\new{
We start by comparing different methods in the weak-coupling regime, where also conventional perturbative 
 approximations can be applied as a benchmark. In particular, for the calculations of physical susceptibilities presented in  Fig.~\ref{fig: fRG vs DMF2RG}, we have exploited: i) the 
random phase approximation (RPA); ii) the \new{ one-loop} fRG,  
where we have also neglected all the $\mathcal{R}^X$ functions \footnote{Evidently, this simplification does not cause significant deviations from the conventional fRG in the weak-coupling regime. We also note that the latter corresponds to the SBE formalism with the inclusion of the rest functions, see App.~\ref{app: fermionic weak coupling}) }; iii) the multiloop extension of the fRG~\cite{Tagliavini2019}, which amounts to the resummation of \new{ all diagrams of the parquet approximation (PA), thus fulfilling the Mermin-Wagner theorem  in 2D \cite{Bickers2004}}; iv) the DMFT; 
and v) the DMF\textsuperscript2RG, where the \emph{nonlocal} $\mathcal{R}^X$ functions are  neglected. For the DMF\textsuperscript2RG, \new{ similarly as for the fRG, neglecting the nonlocal rest functions  does not produce sizable deviations in this first parameter regime considered} (see App.~\ref{app: fermionic weak coupling}).
}
In order to allow for a quantitative comparison of all the approaches considered on an equal footing, these specific weak-coupling fRG and DMF\textsuperscript2RG calculations have been performed including the feedback of the self-energy in the RG flow. 

In the upper panels of Fig.~\ref{fig: fRG vs DMF2RG}, we show the screened interactions $D^X(\bs{q},\Omega)$ for zero transfer frequency $\Omega=0$ and a specific momentum transfer. The lower panels show the corresponding susceptibilities that are connected with the screened interactions via Eq.~\eqref{eq: D}. 
As expected, for the lowest values of $U$,  the different calculations display a very good agreement, which is gradually reduced for larger values of the coupling. 
Consistent to the physics expected in the half-filled Hubbard model, all methods outline predominant antiferromagnetic (AF) correlations for the coupling values considered, though they differ on the quantitative level. %
In particular, the RPA yields, for all values of $U$, the largest magnetic effects. These get suppressed, to different amounts, in the results of the other approaches, because they all capture, differently from RPA, the competition with the other channels.
In particular, a sizable suppression of the AF susceptibility for $U > t$ is clearly observed, within fRG,  already at the level of the $1\ell$ calculations.
Not surprisingly, the suppression due to the channel interference becomes stronger in the PA results, reflecting the inclusion of the corresponding fluctuation effects at all loop orders. Due to the fully two-particle irreducible diagrams which are not included in the PA, it appears to systematically underestimate $\chi_m$ as compared to numerically exact determinant quantum Monte Carlo data  \cite{Hille2020,Review2021} for larger values of $U$. 
At the given temperature $T=0.2t$, 
suppression effects are observed also in the DMFT data.
The 
reduction of the magnetic fluctuations  with respect to the $1\ell$ fRG is ascribable to self-energy effects, which reduce more quickly the electronic coherence in DMFT than in fRG. 
This is consistent with the observation that DMF\textsuperscript2RG results closely resemble the ones of DMFT in this parameter regime. Indeed, the nonlocal correlations included at the $1\ell$ level on top of the DMFT results, induce, as expected, a further reduction of the magnetic correlations, but their quantitative impact remains marginal for $U<3t$. 
We note that nonlocal multiloop corrections, which recover the PA result, have a stronger impact on the suppression than the local DMFT ones at $1\ell$ level. 

We turn now to discuss the other sectors, which are secondary in the physics of the half-filled model. The corresponding results are reported in the central and right panels of Fig.~\ref{fig: fRG vs DMF2RG}, showing that the correlation effects beyond RPA are overall weaker and significantly less dependent by the particular approach chosen than in the magnetic sector. In particular, the slight suppression of the uniform charge and $s$-wave static susceptibilities induced by the interplay with their complementary channels get reflected in a slight increase of corresponding screened interaction  with respect to the RPA values.
%

We note that including the flow of the fermionic rest functions $\mathcal{R}^X$ 
leads only to negligible 
corrections of the results 
shown in Fig.~\ref{fig: fRG vs DMF2RG}, see  Appendix~\ref{app: fermionic weak coupling}.
In the weak-coupling regime, the screened interactions and susceptibilities obtained without $\mathcal{R}^X$ correctly describe the physical behavior, justifying the application of the SBE approximation.

\subsubsection{Intermediate- to strong-coupling \texorpdfstring{DMF\textsuperscript2RG}{DMF2RG} results}

\begin{figure}[t]
    \centering
    \includegraphics[width = 0.475 \textwidth]{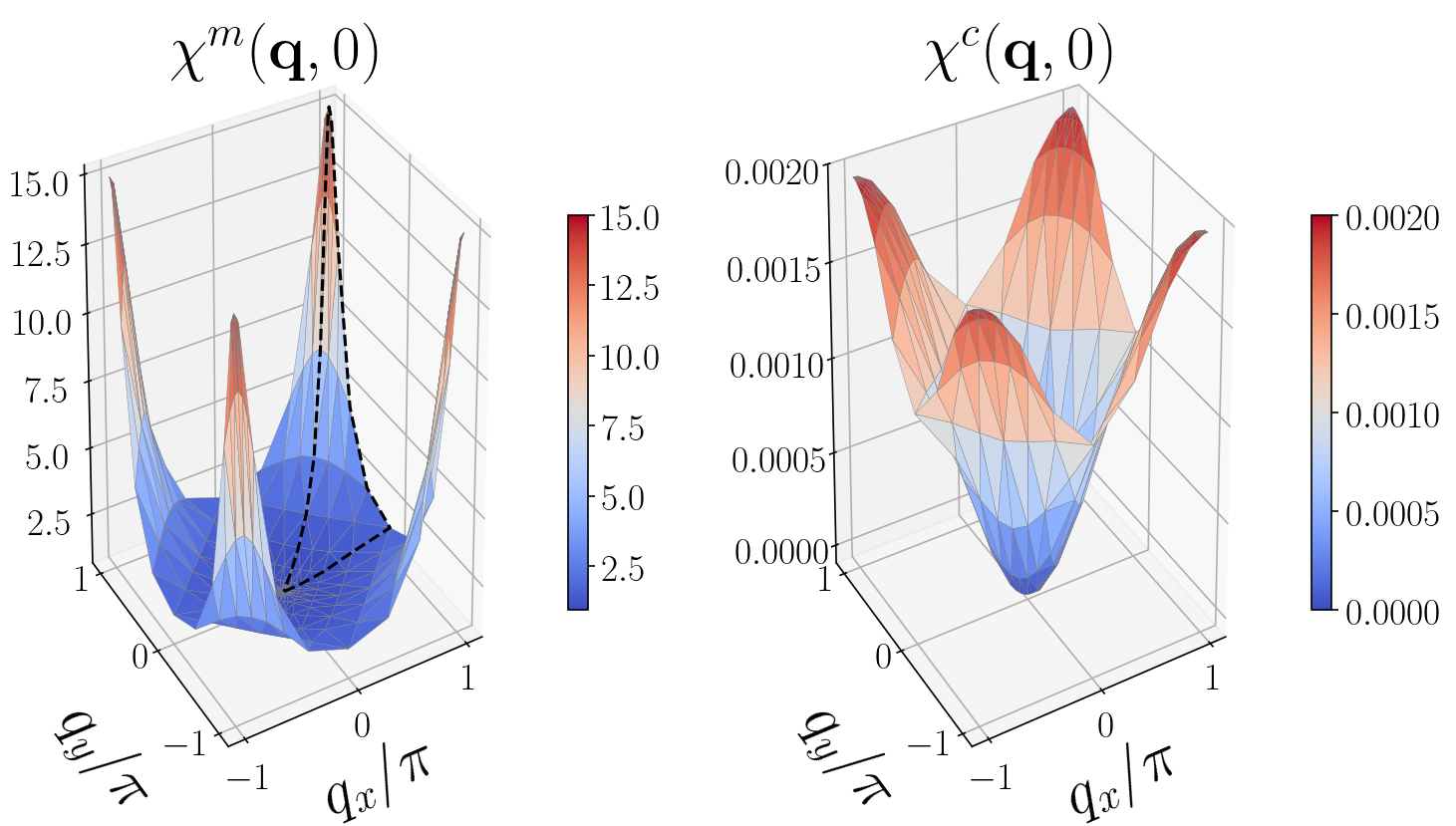}
    \caption{Magnetic and charge susceptibilities in the BZ for transfer frequency, at $n=1$ and for $U=16t$ ($t'=0$), $T=0.286t$.}
    \label{fig: 3d chis}
\end{figure}

After this preliminary comparison, we move towards the more challenging parameter regime of intermediate to strong couplings.
Hence, from now on, we will focus on DMF\textsuperscript2RG results for the physical susceptibilities of the different sectors, whereas, for simplicity,  we turn off the flow of the nonlocal self-energy corrections. Especially in the strong-coupling regime, this simplification is not expected \cite{Rohringer2011,Rohringer2016,Rohringer2018,Vilardi2019} to affect the final results for the response functions in a significant way.
We start our analysis of the strong-coupling regime, by presenting in Fig.~\ref{fig: 3d chis}  our DMF\textsuperscript2RG results for the whole momentum dependence of the static susceptibilities (at zero bosonic frequency) in the magnetic and in the charge sectors for the highest coupling considered $U=16t$, which is  significantly beyond the critical interaction value of the Mott-Hubbard metal to insulator transition of DMFT ($U_{\rm MIT}(T\!=\!0) \sim 12 t$), and for a temperature slightly above 
the DMFT N\'eel temperature (see leftmost panel of Fig.~\ref{fig: rest norest} for the precise location in the DMFT phase-diagram).
Note that at half filling particle-hole symmetry implies $\chi^s(\bf{q},\omega)=\chi^c(\bf{q}+\bf{Q})$ for the pairing susceptibility, with $\bf{Q}=(\pi,\pi)$. 
The magnetic susceptibility exhibits a very pronounced peak around momentum $\bf{q}=(\pi,\pi)$, a hint of strong AF fluctuations. This indicates that, similarly as for weak-coupling data, the nonlocal correlations captured by the $1\ell$ DMF\textsuperscript2RG do not suppress the ordering temperature of DMFT in a significant way. 
At the same time, the static response in the charge (and pairing) sector bears very clear hallmarks of the strong-coupling physics. Except for a residual momentum modulation, these response functions appear massively suppressed (note the different order of magnitude of the scales in the two panels), reflecting the almost insulating nature of the Mott-Hubbard physics at finite $T$.

%
Following the analysis of Ref.~\cite{Chalupa2021}, it is instructive to relate the results presented so far to the behavior of the corresponding generalized susceptibilities, which describe the underlying fermionic scattering processes.
We recall that, in general \footnote{This rigorously holds for the exact solutions of the problem and/or for approximations based on a definite subset of diagrams, such as RPA, PA, DMFT. At the level of a truncated (1$\ell$) fRG/DMF\textsuperscript2RG, deviations between the susceptibilities computed via the flow and those computed by summing the internal frequencies (post-processing) may occur \cite{Kugler2018_I,Tagliavini2019}.}, the susceptibilities $\chi^X(q)$ describing the physical response of the system  can be obtained by the generalized ones $\chi^X_{\nu \nu'}(q)$ by summing over all the fermionic Matsubara frequencies $\nu, \nu'$.
%
In our notation, the explicit definition reads, also referred to as "post-processing": 
\begin{subequations}
    \begin{align}
       &\chi^m_{\nu \nu'}(q) = \beta\Pi^m_\nu(q)\delta_{\nu\nu'} + \Pi^m_\nu(q) L^m_{\nu\nu'}(q) \Pi^m_{\nu'}(q), \\
       &\chi^c_{\nu \nu'}(q) = -\beta\Pi^c_\nu(q)\delta_{\nu\nu'} - \Pi^c_\nu(q) L^c_{\nu\nu'}(q) \Pi^c_{\nu'}(q),  
    \end{align}
    \label{eq: chi generalized}
\end{subequations}
where
\begin{equation}
    L^X_{\nu\nu'}(q) = h^X_\nu(q) D^X(q) \Bar{h}^X_{\nu'}(q) + \mathcal{L}_{\nu\nu'}^X(q).
    \label{eq: chi L}
\end{equation}
From a numerical side, we note that a larg number of 
Matsubara frequencies are required for the internal summation 
in order to account correctly for the 
high-frequency asymptotics. A restricted frequency summation yields unphysical negative values in the charge response function due to the formation of the local moment  \cite{Chalupa2021}.
\begin{figure}[t]
    \includegraphics[width = 0.475 \textwidth]{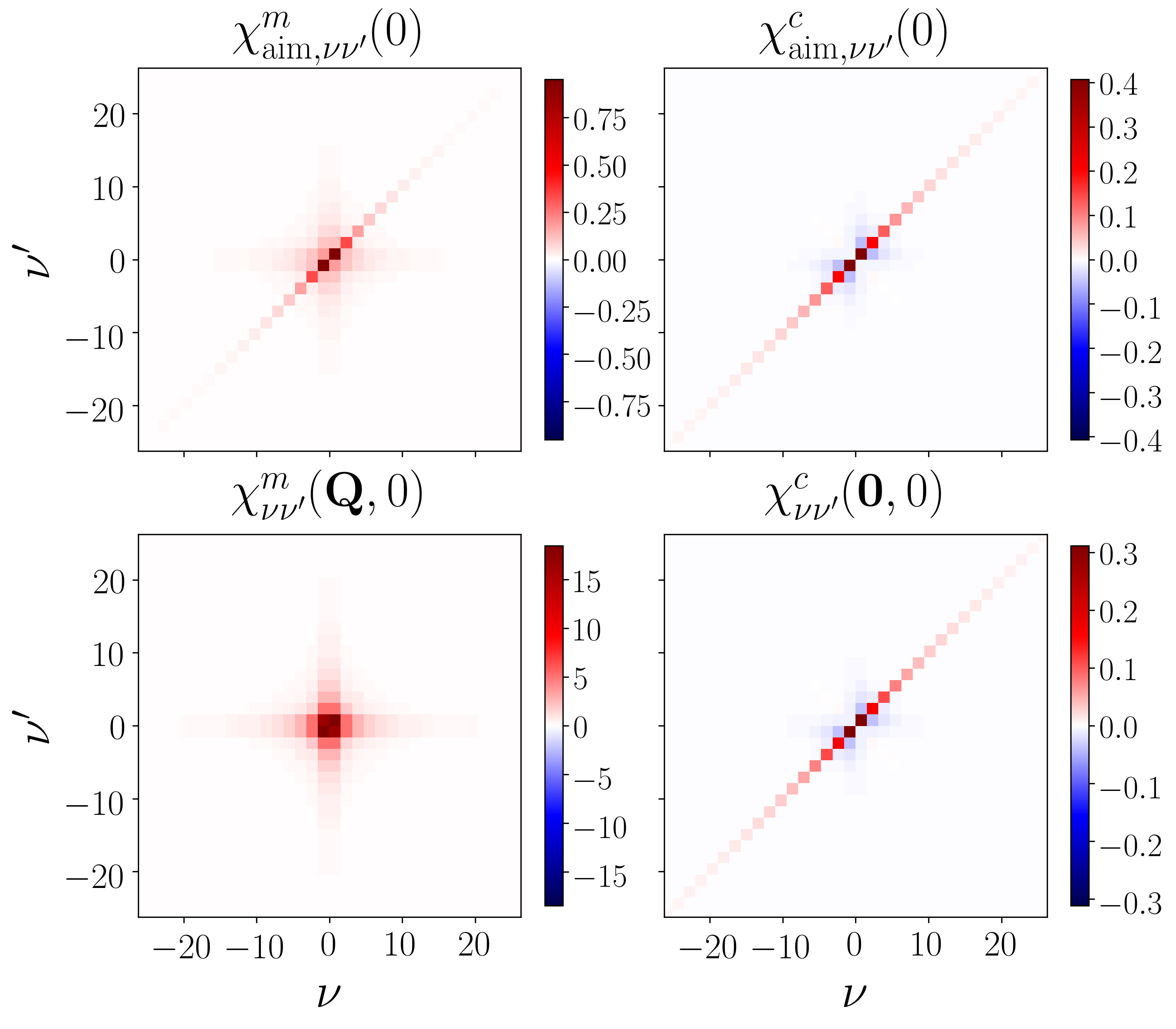}
    \caption{Generalized charge and magnetic susceptibilities as obtained from the self-consistent AIM (top) and the DMF\textsuperscript2RG (bottom), at half filling and $U=4t$ ($t'=0$), $T=0.25t$.}
    \label{fig: chi4 U4t}
\end{figure}
\begin{figure}[b]
    \centering
    \includegraphics[width = 0.475 \textwidth]{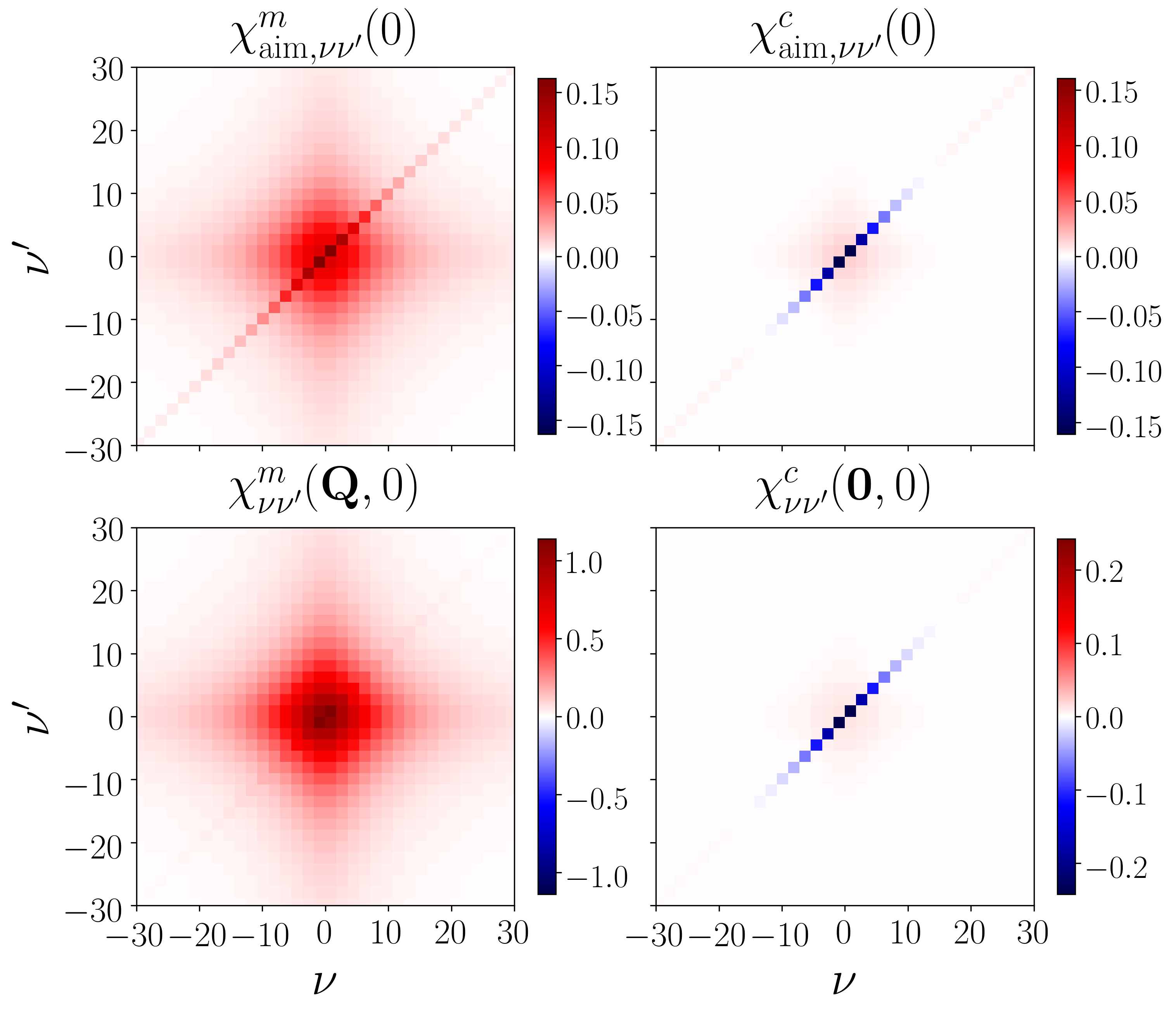}
    \caption{Same as Fig.~\ref{fig: chi4 U4t} for $U=16t$ and $T=0.286t$.}
    \label{fig: chi4 U16t}
\end{figure}

For illustrative reasons, we here explicitly discuss the 
 $\Omega=0$ case 
of the generalized charge and magnetic susceptibilities for two rather different interaction values, namely $U=4t$ and $U=16t$, which correspond, in DMFT, to a Fermi-liquid paramagnetic metallic (PM) and Mott-Hubbard paramagnetic insulating (PI) regime, respectively.
The corresponding data are shown in Figs.~\ref{fig: chi4 U4t} and \ref{fig: chi4 U16t}, where we report the on-site \footnote{Precisely, these are generalized susceptibilities of the auxiliary AIM associated to the corresponding self-consistent DMFT solution. We recall that in the limit of $d\rightarrow \infty$ the physical susceptibilities computed via the double Matsubara summation exactly coincide with the local (momentum-summed) susceptibilities of the lattice.} generalized magnetic (left panels) and charge (right panels)  DMFT solutions used as input of our DMF\textsuperscript2RG calculations in the  upper panels and the corresponding DMF\textsuperscript2RG results for the uniform ($\bf q=0$) generalized susceptibility in the lower ones.
As discussed in Ref.~\cite{Chalupa2021}, the frequency dependence of generalized {\sl charge} susceptibilities represents a sensitive compass for identifying the underlying physics in fundamental many electron models, since it directly describes the impact of correlations on the electronic mobility and unveils its link to the changes of magnetic response.
A first glance to the data of Figs.~\ref{fig: chi4 U4t}-\ref{fig: chi4 U16t} shows that the qualitative modifications in the frequency structures of the generalized susceptibilities occurs indeed in the charge sector. 
In particular, sign changes in relevant frequency structures of the generalized charge susceptibility take place from weak  ($U=4t$) to strong coupling ($U=16t$), reflecting important differences of the dominating physical mechanisms at play in the two regimes.

For $U=4t$ shown in Fig.~\ref{fig: chi4 U4t}, the leading frequency dependence in the generalized charge susceptibilities of both the AIM (DMFT) and DMF\textsuperscript2RG results appears in the main diagonal structure. This assumes {\sl positive} values, larger at low frequencies $\nu=\nu'$ and slowly decaying for larger $\nu=\nu'$ values. 
This feature is a typical hallmark of the metallic physics in the perturbative regime, as it arises from the bubble term contribution $\chi^0_{\nu \nu'}=-\beta \Pi^c_{\nu \nu'}(\Omega=0)= -\beta G(\nu) G(\nu') \delta_{\nu \nu'} >0 $, built upon a metallic $G(\nu)$. The role of vertex corrections, for $U=4t$, appears merely quantitative, yielding an overall slight suppression (enhancement) of all entries of  $\chi^{c}_{\nu\nu'}$ ($\chi^{m}_{\nu\nu'}$), responsible for the emergence of small negative (positive) off-diagonal contributions (faint bluish/reddish color for small $\nu \neq \nu'$ in the right/left upper panels of Fig.~\ref{fig: chi4 U4t}). The moderate size of the vertex corrections is 
highlighted by the comparison of the two channels, whose results are both dominated by a positive diagonal frequency structure.
Due to the proximity to an AF instability, the inclusion of nonlocal correlations in DMF\textsuperscript2RG strongly affects the generalized susceptibility of the predominant magnetic channel, which gets significantly enhanced with respect to the corresponding AIM data. At the same time, the impact of nonlocal correlations appears marginal in the charge sector.

The data computed at $U=16t$ (Fig.~\ref{fig: chi4 U16t}) display relevant differences, which are induced by significant vertex correction effects. 
In the magnetic channels these drive an overall enhancement of the generalized susceptibility (diffuse reddish zone clearly visible in the left panels), which gradually  overcomes the residual diagonal  contribution of the bubble term. Such generalized enhancement of $\chi^{m}_{\nu \nu'}$ encodes the typical Curie behavior of the local magnetic response in this regime.
At the same time, the corresponding vertex corrections in the charge sector strongly suppress the physical response, by flipping the sign of the diagonal entries of $\chi^c_{\nu \nu'}$ up to an energy scale of order $U$. Specifically, the largely negative low-frequency diagonal structure visible in the right panels of Fig.~\ref{fig: chi4 U16t} is directly responsible  \cite{Gunnarsson2016,Chalupa2021} for freezing the local density fluctuations in the Mott PI regime \footnote{For $U=16t$, the negative values in the main diagonal are 
counterbalanced by the positive contributions at high frequencies, giving rise to an overall positive charge response.}.  Physically, it can be interpreted \cite{Chalupa2021} as the charge counterpart of the local moment formation in the magnetic sector. 

We note that the strong differentiation of the charge and magnetic response, which is a typical hallmark the Mott PM phase, has a clear nonperturbative origin: The {\sl negative} diagonal structure emerging at low-frequency is associated to negative eigenvalues of $\chi^c_{\nu\nu'}$. This implies that sign flips  with respect to the (positive) eigenvalues of the weak-coupling regime must occur by increasing $U$. By any vanishing eigenvalue, however, the matrix $\chi^c_{\nu\nu'}$ becomes singular leading to multiple divergences  \cite{Schaefer2013,Janis2014,Gunnarsson2016,Ribic2016,Chalupa2018,Vucicevic2018,Springer2020} of the irreducible vertex function and to the corresponding  \cite{Gunnarsson2017} problem of the multivaluedness  \cite{Kozik2015,Stan2015,Rossi2015,Tarantino2017,Kozik2020} of the Luttinger-Ward functionals, which have been extensively discussed in the recent literature. 
As these singularities cannot be captured by any perturbative approach (including truncated fRG and PA) \cite{Chalupa2021}, for a proper description of the intermediate to strong-coupling regime it is pivotal to include the associated nonperturbative physics  (such as local moment formation and its Kondo screening  \cite{Chalupa2021}) through a DMFT starting point, emphasizing the necessity of a  DMF\textsuperscript2RG treatment.
In this specific case, the DMF\textsuperscript2RG data (lower panels of Fig.~\ref{fig: chi4 U16t}) display a further increase of the generalized  magnetic susceptibility at ${\bf q=} (\pi, \pi)$, due to the proximity to the AF instability in the phase diagram, as well as a further suppression of the generalized charge susceptibility at ${\bf q}=0$ with respect to the AIM results, consistent with the incompressible nature of the Mott insulating ground state. 


\subsection{Comparison with conventional implementations}
\label{subsec: rest}

All calculations presented above have been performed neglecting the flow of the fermionic rest functions $\mathcal{R}^X$, consistently within the SBE framework.
While this approximation allows for a significant reduction of the numerical effort, it is  important to verify to what extent its application is justified in the different coupling regimes.

\begin{figure}[t]
    \centering
    \includegraphics[width = 0.475 \textwidth]{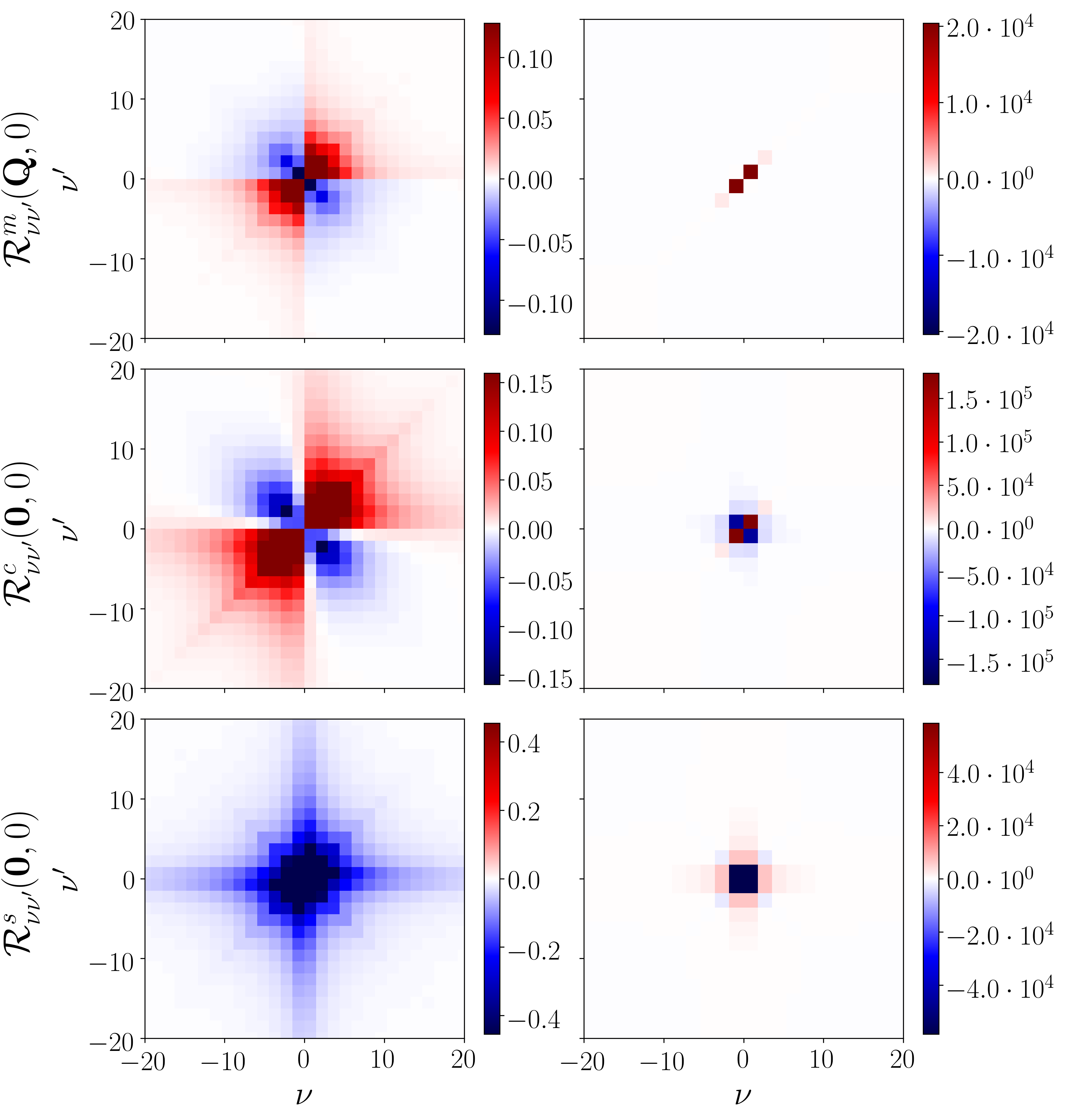}
    \caption{Rest functions magnetic, charge and pairing channels (from top to bottom) at $U=4t$,  $T=0.25t$ (left), and $U=16t$, $T=0.286t$ (right), for $n=1$ (and $t'=0$).}
    \label{fig: rest functions U4 U16}
\end{figure}

We briefly recall here that, when fully considering the flow of the rest functions $\mathcal{R}^X$, the SBE-based implementation becomes formally equivalent to the one used in the conventional fRG based on the 1PI vertex function, see Ref.~\cite{Vilardi2019} for the specific case of DMF\textsuperscript2RG \cite{Taranto2014}.
%
We first analyze the frequency dependence of the (previously neglected) $\mathcal{R}^X$ functions, evaluated for the same bosonic variables ($\bq,\Omega$) of the susceptibilities shown in the preceding subsection, at $U=4t$ and $U=16t$.  
The corresponding results are reported in Fig.~\ref{fig: rest functions U4 U16}. This highlights a key feature of the rest function, namely its characteristic decay to zero at large frequencies. 
Qualitatively, this reflects a general feature of the SBE diagrammatics \cite{Krien2019_I}: the high-frequency asymptotics of the two-particle vertex functions is fully captured by the
effective interactions, $D^X$, and by the Yukawa couplings, $h^X$. 
From a more quantitative perspective, the high-frequency decay appears particularly pronounced at large couplings: In contrast to $U=4t$, for $U=16t$ the rest function exhibits a significant frequency dependence only for the lowest Matsubara frequencies, which, at this temperature, represents the leading frequency dependence of the whole 1PI vertex function. 
Despite the large values of the rest functions at strong coupling, the single-boson contributions $h^X D^X h^X$ are of the same order or even larger over an infinitely broad frequency range. Furthermore, we checked that the $\mathcal{R}^X$ have a marginal feedback effect on the flow of the Yukawa couplings due to cancellations with the fermionic bubbles in the insulating phase. 

\begin{figure*}[t]
    \centering
    \includegraphics[width = 1.0 \textwidth]{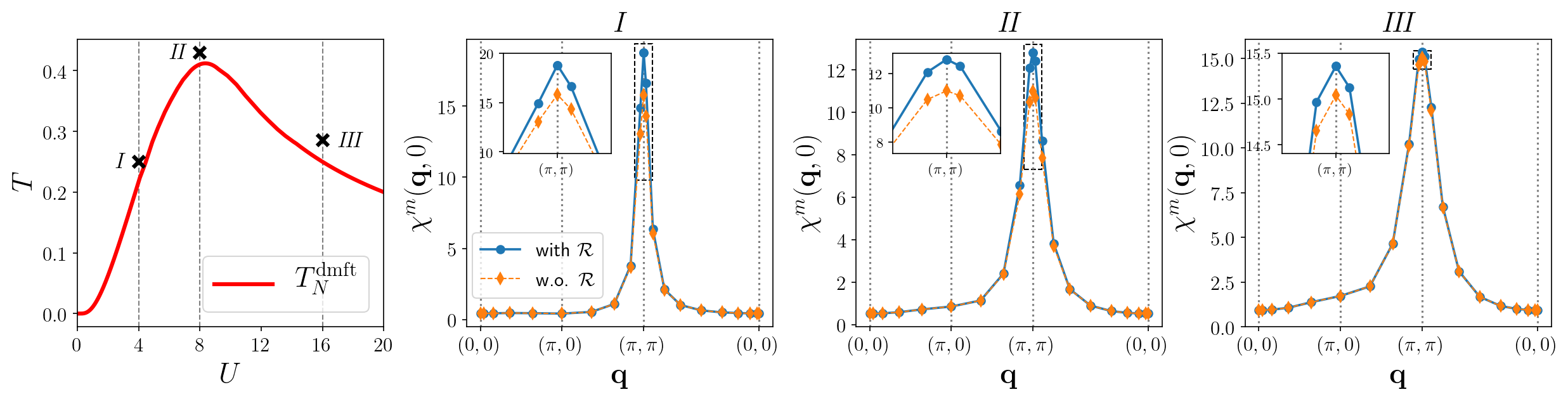}
    \caption{Left panel: N\'eel temperature as obtained from DMFT (left panel). Other panels: Magnetic susceptibility at 
    zero bosonic frequency 
    along the selected BZ path (shown as dashed line in Fig.~\ref{fig: 3d chis}), with and without the flow of the rest function $\mathcal{R}^X$, for $n=1$ ($t'=0$) and different values of $U$ (second, third, and forth panel). The parameters of each of the last three
    panels correspond to the crosses in the first panel, showing their location with respect to the N\'eel temperature.
    \new{For the point indicated as $I$ (shown in the second panel), we have $U=4t$, $T=0.250t$, for $I\!I$ (third panel) $U=8t$, $T=0.430t$, and for $I\!I\!I$ (fourth panel) $U=16t$, $T=0.286t$.}
    }
    \label{fig: rest norest}
\end{figure*}

\begin{figure}[b]
    \centering
    \includegraphics[width = 0.45 \textwidth]{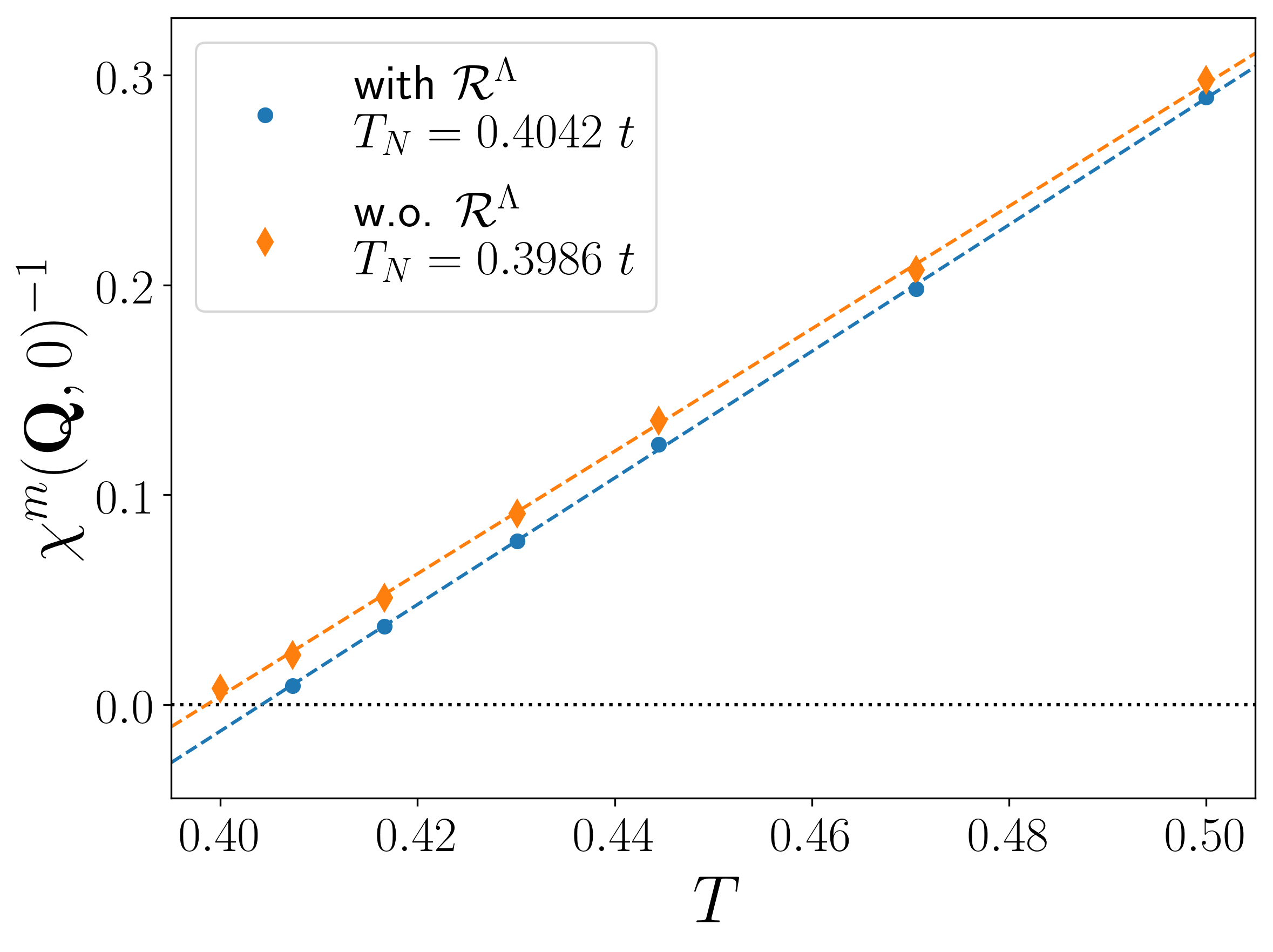}
    \caption{N\'eel temperature $T_{\rm N}$ at $U=8t$.}
    \label{fig: xi vs T}
\end{figure}

After examining the high-frequency decay properties of $\mathcal{R}^X$, which highlight the overall convenience of an SBE-based formalism, we turn now to the analysis of the specific impact that neglecting them has on the final results of our DMF\textsuperscript2RG calculations. 
In Fig.~\ref{fig: rest norest}, we report the susceptibilities of the predominant magnetic channels, computed with and without the inclusion of $\mathcal{R}^X$, for three different values of $U$ depicted on the phase diagram in the inset. The parameter set chosen is particularly relevant due to its proximity to the AF instability of the DMFT calculations, and the expected relevant contribution of large magnetic fluctuations. 
We note immediately that at half filling the magnetic susceptibilities obtained without $\mathcal{R}^X$ 
correctly describe the physical properties expected in the whole parameter region considered, including the strong-coupling Mott-insulating regime at $U=16 t$. 
Furthermore, the corrections to the magnetic response induced by the inclusion of  $\mathcal{R}^X$ in our DMF\textsuperscript2RG calculations appear to be overall rather marginal (see insets of Fig.~\ref{fig: rest norest}) and further decreasing at larger interaction values.
As a consequence of this, the inclusion of the rest function has also a minor impact on the determination of the temperature of the AF instability (N\'eel temperature $T_{\rm N}$), which can be finite in all approaches where the Mermin-Wagner theorem is 
violated. 
In Fig.~\ref{fig: xi vs T} the inverse magnetic susceptibility is shown as a function of the temperature, with and without the inclusion of $\mathcal{R}^X$. Both data-sets clearly display a linear mean-field like critical behavior, resulting in 
$T_{\rm N}=0.4042t$ and $T_{\rm N}=0.3986t$ respectively. These values, both slightly lower than the corresponding DMFT results, are quite close. 

Hence, the overall effect of neglecting  $\mathcal{R}^X$ appears to be marginal, even quantitatively, within the $1\ell$ DMF\textsuperscript2RG scheme, in particular for the dominant magnetic channel.
Not surprisingly, the corrections induced by $\mathcal{R}^X$ are even lower in the other channels: charge and pairing susceptibilities determined with and without the rest function  display almost no difference (not shown).

\subsection{Yukawa couplings}
\begin{figure}[b]
    \centering
    \includegraphics[width = 0.5 \textwidth]{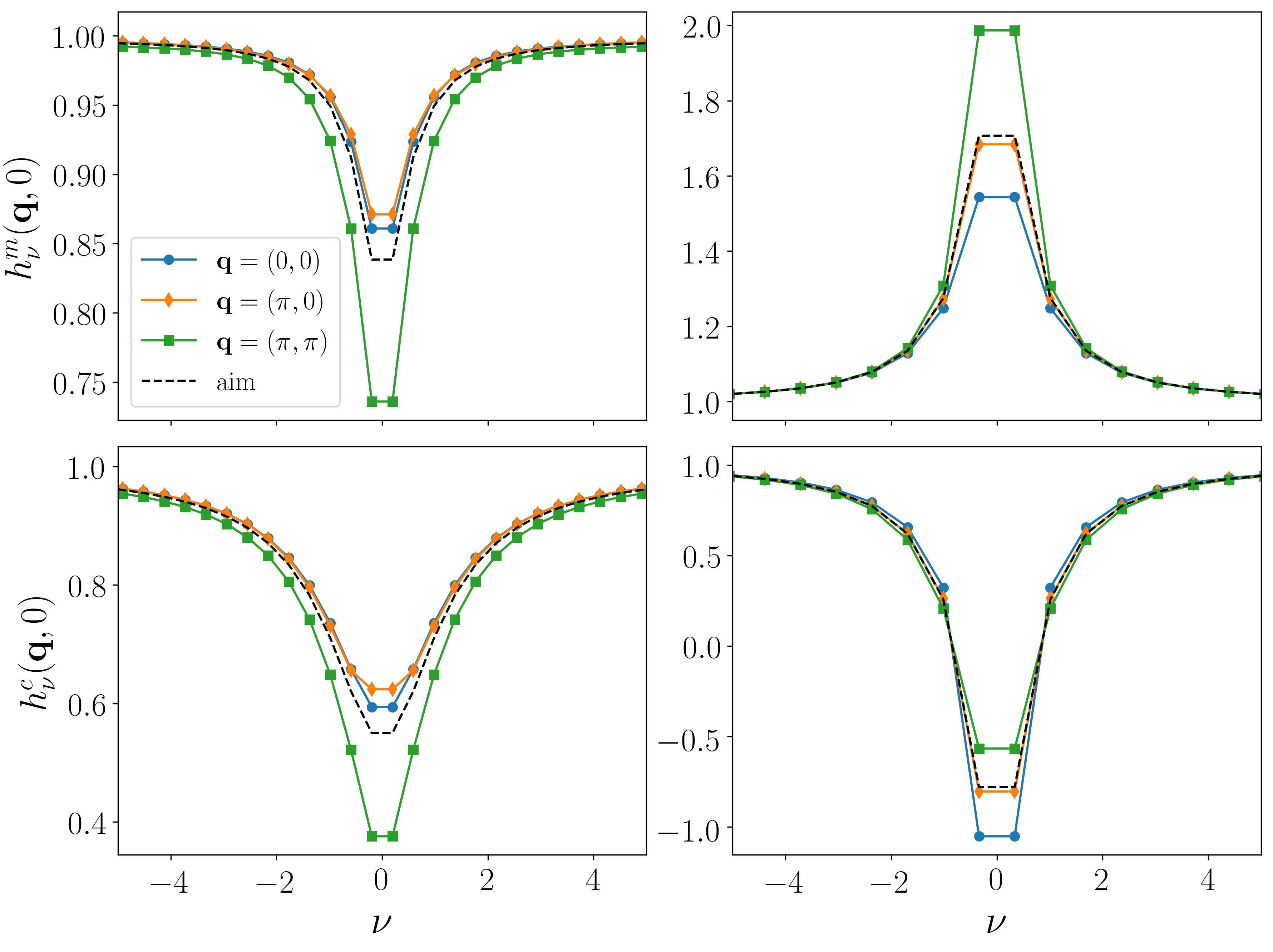}
    \caption{Yukawa couplings at zero bosonic frequency and 
    $U=4t$,  $T=0.25t$ (left), and $U=16t$, $T=0.286t$ (right), for $n=1$ (and $t'=0$).}
    \label{fig: yukawas half filling}
\end{figure}

In this subsection, we focus on the second main ingredient of the SBE formalism, that is the (fermion-boson) Yukawa couplings, s. also Refs.~\cite{VanLoon2018,Krien2019_II,Harkov2021,Harkov2021a}. The corresponding results for the magnetic and the charge sector are shown in Fig.~\ref{fig: yukawas half filling}, while the $s$-wave pairing one can be obtained from the relation $h^s_\nu(\bq,\Omega)=h^c_\nu(\bq+\bQ,\Omega)$, with $\bQ=(\pi,\pi)$, valid when the particle-hole symmetry is realized. 

At weak coupling ($U=4t$) we observe for both channels a moderate dependence of the static ($\Omega\!=\!0$) Yukawa couplings on the fermionic frequency $\nu$. This results in an overall slight suppression of the asymptotic value of the coupling ($h^X\sim 1$) and encodes the metallic screening effects  \cite{Toschi2007,Katanin2009} 
occurring in the proximity of the Fermi level.
The inclusion of nonlocal effects in DMF\textsuperscript2RG induces mild corrections with respect to the DMFT (AIM) results (black dashed line).  In particular, it can be noted that, in both sectors, the fluctuations associated to a transfer momentum of $\bf{q}=(\pi,\pi)$  tend to magnify the frequency dependence displayed by the local DMFT (AIM) calculations, while the uniform ones ${\bf q}=(0,0)$ have the opposite effect. 

At strong-coupling, the situation appears essentially reversed. The DMFT (AIM) calculations performed at $U=16t$ show  a much stronger dependence of the couplings on the fermionic Matsubara frequency and a dycothomic behavior in the two channels -- a typical hallmark \cite{Chalupa2021,Reitner2020} of large vertex corrections.
In particular, the magnetic Yukawa coupling gets significantly enhanced at low-frequencies with respect to the $h^m=1$ value, whereas the largest value is found in DMF\textsuperscript2RG for the AF ordering wavevector. The charge channel, instead, exhibits a strong suppression, that even makes $h^c$ {\sl negative} for the smallest frequencies. Here, the strongest suppression, according to our DMF\textsuperscript2RG results occurs for $\bq=(0,0)$, consistent to a vanishing isothermal compressibility in the Mott-Hubbard ground state  \cite{Georges1996,Reitner2020,VanLoon2020}. Physically, these opposite behaviors should be regarded as the nonperturbative fingerprints \cite{Chalupa2021} of the local moment in the Yukawa couplings, since its formation enhances the system's response to a magnetic perturbation and, at the same time, freezes the fluctuations of the electronic density.


\section{Results at finite doping}
\label{sec:results doped regime}
To showcase the validity of our method in a physically more relevant parameter regime, we have also 
performed DMF\textsuperscript2RG calculations for the non particle-hole symmetric case.
In particular, we consider an intermediate-coupling
of $U=8t$ and a fairly low temperature $T=0.044t$,  doping the system to the filling $n=0.82$, and frustrating the magnetic correlations with a next to nearest neighbor hopping $t^\prime=-0.2t$. 
As the Mermin-Wagner theorem is not fulfilled at the level of the $1\ell$ truncation of the DMF\textsuperscript2RG \cite{Taranto2014,Kugler2018_I,Tagliavini2019,Vilardi2019}, we observe a magnetic instability in the considered parameter region. Hence, in order to highlight the relation between superconductivity and magnetism, we stop the flow before reaching the final scale, as customarily done in the fRG treatment of pseudocritical transitions. 
In particular, the divergence of the magnetic propagator has been prevented, by stopping the flow when the magnetic propagator $D^m(q)$ exceeds a value of $8\times 10^3 t$, which, for $U=8t$, corresponds to a susceptibility of $\sim 120/t$, resulting in a stopping scale $\Lambda_\mathrm{cr} \simeq 0.067 t$. 
\new{Although the criterion for stopping the flow is arbitrary, close to an instability the screened interactions diverge quickly, implying that even large changes in the stopping criterion translate into tiny differences in the stopping scales. This makes the results weakly dependent on the specific choice of the condition to interrupt the flow. \new{ Eventually, as the PA fulfills the Mermin-Wagner theorem in 2D \cite{Bickers2004}, we expect the future inclusion of multiloop corrections to} suppress the magnetic screened interaction, allowing to continue the flow down to lower scales (to $\Lambda=0$ at loop convergence).}

To account for $d$-wave pairing correlations, we have included the flow of the $d$-wave pairing channel (consisting exclusively of its rest function), as explained in Sec.~\ref{Sec: dwave pairing channel methods}. 
We neglect the rest functions of the magnetic, charge and $s$-wave pairing channels since these appear to yield only minor corrections, 
together with nonlocal corrections to the self-energy. 

\begin{figure*}[t]
    \centering
    \includegraphics[width = 1.0 \textwidth]{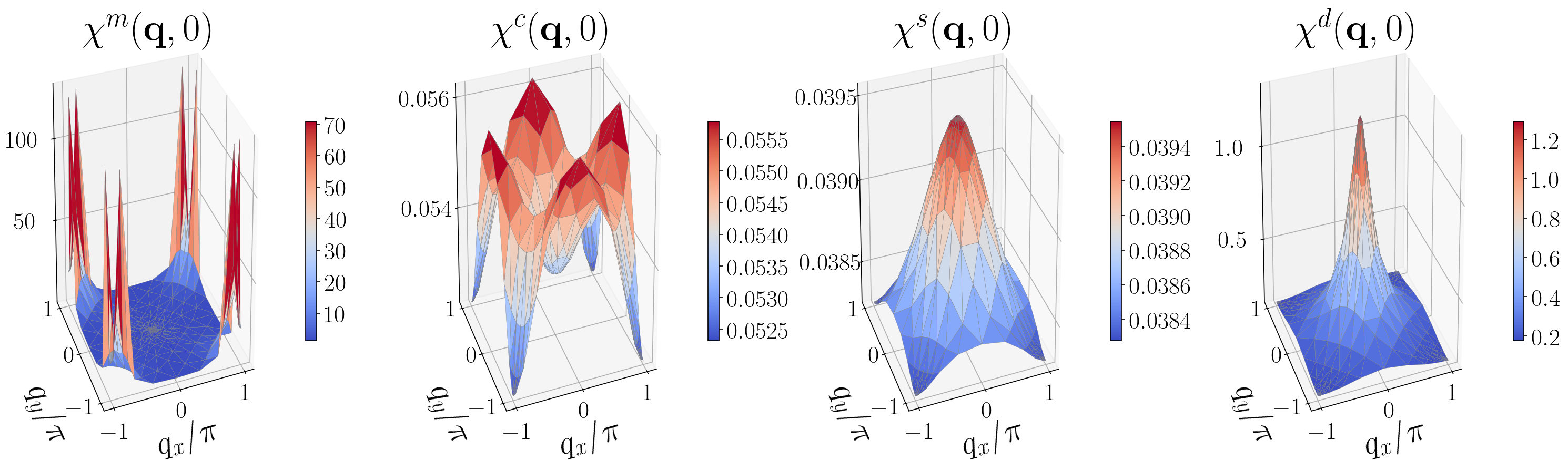}
    \caption{Magnetic, charge, $s$- and $d$-wave pairing susceptibilities (from left to right) at zero bosonic frequency, as functions of the momentum $\bq$, for
    $n=0.82$, $U=8t$, $t^\prime=-0.2t$, and $T=0.044t$, determined at the stopping scale $\Lambda_\mathrm{cr} \simeq 0.067t$. 
    }
    \label{fig: dwave susc 3D}
\end{figure*}

\subsection{Susceptibilities}

Here, we will first discuss the results obtained for the magnetic $\chi^m$, charge $\chi^c$, and $d$-wave pairing $\chi^d$ susceptibilities computed at the stopping scale $\Lambda_\mathrm{cr}$. 
While the first two can be directly extracted from the respective propagators, see Eqs.~\eqref{eq: Dm from chi} and \eqref{eq: Dc from chi}, the $d$-wave pairing one is computed through
\begin{equation}
    \chi^d(q) = T \sum_\nu \Pi^d_\nu(q) + T^2\sum_{\nu\nu'} \Pi^d_\nu(q) L^d_{\nu\nu'}(q) \Pi^d_{\nu'}(q), 
    \label{eq: chid postprocessed}
\end{equation}

where we have neglected terms of the type $T^2\sum_{\nu\nu'} \Pi^{sd}_\nu(q) L^s_{\nu\nu'}(q) \Pi^{sd}_{\nu'}(q)$, with $\Pi^{sd}_\nu(q)$ the mixed $s$-$d$-wave pairing bubble, as it is nonzero only for $q\neq0$ and generally rather small \footnote{S. Heinzelmann, private communication.}.
In Fig.~\ref{fig: dwave susc 3D}, we show the 
static  susceptibilities in the whole Brillouin zone. 
Specifically, 
the magnetic susceptibility is close to a divergence 
at momenta $(\pi, \pi \pm 2\pi \eta)$ (as well as, by symmetry, at $(\pi\pm 2\pi \eta,\pi )$) with $\eta\simeq 0.08$, which indicates the tendency towards an \textit{incommensurate} magnetic instability \cite{Metzner12,Halboth00A,Halboth00B,Yamase16}. 
We recall that magnetic long-range orders with incommensurate wave vectors have been found in several mean-field studies  \cite{Schulz90,Dombre90,Fresard91,Igoshev10} of the Hubbard model at finite doping. Similar conclusions have been also drawn when including fluctuations beyond the static mean-field, for example, by means of expansions in the hole-density  \cite{Shraiman89,Chubukov92,Chubukov95,Kotov04}, or exploiting 
extensions of DMFT 
\cite{Fleck99,Schaefer17}. 
%
DMFT calculations 
suggest that the ordering wave vector is related to the Fermi
surface geometry not only at weak, but also at strong coupling \cite{Vilardi18}.

Differently, the charge susceptibility exhibits a rather weak dependence on $\bq$, that is, only moderate deviations are observed from the local description of DMFT in this sector. We expect this feature not to depend on the fact that we have a small finite $\Lambda_{\textrm{cr}}$. 
At a closer inspection, anyway, one notes two peaks located at $(\pi,0)$ and $(0,\pi)$. These signal the presence of mild charge-stripes correlations.

Eventually, we focus on the $d$-wave pairing susceptibility. Although its absolute values are not excessively large, 
it presents a pronounced $\bq$-dependence with a well-defined peak at $\bq=\bs{0}$. Hence, one can reasonably expect that, if we were able to 
continue the flow below $\Lambda_\mathrm{cr}$, $\chi^d$ would further increase, possibly even diverging at some finite scale. This heuristic expectation is 
supported by the analysis presented in the following sections.

\subsection{Yukawa couplings}
\begin{figure*}[t]
    \centering
    \includegraphics[width = .9 \textwidth]{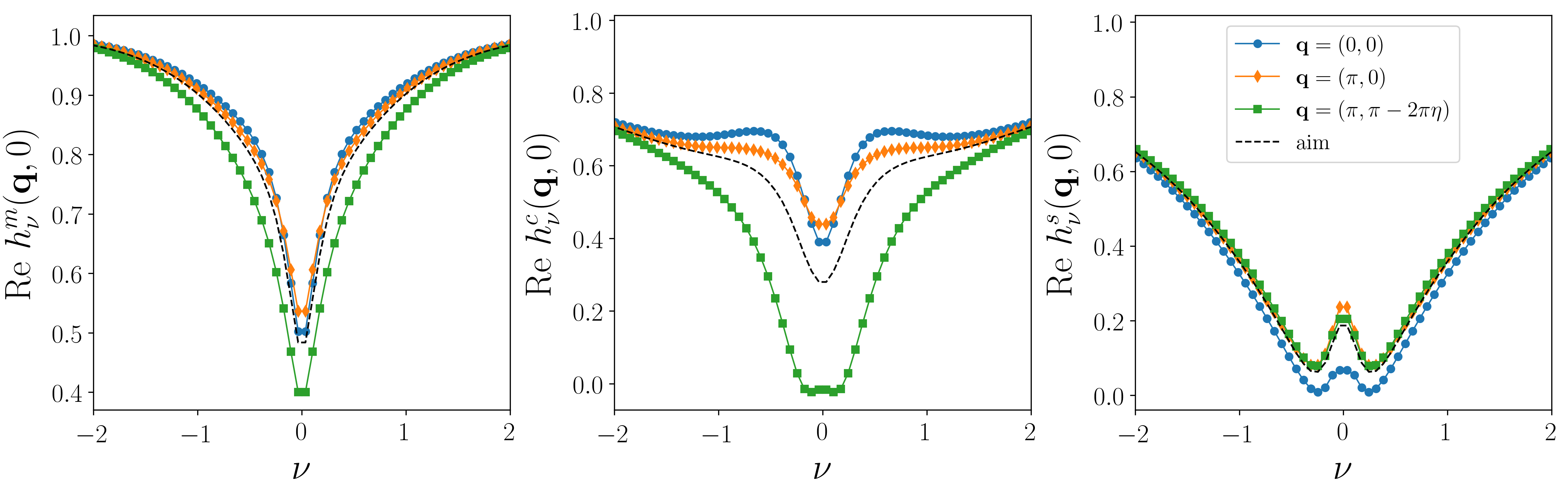}
    \caption{Yukawa couplings in the magnetic, charge, and $s$-wave pairing channel (from left to right) at zero bosonic frequency, as a function of the fermionic Matsubara frequency $\nu$, 
    for the same parameters as in Fig.~\ref{fig: dwave susc 3D} and for various choices of the spatial momentum $\bq$. 
    }
    \label{fig: dwave yukawas}
\end{figure*}

In this section, we analyze the real part of the magnetic, charge, and $s$-wave pairing Yukawa couplings at the stopping scale $\Lambda_\mathrm{cr}$ 
shown in Fig.~\ref{fig: dwave yukawas}.
The magnetic Yukawa coupling exhibits a behavior that qualitatively resembles the one in the weak-coupling regime of the half-filled case (compare with upper left panel of Fig.~\ref{fig: yukawas half filling}) despite the relatively large value of $U=8t$. 
This happens because  the local moment and, thus, its fingerprints in the Yukawa coupling, gets weakened by hole-doping: its low-frequency part is suppressed with respect to the high-frequency ($h^m=1$) value, due to the electronic screening. 

Finally, the $s$-wave pairing Yukawa coupling appears overall suppressed, in spite a relatively small upturn at the lowest frequencies. Not surprisingly, it displays a rather marginal momentum dependence and, thus, small deviations from the AIM result. 

\subsection{\texorpdfstring{$d$-wave}{d-wave} pairing correlations}

\subsubsection{Diagnostics of the correlations}
The aim of this section is to thoroughly inspect 
all terms contributing to the sizable enhancement of the susceptibility $\chi^d$,  
which we briefly discussed above. 
%
%
In the spirit of the post-processing procedures \cite{Gunnarsson2016,Gunnarsson2015,Rohringer2020,Schaefer2021} recently applied \cite{Krien2019_I,Kauch2020,Rohringer2020,Schaefer2021,Delre2021} to several quantum many-body approaches to the Hubbard model, we decompose Eqs.~\eqref{eq: Ld} and \eqref{eq: chid postprocessed}, in order to distinguish the different contributions to the $d$-wave susceptibility. 
In particular, by combining Eq. \eqref{eq: vertex decomposition Dwave} with \eqref{eq: Ld}, we get
\begin{equation}
    \begin{split}
        L^d_{\nu\nu'}(\bq,\Omega) &= \frac{1}{2}\mathcal{M}^d_{\nu,\Omega-\nu}(\nu'-\nu) + \mathcal{M}^d_{\nu,\Omega-\nu}(\Omega-\nu-\nu')\\ 
        &+\frac{1}{2}\mathcal{C}^d_{\nu,\Omega-\nu}(\nu'-\nu)
        +\mathcal{D}_{\nu\nu'}(\bq,\Omega),
    \end{split}
\end{equation}
where we have introduced the functions 
\begin{equation}
    X^d_{\nu\nu'}(\Omega) = -\int_\bq \frac{\cos q_x + \cos q_y}{2} X_{\nu\nu'}(\bq,\Omega),
\end{equation}
%
with $X=\mathcal{M}$ or $\mathcal{C}$, depending only on frequencies.
We can therefore split the function $L^d$ in 
three 
distinct contributions:
\begin{subequations}
    \label{eq: L_d}
    \begin{align}
        & L^{d(m)}_{\nu\nu'}(\Omega) = \frac{1}{2}\mathcal{M}^d_{\nu,\Omega-\nu}(\nu'-\nu) + \mathcal{M}^d_{\nu,\Omega-\nu}(\Omega-\nu-\nu'), \label{eq: L_dm}\\
        & L^{d(c)}_{\nu\nu'}(\Omega) = \frac{1}{2}\mathcal{C}^d_{\nu,\Omega-\nu}(\nu'-\nu) \label{eq: L_dc},
    \end{align}
\end{subequations}
and $\mathcal{D}_{\nu\nu'}(q)$.  

\begin{figure}[b]
    \centering
    \includegraphics[width = 0.49 \textwidth]{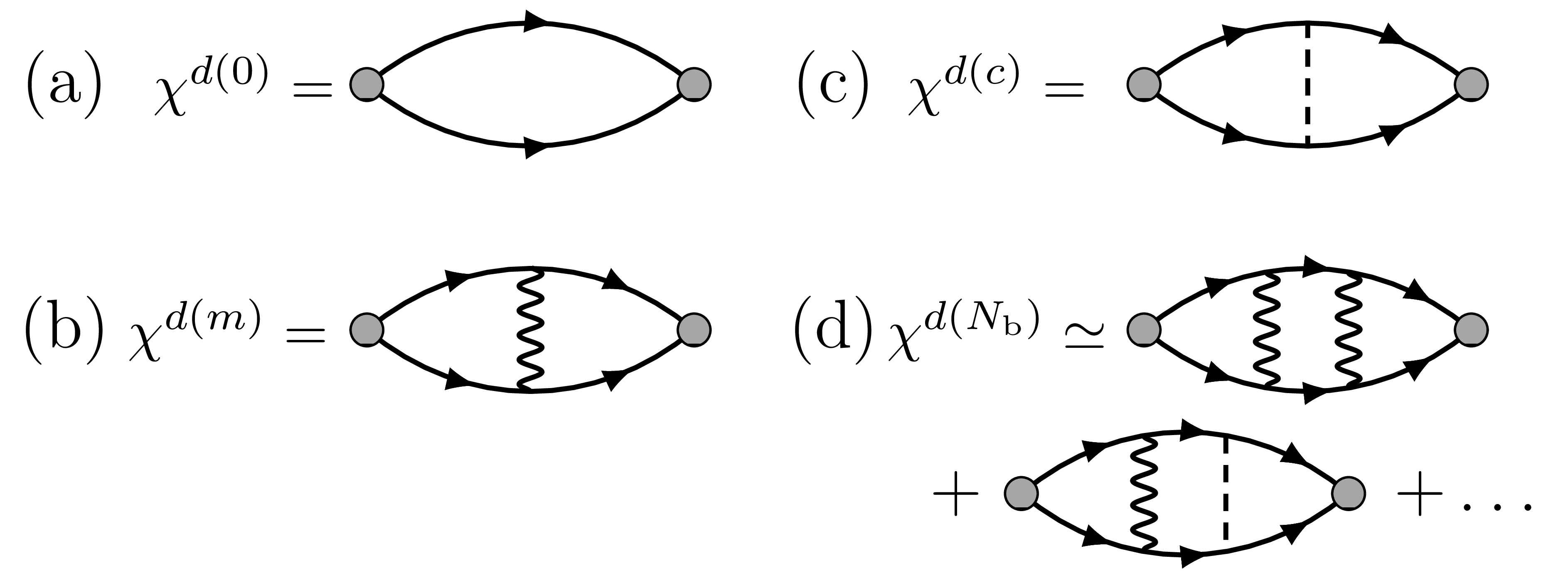}
    \caption{Different diagrammatic contributions to the $d$-wave pairing susceptibility. The solid lines represent fermionic propagators, the wavy ones the magnetic channel $\mathcal{M}$, the dashed ones the charge channel $\mathcal{C}$, and the gray circles the $d$-wave form factors $f_\bk^d$. (a) Bare bubble term. (b-c) Maki-Thompson contributions 
    with one exchanged magnetic and charge boson respectively. (d) $N\geq2$ boson contributions: due to the inaccuracy of the $1\ell$ truncation, the equivalence of $\chi^{d (N_\mathrm{b})}$ with these diagrams 
    is not exact. Notice that we have not drawn the Yukawa couplings, even though they display a nontrivial structure.}
    \label{fig: chi diagrams}
\end{figure}

By inserting this decomposition into the expression for the susceptibility \eqref{eq: chid postprocessed}, we obtain 
\begin{equation}
\chi^d(q) \! = \! \chi^{d(0)}(q) \! +\! \chi^{d(m)}(q) \! + \! \chi^{d(c)}(q) \!+ \! \chi^{d(N_\mathrm{b})}(q)
\label{eq:suscdec}   
\end{equation}
with
%
\begin{subequations}
    \begin{align}
         \chi^{d(0)}(q) &= T\sum_\nu \Pi^d_{\nu}(q),\\
         \chi^{d(m)}(q) &= T^2\sum_{\nu\nu'}\Pi^d_\nu(q) L^{d(m)}_{\nu\nu'}(\Omega) \Pi^d_{\nu'}(q),\\
         \chi^{d(c)}(q) &= T^2\sum_{\nu\nu'}\Pi^d_\nu(q) L^{d(c)}_{\nu\nu'}(\Omega) \Pi^d_{\nu'}(q),\\
         \chi^{d(N_\mathrm{b})}(q) &= T^2\sum_{\nu\nu'}\Pi^d_\nu(q) \mathcal{D}_{\nu\nu'}(q) \Pi^d_{\nu'}(q),
    \end{align}
\end{subequations}
%
Here, $\chi^{d(0)}$ is the bare bubble term, while
$\chi^{d(m)}$ and $\chi^{d(c)}$ represent the Maki-Thompson contributions to the susceptibility, as diagrammatically shown in Fig.~\ref{fig: chi diagrams}. Finally, since the $d$-wave pairing channel entails all diagrams which are two-particle ($pp$) reducible but $U$-irreducible,  the remaining $\chi^{d(N_\mathrm{b})}$ term can be interpreted, as the sum of $N$-boson processes. The physics encoded in the latter processes will be analyzed separately in the final part of this section. Here, we only remark that
this representation becomes exact only in the multiloop extension.

Before commenting on the results, we note that, from an fRG perspective, the AF susceptibility gradually evolves from the beginning of the flow~\cite{Halboth00A,Husemann2009,Katanin2009,Vilardi2019}, while the superconducting $d$-wave susceptibility emerges only in proximity of the critical scale \cite{Halboth00B,Eberlein14}.

\begin{figure}[b]
    \centering
    \includegraphics[width = 0.45 \textwidth]{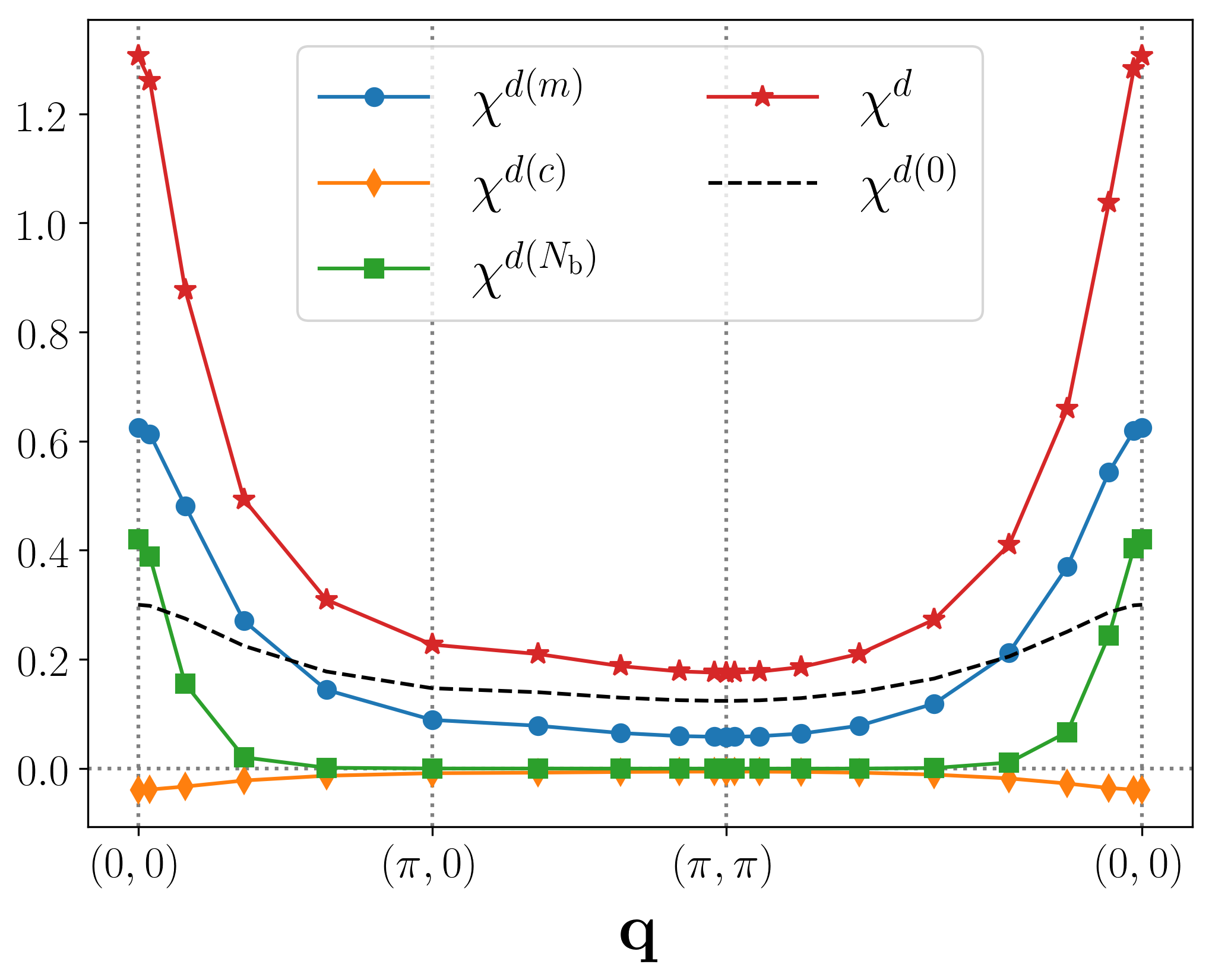}
    \caption{Contributions to the $d$-wave static susceptibility for $n=0.82$, $U=8t$, $t^\prime=-0.2t$, and $T=0.044t$ along a path in the BZ: $\chi^{d(m)}$ and $\chi^{d(c)}$ indicate the Maki-Thompson term with the insertion of a magnetic and a charge bosonic line respectively, and  $\chi^{d(N_\mathrm{b})}$ the $N$-boson processes, with $N\geq2$. In addition, the bare $d$-wave pairing bubble $\chi^{d(0)}$ as well as the total susceptibility $\chi^d$ are shown.}
    \label{fig: dwave chid contributions}
\end{figure}
In Fig.~\ref{fig: dwave chid contributions}, we plot the different contributions to the $d$-wave static susceptibility defined above, 
as obtained at the stopping scale $\Lambda_\mathrm{cr}$. We observe that in the considered parameter region the magnetic Maki-Thompson 
processes yield the most relevant contribution. 
Moreover, we find a fairly large contribution of the multiboson terms. Indeed, we expect that, by approaching a $d$-wave pairing instability, 
the multiboson term would dominate over the Maki-Thompson ones, eventually driving the divergence of the associated susceptibility at the thermodynamic instability. More in general, if one were able to separate the $N<\bar{N}$ ($\chi^{d(N_\mathrm{b}<\bar{N})}$) from the $N\geq\bar{N}$ ($\chi^{d(N_\mathrm{b}\geq \bar{N})}$) boson processes in the susceptibility, arbitrarily close to the critical point one would detect
\begin{equation}
    \chi^{d(N_\mathrm{b}\geq \bar{N})}(\bq\simeq\bs{0},0) > \chi^{d(N_\mathrm{b}<\bar{N})}(\bq\simeq\bs{0},0),
    \label{eq: N boson inequality}
\end{equation}
for every finite 
$\bar{N}$. 
This happens because the divergence of the susceptibility is due to a term $(1-\lambda_d)^{-1}$, with $\lambda_d$, the maximum eigenvalue of the matrix product between the $d$-wave bubble and the two-particle irreducible vertex in the $d$-wave pairing channel, approaching $1$. 
Since the term $(1-\lambda_d)^{-1}$ results from a resummation of infinite order diagrams, it can only be 
encoded in the $N\geq \bar{N}$ term, with the $N<\bar{N}$ one scaling as $\bar{N}\lambda_d^{\bar{N}}\sim \bar{N}$ close to the instability. 
In general, a measure of the maximum $\bar{N}$ for which Eq.~\eqref{eq: N boson inequality} is fulfilled, 
might be exploited to quantify 
the actual proximity to a thermodynamic ($d$-wave pairing) instability.
%
%

As anticipated above, we eventually draw our attention on the last term in Eq.~\eqref{eq:suscdec} identified by the presence of $\mathcal{D}_{\nu\nu'}(q)$, which entails all the reducible scattering processes in the $d$-wave pairing channel. 
Due to the $d$-wave symmetry in momentum space, $\mathcal{D}_{\nu\nu'}(q)$ arises exclusively from scattering events involving the exchange of {\sl two} or {\sl  more} bosons. A refined inspection of these scattering processes is then possible via a corresponding decomposition of $\mathcal{D}_{\nu\nu'}(q)$. 

%
According to Eqs.~\eqref{eq: d-wave flow equation} and \eqref{eq: L_d},
we can classify the different contributions to $\mathcal{D}_{\nu\nu'}(q)$  directly from the corresponding flow equation:
\begin{subequations}
    \begin{align}
        \partial_\Lambda \mathcal{D}^{mm}_{\nu\nu'}(q) &= T\sum_{\nu^{\prime\prime}} L^{d(m)}_{\nu\nu^{\prime\prime}}(q)\left[\widetilde{\partial}_\Lambda \Pi^d_{\nu^{\prime\prime}}(q)\right] L^{d(m)}_{\nu^{\prime\prime}\nu'}(q), \label{eq: Dmm}\\
        \partial_\Lambda \mathcal{D}^{cc}_{\nu\nu'}(q)& = T\sum_{\nu^{\prime\prime}} L^{d(c)}_{\nu\nu^{\prime\prime}}(q)\left[\widetilde{\partial}_\Lambda \Pi^d_{\nu^{\prime\prime}}(q)\right] L^{d(c)}_{\nu^{\prime\prime}\nu'}(q),\label{eq: Dcc}\\
        \partial_\Lambda \mathcal{D}^{mc}_{\nu\nu'}(q) &= T\sum_{\nu^{\prime\prime}} L^{d(m)}_{\nu\nu^{\prime\prime}}(q)\left[\widetilde{\partial}_\Lambda \Pi^d_{\nu^{\prime\prime}}(q)\right] L^{d(c)}_{\nu^{\prime\prime}\nu'}(q) \label{eq: Dmc}\nonumber\\
        &
        +T\sum_{\nu^{\prime\prime}} L^{d(c)}_{\nu\nu^{\prime\prime}}(q)\left[\widetilde{\partial}_\Lambda \Pi^d_{\nu^{\prime\prime}}(q)\right] L^{d(m)}_{\nu^{\prime\prime}\nu'}(q),\\
        \partial_\Lambda \mathcal{D}^{N_\mathrm{b}\geq 3}_{\nu\nu'}(q) &= \partial_\Lambda \mathcal{D}_{\nu\nu'}(q) - \partial_\Lambda \mathcal{D}^{mm}_{\nu\nu'}(q) \label{eq: DN3}  \nonumber\\
        &
        - \partial_\Lambda \mathcal{D}^{cc}_{\nu\nu'}(q) - \partial_\Lambda \mathcal{D}^{mc}_{\nu\nu'}(q).
    \end{align}
    \label{eq: Dal}
\end{subequations}
%
\begin{figure}[b]
    \centering
    \includegraphics[width = 0.5 \textwidth]{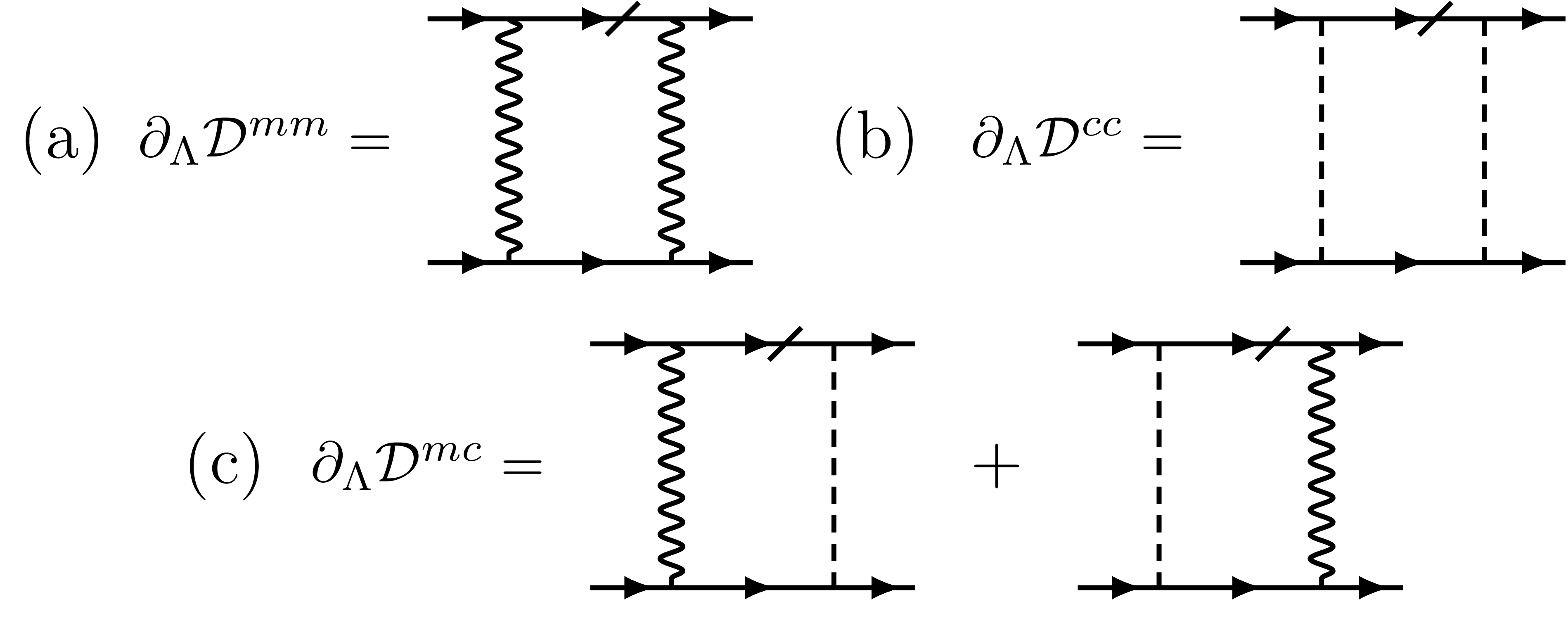}
    \caption{Diagrammatic representation of the two-boson contributions to the flow equation for $\mathcal{D}_{\nu\nu'}(q)$, defined in Eqs.~\eqref{eq: Dmm}-\eqref{eq: Dmc}. The symbols are the same as in Fig.~\ref{fig: chi diagrams}, where the "ticks" on the fermionic lines represent the single scale derivatives $\widetilde{\partial}_\Lambda$. Note that for each of the diagrams 
    there is another contribution with the tick on the lower fermionic line.}
    \label{fig: dwave ch diagrams}
\end{figure}
The diagrammatic representation of the terms  $\mathcal{D}^{mm}$, $\mathcal{D}^{cc}$, and $\mathcal{D}^{mc}$
is shown in Fig.~\ref{fig: dwave ch diagrams}. These two-boson processes can be associated to the so-called Aslamazov-Larkin diagrams. As one might expect, a closer inspection of Eqs.~\eqref{eq: Dmm}-\eqref{eq: Dmc} shows that they are {\sl not} fully reconstructed during the flow 
because the functions $L^{d(m)}$, and $L^{d(c)}$ (as well as the nonlocal self-energy if included) also depend on the fRG scale. 
This feature 
is a typical artifact of the $1\ell$ truncation, which can be fully resolved in the framework of future multiloop extensions of the approach. 
With this caveat, it is nonetheless possible to interpret  
$\mathcal{D}^{mm}$, $\mathcal{D}^{cc}$, and $\mathcal{D}^{mc}$ as 
the two-boson contributions to the $d$-wave pairing channel, and 
the remainder $\mathcal{D}^{N_\mathrm{b}\geq 3}$ in terms of higher order processes in the number of exchanged bosons. 
\begin{figure}[t]
    \centering
    \includegraphics[width = 0.45 \textwidth]{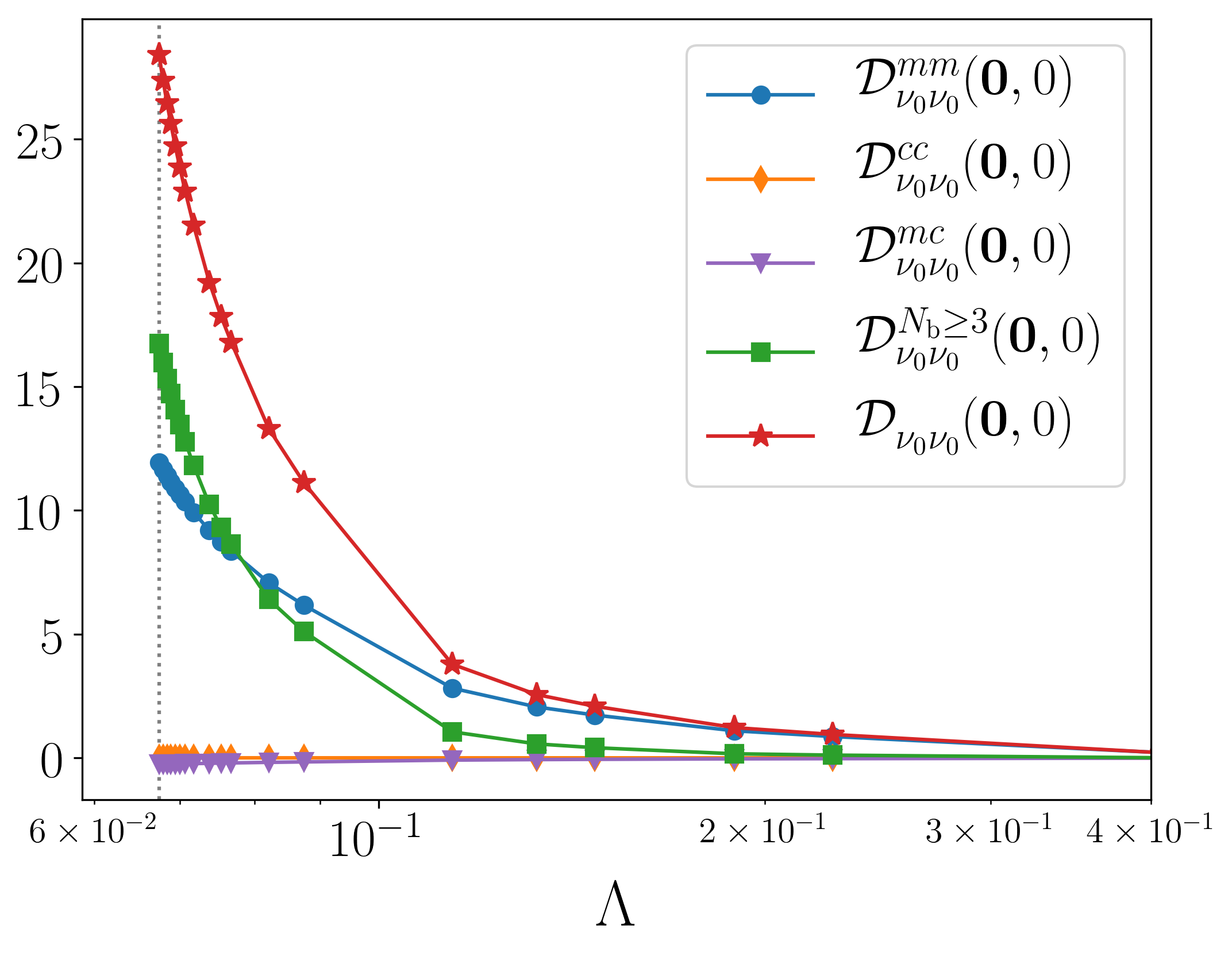}
    \caption{Different contributions to the $d$-wave pairing channel for the same parameters as in Fig.~\ref{fig: dwave chid contributions}, as obtained from Eq.~\eqref{eq: Dal} at zero bosonic frequency and spatial momentum and evaluated at $\nu=\nu^\prime=\nu_0\equiv\pi T$ (s. also Fig.~\ref{fig: dwave diagnostics}), as a function of the fRG scale $\Lambda$.
    }
    \label{fig: dwave flow}
\end{figure}
In Fig.~\ref{fig: dwave flow} we plot the various terms contributing to $\mathcal{D}_{\nu\nu'}(q)$
as a function of the fRG scale $\Lambda$. We observe that the multiboson ($N_b \ge 3$) term 
develops at significantly lower scales as compared to the two-boson ones. At the same time, it increases considerably by approaching the stopping scale. 
This behavior is consistent to the following general consideration: If the system is close enough (in temperature, doping, or other parameters) to a thermodynamic instability, at some point
$\mathcal{D}^{N_\mathrm{b}\geq 3}$ will 
 overtake 
the other terms in Eq.~(\ref{eq:suscdec}), and eventually diverge at the transition itself. 
In fact, similar as 
discussed for the susceptibility, 
sufficiently close to the transition the $N\geq\Bar{N}$ term is going to exceed the sum of the $N<\bar{N}$ terms for every finite $\bar{N}$.
Consistent with these general consideration, in the framework of our $1\ell$ DMF\textsuperscript2RG, we indeed observe that the multiboson term $\mathcal{D}^{N_\mathrm{b}\geq 3}$ becomes dominating at a finite scale just before the end of the flow. Hence, for the selected parameter choice, an important {\sl precondition} for the onset of a $d$-wave pairing instability has been already realized. This represents a promising hint that a true superconducting transition may be unveiled at lower temperatures by means of higher loop-order calculations.
%
\begin{figure}[t]
    \centering
    \includegraphics[width = 0.5 \textwidth]{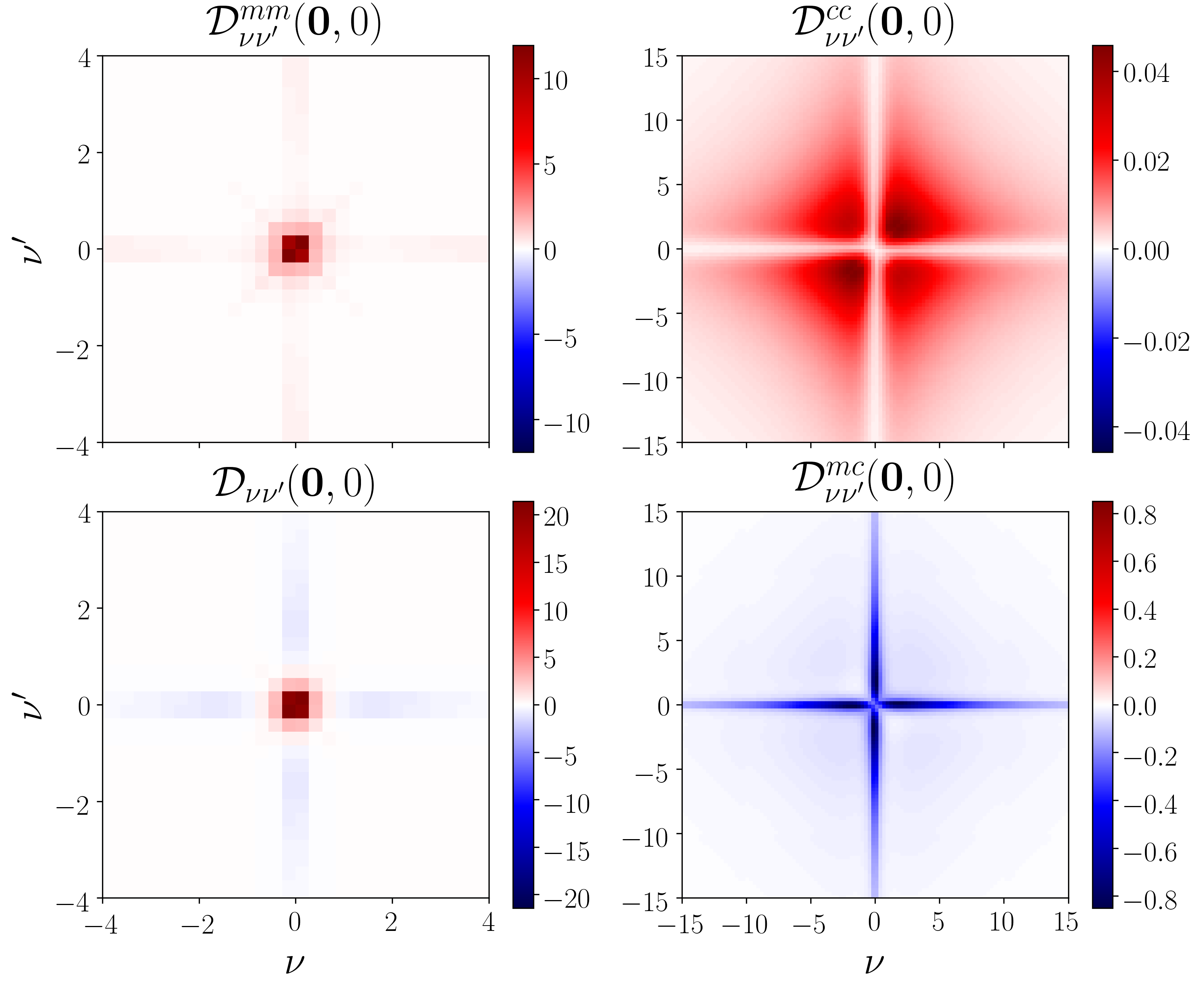}
    \caption{Different contributions to the $d$-wave pairing channel at the stopping scale $\Lambda_\mathrm{cr}$, for zero bosonic frequency and spatial momentum and the same parameters as in Fig.~\ref{fig: dwave chid contributions}. 
    Aslamazov-Larkin diagram (see text) with two exchanged magnetic bosons (\textit{top left}), two exchanged charge bosons (\textit{top right}), one charge and one magnetic boson (\textit{bottom right}), and total $d$-wave pairing channel (\textit{bottom left}) total $d$-wave pairing channel.
    Note that the scale in $\nu$,$\nu^\prime$
    is different
    in order to better resolve the frequency structures.}
    \label{fig: dwave diagnostics}
\end{figure}

In Fig.~\ref{fig: dwave diagnostics} we show the frequency structure of the two-boson contributions to the $d$-wave pairing channel as well as $\mathcal{D}$ itself, at $q=(\bs{0},0)$ as functions of the fermionic Matsubara frequencies $\nu$, $\nu'$. Being associated only to 
$U$-irreducible diagrams, the $d$-wave pairing channel is a rapidly decaying function of the Matsubara frequencies. It exhibits a structure centered around the frequencies $\nu=\pm\nu_0$, $\nu'=\pm\nu_0$ ($\nu_0=\pi T$), where it assumes fairly large values. About 50\% of this structure is generated by two magnetic boson processes and (most of) the rest by multiboson ones, as 
the $\mathcal{D}^{cc}$ and $\mathcal{D}^{mc}$ terms play a %
very marginal role in the formation of $d$-wave pairing correlations. It appears hence natural to conclude that among all multiboson terms, the one consisting only of multiple magnetic boson processes will have the largest weight. This observation suggests that $d$-wave pairing correlations in the normal phase of the Hubbard model are mostly generated by (incommensurate or not) AF fluctuations, 
consistent with previous findings in fRG~\cite{Halboth00B,Husemann2009,Katanin2009,Vilardi2019} and DMF\textsuperscript2RG~\cite{Taranto2014,Vilardi2019}, as well as in recent numerical analyses 
\cite{Gunnarsson2015,Wu2017,Schaefer2021}.

%

\subsubsection{Remarks on the bosonization of the d-wave pairing channel}

In view of further reductions of the numerical complexity in future applications, it might
be helpful to describe {\sl  also}  the $d$-wave pairing channel $\mathcal{D}_{\nu\nu'}(q)$ in terms of single boson processes. Since the bare interaction $U$ 
has
$s$-wave symmetry, the diagrammatic argument of the SBE decomposition does not hold in this case. 
Hence, a proper decomposition of the $d$-wave pairing channel that factorizes the dependence on the fermionic frequencies is needed.


Inspired by earlier fRG works where the Yukawa coupling has been calculated from the channel functions at given values of the fermionic frequencies~\cite{Husemann2012}, we can define the $d$-wave screened interaction and boson-fermion coupling as 
\begin{subequations}
    \label{eq: rebosonized dwave}
    \begin{align}
        &D^d(q) = \frac{\mathcal{D}_{\bar{\nu}_0, \bar{\nu}_0}(q)+\mathcal{D}_{\bar{\nu}_0, -\bar{\nu}_0}(q)}{2},\\
        &h^d_\nu(q) = \frac{\mathcal{D}_{\nu, \bar{\nu}_0}(q)+\mathcal{D}_{\nu, -\bar{\nu}_0}(q)}{2D^d(q)},
    \end{align}
\end{subequations}
where $\bar{\nu}_0$ is a fixed fermionic frequency (eventually $q$-dependent). We notice that in Eq.~\eqref{eq: rebosonized dwave} the symmetrization over $\pm\bar{\nu}_0$ is necessary to guarantee the correct 
symmetries for $D^d$ and $h^d$. In this way the $d$-wave pairing channel can be expressed in a similar form as the other channels:
\begin{equation}
    \mathcal{D}_{\nu\nu'}(q) = h^d_\nu(q)\,D^d(q)\,h^d_{\nu'}(q) + \mathcal{R}^d_{\nu\nu'}(q).
\end{equation}
Setting $\bar{\nu}_0=\infty$ in Eq.~\eqref{eq: rebosonized dwave}, as for the $s$-wave channels, would lead to the conditions $h^d=D^d=0$ and $\mathcal{D}=\mathcal{R}^d$, for which an effective decomposition of $d$-wave pairing channel would not be possible. Hence, another choice of $\bar{\nu}_0$ is necessary, e.g., $\bar{\nu}_0=\pm \pi T$.
The resulting flow equation for $D^d$ and $h^d$ can be then extracted by applying Eq.~\eqref{eq: rebosonized dwave} to the flow of the interaction $\mathcal{D}$ in Eq.~\eqref{eq: channel flow equation}. 
A key point to keep in mind when choosing $\bar{\nu}_0$ is that, once a $d$-wave pairing instability occurs, the divergence of the effective interaction must be reabsorbed into the bosonic propagator, while the Yukawa coupling and the rest function should remain finite, similarly to what happens in the other competing channels. 

\section{Conclusions}
\label{sec:concl}

We have applied the recently introduced
SBE representation to the fRG and DMF\textsuperscript2RG, which relies on a diagrammatic decomposition in contributions mediated by the exchange of a single boson in the different channels.
Specifically, the ($1\ell$) flow equations for the two-particle vertex are recast into SBE contributions and a residual four-point fermion vertex.

The SBE-based formulation leads to a {\sl substantial} reduction of the numerical effort, since the corresponding rest function is significantly localized in frequency space, especially in the strong-coupling regime. This justifies the approximation to significantly restrict the total number of frequencies taken into account 
in the RG flow or even to fully neglect the \new{ nonlocal} \new{  corrections to the DMFT} rest function. 
The reduced numerical effort 
facilitates
the applicability of the fRG and DMF\textsuperscript2RG to the most interesting regime of intermediate to strong correlations and/or low $T$. 
The advantage of this implementation is well illustrated by hands of our DMF\textsuperscript2RG calculations of the 2D Hubbard model performed up to very strong interaction ($U = 16$t) at and out of half-filling.
\new{
In this case, we specifically analyze the impact of \new{ including/excluding} the rest function \new{ flow} and observed a marginal effect from weak to strong coupling. 
}
Moreover, the SBE decomposition naturally allows for a clear physical identification of the relevant degrees of freedoms. As pertinent example, we exploited this specific feature of our approach to diagnose the tendency towards a $d$-wave superconducting instability of the doped Hubbard model in terms of magnetic and charge driven processes.

The derivation of the SBE-based RG flow within the $1\ell$ truncation represents the natural starting point for future multiloop extensions  \cite{Kugler2018_I,Kugler2018_II,Tagliavini2019,Hille2020}, as well as for the systematic inclusion of multiboson contributions. 
Within these methodological extensions, fRG- and DMF\textsuperscript2RG-based computation schemes can be brought to a quantitative level for \emph{all} coupling strengths. Physically, we expect that nonlocal correlation effects, associated to higher loop order processes, might become progressively more pronounced, especially in the most challenging low-temperature regime. 

Finally, the present formalism offers the possibility to explicitly introduce bosonic fields and study the flow of a mixed boson-fermion system. This would allow, for example, to study the effect of bosonic fluctuations on top of mean-field solutions~\cite{Wang2014,Yamase16,Vilardi20,Bonetti20} below the (pseudo) critical scale, where symmetry breaking occurs. 

%
\section*{Acknowledgments}
The authors thank P.~Chalupa, J.~Hauck, S.~Heinzelmann, C.~Honerkamp,  
F.~Krien, F.~Kugler, W.~Metzner, 
and T.~Sch\"afer for valuable discussions and W.~Metzner for a careful reading of the manuscript. We acknowledge financial support from the Deutsche Forschungsgemeinschaft (DFG) through Project No. AN 815/6-1 and from the Austrian Science Fund (FWF) through Project No. I 2794-N35. 

\appendix

\section{Comparison with conventional fermionic formalism}
\label{app: fermionic weak coupling}

\begin{figure}[t]
    \centering
    \includegraphics[width=0.475\textwidth]{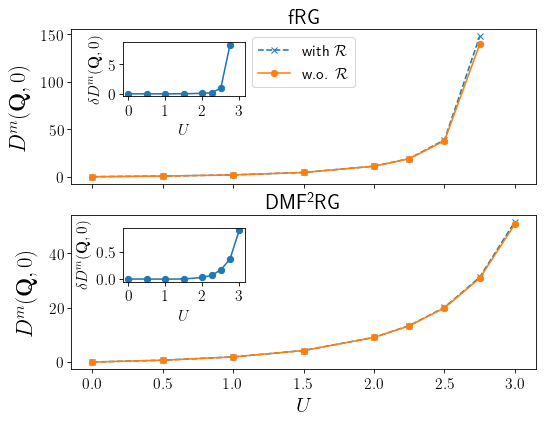}
    \caption{Comparison between the SBE and the conventional fermionic formalism, both for the fRG (upper panels) and DMF\textsuperscript2RG (lower panels), for the same parameters as in Fig.~\ref{fig: fRG vs DMF2RG}. 
    \new{Main panels}: magnetic screened interactions, computed with and without the inclusion of the rest functions. 
    \new{Insets}: difference between the two above mentioned screened interactions. 
    }
    \label{fig: fRG vs DMF2RG 2}
\end{figure}

We here discuss the relation of the SBE formalism with the channel asymptotics introduced in Ref.~\cite{Wentzell2016} (see also Ref.~\cite{Harkov2021a}). Both approaches rely on a similar classification of the diagrams contributing to the two-particle reducible contribution to the vertex function. For this reason, it is possible to connect the screened interactions $D^X$ to the so-called $\mathcal{K}^{(1)X}$ functions via
\begin{equation}
    D^X(q)=U + \mathcal{K}^{(1)X}(q).
\end{equation}
%
$\mathcal{K}^{(1)X}$ is defined as ~\cite{Wentzell2016}:
\begin{equation}
    \mathcal{K}^{(1)X}(q)=\lim_{\nu,\nu'\to\infty} \phi^X_{(\bk,\nu),(\bk',\nu')}(q),
\end{equation}
with $\phi^X$ the sum of all two-particle reducible diagrams in the $m$, $c$, or $s$ channel, as in Eq.~\eqref{eq: channel flow equation}. Similarly, the Yukawa coupling is related to the $\mathcal{K}^{(2)X}$ asymptotic function by 
\begin{equation}
    h^X_k(q) = 1 + \frac{\mathcal{K}^{(2)X}_k(q)}{D^X(q)},
\end{equation}
with 
\begin{equation}
    \mathcal{K}^{(2)X}_k(q)=\lim_{\nu'\to\infty} \phi^X_{k,(\bk',\nu')}(q) - \mathcal{K}^{(1)X}(q).
\end{equation}
At this stage the SBE decomposition seems to offer no substantial computational gain as compared to the channel asymptotics, the only exception being in the vicinity of a critical point, where in the asymptotic formalism both $\mathcal{K}^{(1)X}$ and $\mathcal{K}^{(2)X}$ 
acquire large values, while in the SBE one this 
occurs only for $D^X$, the Yukawa coupling being always finite. However, the most important 
difference is visible in 
the rest functions. Indeed, the SBE and the asymptotic rest functions are related via (see also Ref.~\cite{Bonetti20})
\begin{equation}
    \mathcal{R}^{\mathrm{SBE},X}_{kk'}(q)=\mathcal{R}^{\mathrm{asym},X}_{kk'}(q)-[h^X_k(q)-1]D^X(q)[h^X_{k'}(q)-1],
    \label{eq: R_SBE vs R_asym}
\end{equation}
with 
\begin{equation}
    \mathcal{R}^{\mathrm{asym},X}_{kk'}(q) = \phi_{kk'}(q) - \mathcal{K}^{(1)X}(q) - \mathcal{K}^{(2)X}_{k}(q) -\mathcal{K}^{(2)X}_{k'}(q).
\end{equation}
From Eq.~\eqref{eq: R_SBE vs R_asym} we notice two important facts: First, in the asymptotics formalism some rest functions can diverge at a given critical point as they contain $D^X$. Therefore, in this regime it is not safe to neglect them.
Second, $\mathcal{R}^{\mathrm{SBE},X}$ contains \emph{less} diagrams than $\mathcal{R}^{\mathrm{asym}X}$, as the latter still includes some $U$ reducible contribution. 
An approximation which neglects $\mathcal{R}^{\mathrm{SBE},X}$ can be always justified as the selection of a well-defined class of diagrams.

For completeness we report the fRG and DMF\textsuperscript2RG results of Fig.~\ref{fig: fRG vs DMF2RG} obtained by including the flow of the rest functions $\mathcal{R}^X$, where the SBE-based implementation is equivalent to the fRG and DMF\textsuperscript2RG respectively. In Fig.~\ref{fig: fRG vs DMF2RG 2} we show the impact of the inclusion/neglection of the rest function in the magnetic screened interaction, which displays the largest difference, for both conventional fRG and DMF\textsuperscript2RG. We notice that, as discussed in the main text for stronger coupling, sizable differences between the two approaches (with and without $\mathcal{R}^X$) arise only in the vicinity of the N\'{e}el transition. We then conclude that, away from the AF transition point, the tiny differences fully justify the approximation to neglect the rest function, while close to the critical transition the inclusion/neglection of the rest functions will result in a (generally small) change of the values for the critical temperature or coupling.

\section{Alternative derivation of the flow equations}
\label{app: Hubbard-Stratonovich}

In this Appendix we derive the flow equations presented in the main text as obtained by an alternative derivation based on an explicit introduction of the order parameter fluctuation field via the Hubbard-Stratonovich transformation (HST). In particular, after having split the bare Hubbard interaction into three equal terms ($Un_{\up}n_{\down}=3Un_{\up}n_{\down} - 2 Un_{\up}n_{\down}$), we apply three different HST on the first three terms, one on each physical channel. In formulas   
\begin{equation}
    \begin{split}
        \mathcal{Z}_\text{Hubbard}&=\int \mathcal{D}\left(\psi,\Bar{\psi}\right)e^{-\mathcal{S}_\text{Hubbard}\left[\psi,\Bar{\psi}\right]}\\&=\int\mathcal{D}\Phi\mathcal{D}\left(\psi,\Bar{\psi}\right)e^{-\mathcal{S}_\text{bos}\left[\psi,\Bar{\psi},\Phi\right]},
    \end{split}
    \label{appeq:Z}
\end{equation}
where $\Phi=(\phi_c,\Vec{\phi}_m,\phi_p,\phi_p^*)$ collects all bosonic fields, 
\begin{equation}
\begin{split}
    \mathcal{S}_\text{Hubbard}\left[\psi,\Bar{\psi}\right] &= -\int_{k,\sigma} \Bar{\psi}_{k,\sigma} \left(i\nu+\mu-\epsilon_\bk\right) \psi_{k,\sigma} \\&+ U \int_0^\beta d\tau \sum_i n_{\uparrow,i}(\tau) n_{\downarrow,i}(\tau),
    \end{split}
\end{equation}
and
\begin{equation}
    \begin{split}
        \mathcal{S}_\text{bos}&\left[\psi,\Bar{\psi},\Phi\right] = 
        -\int_{k,\sigma} \Bar{\psi}_{k,\sigma} \left(i\nu+\mu-\epsilon_k\right) \psi_{k,\sigma}\\
        &+\frac{1}{2}\int_{q}\phi_c(-q) \frac{1}{U} \phi_c(q)
        +\frac{1}{2}\int_{q}\Vec{\phi}_m(-q) \cdot \frac{1}{U} \Vec{\phi}_m(q)\\
        &+\int_{q}\phi^*_p(q)\frac{1}{U} \phi_p(q)
       +\int_{k,q,\sigma}\phi_c(q) \,\Bar{\psi}_{k+\frac{q}{2},\sigma}\psi_{k-\frac{q}{2},\sigma}\\
       &+\int_{k,q,\sigma,\sigma'}\Vec{\phi}_m(q)\cdot \,\Bar{\psi}_{k+\frac{q}{2},\sigma}\Vec{\tau}_{\sigma\sigma'}\psi_{k-\frac{q}{2},\sigma'}\\
       &+\int_{k,q}\left[\phi_p(q) \,\Bar{\psi}_{\frac{q}{2}+k,\uparrow}\Bar{\psi}_{\frac{q}{2}-k,\downarrow}+\phi^*_p(q) \,\psi_{\frac{q}{2}-k,\downarrow}\psi_{\frac{q}{2}+k,\uparrow}\right]\\       &
       -2U \int_0^\beta d\tau \sum_i n_{\uparrow,i}(\tau) n_{\downarrow,i}(\tau).
       \label{eq: HS action}
    \end{split}
\end{equation}
%
The remaining (not bosonized) $-2U$ term in $\mathcal{S}_\text{bos}$, avoids double counting of the bare interaction.

To derive the functional flow equations, a regulator term is added to the definition of the generating functionals by replacing the bare action as
\begin{equation}
    \begin{split}
    Z^\Lambda\left[\eta,\Bar{\eta},J\right]&=
    \int\mathcal{D}\Phi\int \mathcal{D}\left(\psi,\Bar{\psi}\right)
    \\ 
    &\times e^{-S^\Lambda_\text{bos}\left[\psi,\Bar{\psi},\Phi\right]+\left(\Bar{\psi},\eta\right)+\left(\Bar{\eta},\psi\right) + \Phi J},
    \label{appeq:ZL}
    \end{split}
\end{equation}
where
\begin{equation}
    S^\Lambda_\text{bos}\left[\psi,\Bar{\psi},\Phi\right]=S_\text{bos}\left[\psi,\Bar{\psi},\Phi\right]+\int_{k,\sigma} \Bar{\psi}_{k,\sigma}\,R^\Lambda(k)\,\psi_{k,\sigma}.
\label{eqbos: reg}
\end{equation}
Its value at some initial scale $\Lambda_\text{ini}$ depends on the formalism used. For instance, in the plain fRG we impose $R^{\Lambda\rightarrow\Lambda_\text{ini}}(k)\rightarrow \infty$, such that, from the saddle point equation of the functional integral, the initial conditions are determines by the bare ones. 

Differently, in the DMF\textsuperscript 2RG, the cutoff must fulfill
\begin{equation}
    R^{\Lambda_{\text{ini}}}(k) = \epsilon_{\boldsymbol{k}} - \Delta_\text{AIM}\left(\nu\right),
\end{equation}
so that the rescaled action at $\Lambda_\text{ini}$ is
\begin{equation}
    \mathcal{S}^{\Lambda_\text{ini}}\left[\psi,\Bar{\psi},\Phi\right] =
    \mathcal{S}_\text{AIM}\left[\psi,\Bar{\psi},\Phi\right],
\end{equation}
where $\mathcal{S}_\text{AIM}\left[\psi,\Bar{\psi}\right]$ is the action of the self-consistent AIM. Here the same procedure involving HST for the Hubbard interaction at the impurity has been performed. In this way, the initial condition for the resulting effective action reads 
\begin{equation}
    \Gamma^{\Lambda_\text{ini}}\left[\psi,\Bar{\psi},\Phi\right]=\Gamma_\text{AIM}\left[\psi,\Bar{\psi},\Phi\right],
\end{equation}
where $\Gamma_\text{AIM}\left[\psi,\Bar{\psi},\Phi\right]$ is the effective action of the self-consistent AIM. By expanding it in terms of 1PI functions, one recovers the initial conditions given in the main text, where the screened interactions $D^X$ and the Yukawa couplings $h^X$ play the role of bosonic propagators and fermion-boson interactions respectively. 

Note that we do not explicitly introduce a regulator for bosonic fluctuations. Indeed, the bosonic path integral at this stage has to be performed in some way to recover the fermionic formalism. For this scope, a second scale $\Lambda_b$ can be introduced, responsible for the integration over $\Phi$ and which is associated to a second cutoff function $R_b^{\Lambda_b}$ regularizing the bare bosonic propagator. In this context, the bosonic flow is shown to be ineffective when integrated before the fermionic flow integration~\cite{Strack2008}.

\section{Numerical implementation}
\label{app: technical aspects}
In this Appendix we discuss a few numerical details. Regarding the DMFT calculation, we solve the AIM with the exact diagonalization (ED) by discretizing the bath into $4$ sites. The local magnetic, charge and pairing screened interactions $D^X$ and Yukawa couplings $h^X$ are computed by Lehmann representation as in Ref.~\cite{Tagliavini2019} in a relatively large frequency range as well as the two-particle Green's function needed to extract the $U$-irreducible vertex $\Lambda_{U\mathrm{irr}}^\mathrm{loc}$. Regarding the flow equations of $D^X$ and $h^X$, we use a frequency domain ranging from 30 to 64 positive values for both the bosonic and the fermionic Matsubara frequencies, depending on the convergence of the calculation. 
The treatment of the frequency asymptotics here simplifies since the screened interactions $D^X$ and Yukawa couplings $h^X$ 
tend to 1 and $U$, respectively, 
at large frequencies, see Eqs.~\eqref{eq: init frg} or~\eqref{eq: init dmf2rg}. 

Regarding the Matsubara summation, in general we select a range from 100 to 256 positive frequencies, depending on the physical regime. Contrary to the fRG, in the DMF$^2$RG the single-scale propagator decays faster at large frequency, simplifying the convergence in the Matsubara summation. 
Moreover, it is noteworthy to mention that, as stated in Sec.~\ref{subsec: susc}, the computation of the charge susceptibility in the Mott phase requires a larger number of frequencies in the Matsubara summation, which must be then extended, despite the relatively high temperature range, to 1000 to recover its physical value~\cite{Chalupa2021}.

While part of the momentum dependence is projected onto form-factors as explained in the main text, the transfer momentum dependence has been patched similarly to Ref.~\cite{Vilardi20}, retaining 38 patches in the reduced Brillouin zone $\mathcal{B}_\mathrm{red}=\{(k_x,k_y): 0\leq k_y\leq k_x \leq \pi\}$. 

Finally, the flow equations have been solved using the adaptive Runge-Kutta Cash-Karp 54 method and the momentum integration over the Brillouin zone is carried out via an adaptive cubature technique. 
\bibliography{biblio.bib}

\begin{thebibliography}{110}%
\makeatletter
\providecommand \@ifxundefined [1]{%
 \@ifx{#1\undefined}
}%
\providecommand \@ifnum [1]{%
 \ifnum #1\expandafter \@firstoftwo
 \else \expandafter \@secondoftwo
 \fi
}%
\providecommand \@ifx [1]{%
 \ifx #1\expandafter \@firstoftwo
 \else \expandafter \@secondoftwo
 \fi
}%
\providecommand \natexlab [1]{#1}%
\providecommand \enquote  [1]{``#1''}%
\providecommand \bibnamefont  [1]{#1}%
\providecommand \bibfnamefont [1]{#1}%
\providecommand \citenamefont [1]{#1}%
\providecommand \href@noop [0]{\@secondoftwo}%
\providecommand \href [0]{\begingroup \@sanitize@url \@href}%
\providecommand \@href[1]{\@@startlink{#1}\@@href}%
\providecommand \@@href[1]{\endgroup#1\@@endlink}%
\providecommand \@sanitize@url [0]{\catcode `\\12\catcode `\$12\catcode
  `\&12\catcode `\#12\catcode `\^12\catcode `\_12\catcode `\%12\relax}%
\providecommand \@@startlink[1]{}%
\providecommand \@@endlink[0]{}%
\providecommand \url  [0]{\begingroup\@sanitize@url \@url }%
\providecommand \@url [1]{\endgroup\@href {#1}{\urlprefix }}%
\providecommand \urlprefix  [0]{URL }%
\providecommand \Eprint [0]{\href }%
\providecommand \doibase [0]{https://doi.org/}%
\providecommand \selectlanguage [0]{\@gobble}%
\providecommand \bibinfo  [0]{\@secondoftwo}%
\providecommand \bibfield  [0]{\@secondoftwo}%
\providecommand \translation [1]{[#1]}%
\providecommand \BibitemOpen [0]{}%
\providecommand \bibitemStop [0]{}%
\providecommand \bibitemNoStop [0]{.\EOS\space}%
\providecommand \EOS [0]{\spacefactor3000\relax}%
\providecommand \BibitemShut  [1]{\csname bibitem#1\endcsname}%
\let\auto@bib@innerbib\@empty
\bibitem [{\citenamefont {Taranto}\ \emph {et~al.}(2014)\citenamefont
  {Taranto}, \citenamefont {Andergassen}, \citenamefont {Bauer}, \citenamefont
  {Held}, \citenamefont {Katanin}, \citenamefont {Metzner}, \citenamefont
  {Rohringer},\ and\ \citenamefont {Toschi}}]{Taranto2014}%
  \BibitemOpen
  \bibfield  {author} {\bibinfo {author} {\bibfnamefont {C.}~\bibnamefont
  {Taranto}}, \bibinfo {author} {\bibfnamefont {S.}~\bibnamefont
  {Andergassen}}, \bibinfo {author} {\bibfnamefont {J.}~\bibnamefont {Bauer}},
  \bibinfo {author} {\bibfnamefont {K.}~\bibnamefont {Held}}, \bibinfo {author}
  {\bibfnamefont {A.}~\bibnamefont {Katanin}}, \bibinfo {author} {\bibfnamefont
  {W.}~\bibnamefont {Metzner}}, \bibinfo {author} {\bibfnamefont
  {G.}~\bibnamefont {Rohringer}},\ and\ \bibinfo {author} {\bibfnamefont
  {A.}~\bibnamefont {Toschi}},\ }\bibfield  {title} {\bibinfo {title} {{From
  Infinite to Two Dimensions through the Functional Renormalization Group}},\
  }\href {https://doi.org/10.1103/PhysRevLett.112.196402} {\bibfield  {journal}
  {\bibinfo  {journal} {Phys. Rev. Lett.}\ }\textbf {\bibinfo {volume} {112}},\
  \bibinfo {pages} {196402} (\bibinfo {year} {2014})}\BibitemShut {NoStop}%
\bibitem [{\citenamefont {Wentzell}\ \emph {et~al.}(2015)\citenamefont
  {Wentzell}, \citenamefont {Taranto}, \citenamefont {Katanin}, \citenamefont
  {Toschi},\ and\ \citenamefont {Andergassen}}]{Wentzell2015}%
  \BibitemOpen
  \bibfield  {author} {\bibinfo {author} {\bibfnamefont {N.}~\bibnamefont
  {Wentzell}}, \bibinfo {author} {\bibfnamefont {C.}~\bibnamefont {Taranto}},
  \bibinfo {author} {\bibfnamefont {A.}~\bibnamefont {Katanin}}, \bibinfo
  {author} {\bibfnamefont {A.}~\bibnamefont {Toschi}},\ and\ \bibinfo {author}
  {\bibfnamefont {S.}~\bibnamefont {Andergassen}},\ }\bibfield  {title}
  {\bibinfo {title} {Correlated starting points for the functional
  renormalization group},\ }\href {https://doi.org/10.1103/PhysRevB.91.045120}
  {\bibfield  {journal} {\bibinfo  {journal} {Phys. Rev. B}\ }\textbf {\bibinfo
  {volume} {91}},\ \bibinfo {pages} {045120} (\bibinfo {year}
  {2015})}\BibitemShut {NoStop}%
\bibitem [{\citenamefont {Vilardi}\ \emph {et~al.}(2019)\citenamefont
  {Vilardi}, \citenamefont {Taranto},\ and\ \citenamefont
  {Metzner}}]{Vilardi2019}%
  \BibitemOpen
  \bibfield  {author} {\bibinfo {author} {\bibfnamefont {D.}~\bibnamefont
  {Vilardi}}, \bibinfo {author} {\bibfnamefont {C.}~\bibnamefont {Taranto}},\
  and\ \bibinfo {author} {\bibfnamefont {W.}~\bibnamefont {Metzner}},\
  }\bibfield  {title} {\bibinfo {title} {{Antiferromagnetic and $d$-wave
  pairing correlations in the strongly interacting two-dimensional Hubbard
  model from the functional renormalization group}},\ }\href
  {https://doi.org/10.1103/PhysRevB.99.104501} {\bibfield  {journal} {\bibinfo
  {journal} {Phys. Rev. B}\ }\textbf {\bibinfo {volume} {99}},\ \bibinfo
  {pages} {104501} (\bibinfo {year} {2019})}\BibitemShut {NoStop}%
\bibitem [{\citenamefont {Metzner}\ and\ \citenamefont
  {Vollhardt}(1989)}]{Metzner1989}%
  \BibitemOpen
  \bibfield  {author} {\bibinfo {author} {\bibfnamefont {W.}~\bibnamefont
  {Metzner}}\ and\ \bibinfo {author} {\bibfnamefont {D.}~\bibnamefont
  {Vollhardt}},\ }\bibfield  {title} {\bibinfo {title} {{Correlated Lattice
  Fermions in $d=\ensuremath{\infty}$ Dimensions}},\ }\href
  {https://doi.org/10.1103/PhysRevLett.62.324} {\bibfield  {journal} {\bibinfo
  {journal} {Phys. Rev. Lett.}\ }\textbf {\bibinfo {volume} {62}},\ \bibinfo
  {pages} {324} (\bibinfo {year} {1989})}\BibitemShut {NoStop}%
\bibitem [{\citenamefont {Georges}\ \emph {et~al.}(1996)\citenamefont
  {Georges}, \citenamefont {Kotliar}, \citenamefont {Krauth},\ and\
  \citenamefont {Rozenberg}}]{Georges1996}%
  \BibitemOpen
  \bibfield  {author} {\bibinfo {author} {\bibfnamefont {A.}~\bibnamefont
  {Georges}}, \bibinfo {author} {\bibfnamefont {G.}~\bibnamefont {Kotliar}},
  \bibinfo {author} {\bibfnamefont {W.}~\bibnamefont {Krauth}},\ and\ \bibinfo
  {author} {\bibfnamefont {M.~J.}\ \bibnamefont {Rozenberg}},\ }\bibfield
  {title} {\bibinfo {title} {Dynamical mean-field theory of strongly correlated
  fermion systems and the limit of infinite dimensions},\ }\href
  {https://doi.org/10.1103/RevModPhys.68.13} {\bibfield  {journal} {\bibinfo
  {journal} {Rev. Mod. Phys.}\ }\textbf {\bibinfo {volume} {68}},\ \bibinfo
  {pages} {13} (\bibinfo {year} {1996})}\BibitemShut {NoStop}%
\bibitem [{\citenamefont {Salmhofer}(1999)}]{Salmhofer1999}%
  \BibitemOpen
  \bibfield  {author} {\bibinfo {author} {\bibfnamefont {M.}~\bibnamefont
  {Salmhofer}},\ }\href {https://doi.org/10.1007/978-3-662-03873-4} {\emph
  {\bibinfo {title} {Renormalization}}},\ Theoretical and Mathematical Physics\
  (\bibinfo  {publisher} {Springer Berlin Heidelberg},\ \bibinfo {year}
  {1999})\BibitemShut {NoStop}%
\bibitem [{\citenamefont {Berges}\ \emph {et~al.}(2002)\citenamefont {Berges},
  \citenamefont {Tetradis},\ and\ \citenamefont {Wetterich}}]{Berges2002}%
  \BibitemOpen
  \bibfield  {author} {\bibinfo {author} {\bibfnamefont {J.}~\bibnamefont
  {Berges}}, \bibinfo {author} {\bibfnamefont {N.}~\bibnamefont {Tetradis}},\
  and\ \bibinfo {author} {\bibfnamefont {C.}~\bibnamefont {Wetterich}},\
  }\bibfield  {title} {\bibinfo {title} {Non-perturbative renormalization flow
  in quantum field theory and statistical physics},\ }\href
  {https://doi.org/10.1016/S0370-1573(01)00098-9} {\bibfield  {journal}
  {\bibinfo  {journal} {Physics Reports}\ }\textbf {\bibinfo {volume} {363}},\
  \bibinfo {pages} {223 } (\bibinfo {year} {2002})}\BibitemShut {NoStop}%
\bibitem [{\citenamefont {Kopietz}\ \emph {et~al.}(2010)\citenamefont
  {Kopietz}, \citenamefont {Bartosch},\ and\ \citenamefont
  {Sch\"{u}tz}}]{Kopietz2010}%
  \BibitemOpen
  \bibfield  {author} {\bibinfo {author} {\bibfnamefont {P.}~\bibnamefont
  {Kopietz}}, \bibinfo {author} {\bibfnamefont {L.}~\bibnamefont {Bartosch}},\
  and\ \bibinfo {author} {\bibfnamefont {F.}~\bibnamefont {Sch\"{u}tz}},\
  }\href {https://doi.org/10.1007/978-3-642-05094-7} {\emph {\bibinfo {title}
  {Introduction to the Functional Renormalization Group}}}\ (\bibinfo
  {publisher} {Springer Berlin Heidelberg},\ \bibinfo {year}
  {2010})\BibitemShut {NoStop}%
\bibitem [{\citenamefont {Metzner}\ \emph {et~al.}(2012)\citenamefont
  {Metzner}, \citenamefont {Salmhofer}, \citenamefont {Honerkamp},
  \citenamefont {Meden},\ and\ \citenamefont {Sch\"onhammer}}]{Metzner12}%
  \BibitemOpen
  \bibfield  {author} {\bibinfo {author} {\bibfnamefont {W.}~\bibnamefont
  {Metzner}}, \bibinfo {author} {\bibfnamefont {M.}~\bibnamefont {Salmhofer}},
  \bibinfo {author} {\bibfnamefont {C.}~\bibnamefont {Honerkamp}}, \bibinfo
  {author} {\bibfnamefont {V.}~\bibnamefont {Meden}},\ and\ \bibinfo {author}
  {\bibfnamefont {K.}~\bibnamefont {Sch\"onhammer}},\ }\bibfield  {title}
  {\bibinfo {title} {Functional renormalization group approach to correlated
  fermion systems},\ }\href {https://doi.org/10.1103/RevModPhys.84.299}
  {\bibfield  {journal} {\bibinfo  {journal} {Rev. Mod. Phys.}\ }\textbf
  {\bibinfo {volume} {84}},\ \bibinfo {pages} {299} (\bibinfo {year}
  {2012})}\BibitemShut {NoStop}%
\bibitem [{\citenamefont {Dupuis}\ \emph {et~al.}(2021)\citenamefont {Dupuis},
  \citenamefont {Canet}, \citenamefont {Eichhorn}, \citenamefont {Metzner},
  \citenamefont {Pawlowski}, \citenamefont {Tissier},\ and\ \citenamefont
  {Wschebor}}]{Dupuis2021}%
  \BibitemOpen
  \bibfield  {author} {\bibinfo {author} {\bibfnamefont {N.}~\bibnamefont
  {Dupuis}}, \bibinfo {author} {\bibfnamefont {L.}~\bibnamefont {Canet}},
  \bibinfo {author} {\bibfnamefont {A.}~\bibnamefont {Eichhorn}}, \bibinfo
  {author} {\bibfnamefont {W.}~\bibnamefont {Metzner}}, \bibinfo {author}
  {\bibfnamefont {J.}~\bibnamefont {Pawlowski}}, \bibinfo {author}
  {\bibfnamefont {M.}~\bibnamefont {Tissier}},\ and\ \bibinfo {author}
  {\bibfnamefont {N.}~\bibnamefont {Wschebor}},\ }\bibfield  {title} {\bibinfo
  {title} {The nonperturbative functional renormalization group and its
  applications},\ }\href {https://doi.org/10.1016/j.physrep.2021.01.001}
  {\bibfield  {journal} {\bibinfo  {journal} {Physics Reports}\ }\textbf
  {\bibinfo {volume} {910}},\ \bibinfo {pages} {1–114} (\bibinfo {year}
  {2021})}\BibitemShut {NoStop}%
\bibitem [{\citenamefont {Rohringer}\ \emph {et~al.}(2018)\citenamefont
  {Rohringer}, \citenamefont {Hafermann}, \citenamefont {Toschi}, \citenamefont
  {Katanin}, \citenamefont {Antipov}, \citenamefont {Katsnelson}, \citenamefont
  {Lichtenstein}, \citenamefont {Rubtsov},\ and\ \citenamefont
  {Held}}]{Rohringer2018}%
  \BibitemOpen
  \bibfield  {author} {\bibinfo {author} {\bibfnamefont {G.}~\bibnamefont
  {Rohringer}}, \bibinfo {author} {\bibfnamefont {H.}~\bibnamefont
  {Hafermann}}, \bibinfo {author} {\bibfnamefont {A.}~\bibnamefont {Toschi}},
  \bibinfo {author} {\bibfnamefont {A.~A.}\ \bibnamefont {Katanin}}, \bibinfo
  {author} {\bibfnamefont {A.~E.}\ \bibnamefont {Antipov}}, \bibinfo {author}
  {\bibfnamefont {M.~I.}\ \bibnamefont {Katsnelson}}, \bibinfo {author}
  {\bibfnamefont {A.~I.}\ \bibnamefont {Lichtenstein}}, \bibinfo {author}
  {\bibfnamefont {A.~N.}\ \bibnamefont {Rubtsov}},\ and\ \bibinfo {author}
  {\bibfnamefont {K.}~\bibnamefont {Held}},\ }\bibfield  {title} {\bibinfo
  {title} {Diagrammatic routes to nonlocal correlations beyond dynamical mean
  field theory},\ }\href {https://doi.org/10.1103/RevModPhys.90.025003}
  {\bibfield  {journal} {\bibinfo  {journal} {Rev. Mod. Phys.}\ }\textbf
  {\bibinfo {volume} {90}},\ \bibinfo {pages} {025003} (\bibinfo {year}
  {2018})}\BibitemShut {NoStop}%
\bibitem [{\citenamefont {Chalupa}\ \emph {et~al.}(2021)\citenamefont
  {Chalupa}, \citenamefont {Sch\"afer}, \citenamefont {Reitner}, \citenamefont
  {Springer}, \citenamefont {Andergassen},\ and\ \citenamefont
  {Toschi}}]{Chalupa2021}%
  \BibitemOpen
  \bibfield  {author} {\bibinfo {author} {\bibfnamefont {P.}~\bibnamefont
  {Chalupa}}, \bibinfo {author} {\bibfnamefont {T.}~\bibnamefont {Sch\"afer}},
  \bibinfo {author} {\bibfnamefont {M.}~\bibnamefont {Reitner}}, \bibinfo
  {author} {\bibfnamefont {D.}~\bibnamefont {Springer}}, \bibinfo {author}
  {\bibfnamefont {S.}~\bibnamefont {Andergassen}},\ and\ \bibinfo {author}
  {\bibfnamefont {A.}~\bibnamefont {Toschi}},\ }\bibfield  {title} {\bibinfo
  {title} {{Fingerprints of the Local Moment Formation and its Kondo Screening
  in the Generalized Susceptibilities of Many-Electron Problems}},\ }\href
  {https://doi.org/10.1103/PhysRevLett.126.056403} {\bibfield  {journal}
  {\bibinfo  {journal} {Phys. Rev. Lett.}\ }\textbf {\bibinfo {volume} {126}},\
  \bibinfo {pages} {056403} (\bibinfo {year} {2021})}\BibitemShut {NoStop}%
\bibitem [{\citenamefont {Rohringer}\ \emph {et~al.}(2012)\citenamefont
  {Rohringer}, \citenamefont {Valli},\ and\ \citenamefont
  {Toschi}}]{Rohringer2012}%
  \BibitemOpen
  \bibfield  {author} {\bibinfo {author} {\bibfnamefont {G.}~\bibnamefont
  {Rohringer}}, \bibinfo {author} {\bibfnamefont {A.}~\bibnamefont {Valli}},\
  and\ \bibinfo {author} {\bibfnamefont {A.}~\bibnamefont {Toschi}},\
  }\bibfield  {title} {\bibinfo {title} {{Local electronic correlation at the
  two-particle level}},\ }\href {https://doi.org/10.1103/PhysRevB.86.125114}
  {\bibfield  {journal} {\bibinfo  {journal} {Phys. Rev. B}\ }\textbf {\bibinfo
  {volume} {86}},\ \bibinfo {pages} {125114} (\bibinfo {year}
  {2012})}\BibitemShut {NoStop}%
\bibitem [{\citenamefont {Wentzell}\ \emph {et~al.}(2020)\citenamefont
  {Wentzell}, \citenamefont {Li}, \citenamefont {Tagliavini}, \citenamefont
  {Taranto}, \citenamefont {Rohringer}, \citenamefont {Held}, \citenamefont
  {Toschi},\ and\ \citenamefont {Andergassen}}]{Wentzell2016}%
  \BibitemOpen
  \bibfield  {author} {\bibinfo {author} {\bibfnamefont {N.}~\bibnamefont
  {Wentzell}}, \bibinfo {author} {\bibfnamefont {G.}~\bibnamefont {Li}},
  \bibinfo {author} {\bibfnamefont {A.}~\bibnamefont {Tagliavini}}, \bibinfo
  {author} {\bibfnamefont {C.}~\bibnamefont {Taranto}}, \bibinfo {author}
  {\bibfnamefont {G.}~\bibnamefont {Rohringer}}, \bibinfo {author}
  {\bibfnamefont {K.}~\bibnamefont {Held}}, \bibinfo {author} {\bibfnamefont
  {A.}~\bibnamefont {Toschi}},\ and\ \bibinfo {author} {\bibfnamefont
  {S.}~\bibnamefont {Andergassen}},\ }\bibfield  {title} {\bibinfo {title}
  {{High-frequency asymptotics of the vertex function: Diagrammatic
  parametrization and algorithmic implementation}},\ }\href
  {https://doi.org/10.1103/PhysRevB.102.085106} {\bibfield  {journal} {\bibinfo
   {journal} {Phys. Rev. B}\ }\textbf {\bibinfo {volume} {102}},\ \bibinfo
  {pages} {085106} (\bibinfo {year} {2020})}\BibitemShut {NoStop}%
\bibitem [{\citenamefont {Krien}\ \emph {et~al.}(2019)\citenamefont {Krien},
  \citenamefont {Valli},\ and\ \citenamefont {Capone}}]{Krien2019_I}%
  \BibitemOpen
  \bibfield  {author} {\bibinfo {author} {\bibfnamefont {F.}~\bibnamefont
  {Krien}}, \bibinfo {author} {\bibfnamefont {A.}~\bibnamefont {Valli}},\ and\
  \bibinfo {author} {\bibfnamefont {M.}~\bibnamefont {Capone}},\ }\bibfield
  {title} {\bibinfo {title} {{Single-boson exchange decomposition of the vertex
  function}},\ }\href {https://doi.org/10.1103/PhysRevB.100.155149} {\bibfield
  {journal} {\bibinfo  {journal} {Phys. Rev. B}\ }\textbf {\bibinfo {volume}
  {100}},\ \bibinfo {pages} {155149} (\bibinfo {year} {2019})}\BibitemShut
  {NoStop}%
\bibitem [{\citenamefont {de~Dominicis}\ and\ \citenamefont
  {Martin}(1964)}]{Dedominicis1964}%
  \BibitemOpen
  \bibfield  {author} {\bibinfo {author} {\bibfnamefont {C.}~\bibnamefont
  {de~Dominicis}}\ and\ \bibinfo {author} {\bibfnamefont {P.~C.}\ \bibnamefont
  {Martin}},\ }\bibfield  {title} {\bibinfo {title} {{Stationary Entropy
  Principle and Renormalization in Normal and Superfluid Systems. I. Algebraic
  Formulation}},\ }\href {https://doi.org/10.1063/1.1704062} {\bibfield
  {journal} {\bibinfo  {journal} {J. Math. Phys.}\ }\textbf {\bibinfo {volume}
  {5}},\ \bibinfo {pages} {14} (\bibinfo {year} {1964})}\BibitemShut {NoStop}%
\bibitem [{\citenamefont {{De Dominicis}}\ and\ \citenamefont
  {{Martin}}(1964)}]{Dedominicis1964A}%
  \BibitemOpen
  \bibfield  {author} {\bibinfo {author} {\bibfnamefont {C.}~\bibnamefont {{De
  Dominicis}}}\ and\ \bibinfo {author} {\bibfnamefont {P.~C.}\ \bibnamefont
  {{Martin}}},\ }\bibfield  {title} {\bibinfo {title} {{Stationary Entropy
  Principle and Renormalization in Normal and Superfluid Systems. II.
  Diagrammatic Formulation}},\ }\href {https://doi.org/10.1063/1.1704064}
  {\bibfield  {journal} {\bibinfo  {journal} {J. Math. Phys.}\ }\textbf
  {\bibinfo {volume} {5}},\ \bibinfo {pages} {31} (\bibinfo {year}
  {1964})}\BibitemShut {NoStop}%
\bibitem [{\citenamefont {Bickers}(2004)}]{Bickers2004}%
  \BibitemOpen
  \bibfield  {author} {\bibinfo {author} {\bibfnamefont {N.~E.}\ \bibnamefont
  {Bickers}},\ }\bibfield  {title} {\bibinfo {title} {{Self-Consistent
  Many-Body Theory for Condensed Matter Systems}},\ }in\ \href
  {https://doi.org/10.1007/0-387-21717-7_6} {\emph {\bibinfo {booktitle}
  {Theoretical Methods for Strongly Correlated Electrons}}},\ \bibinfo {editor}
  {edited by\ \bibinfo {editor} {\bibfnamefont {D.}~\bibnamefont
  {S{\'e}n{\'e}chal}}, \bibinfo {editor} {\bibfnamefont {A.-M.}\ \bibnamefont
  {Tremblay}},\ and\ \bibinfo {editor} {\bibfnamefont {C.}~\bibnamefont
  {Bourbonnais}}}\ (\bibinfo  {publisher} {Springer New York},\ \bibinfo
  {address} {New York, NY},\ \bibinfo {year} {2004})\ pp.\ \bibinfo {pages}
  {237--296}\BibitemShut {NoStop}%
\bibitem [{\citenamefont {Yang}\ \emph {et~al.}(2009)\citenamefont {Yang},
  \citenamefont {Fotso}, \citenamefont {Liu}, \citenamefont {Maier},
  \citenamefont {Tomko}, \citenamefont {D'Azevedo}, \citenamefont {Scalettar},
  \citenamefont {Pruschke},\ and\ \citenamefont {Jarrell}}]{Yang2009}%
  \BibitemOpen
  \bibfield  {author} {\bibinfo {author} {\bibfnamefont {S.~X.}\ \bibnamefont
  {Yang}}, \bibinfo {author} {\bibfnamefont {H.}~\bibnamefont {Fotso}},
  \bibinfo {author} {\bibfnamefont {J.}~\bibnamefont {Liu}}, \bibinfo {author}
  {\bibfnamefont {T.~A.}\ \bibnamefont {Maier}}, \bibinfo {author}
  {\bibfnamefont {K.}~\bibnamefont {Tomko}}, \bibinfo {author} {\bibfnamefont
  {E.~F.}\ \bibnamefont {D'Azevedo}}, \bibinfo {author} {\bibfnamefont {R.~T.}\
  \bibnamefont {Scalettar}}, \bibinfo {author} {\bibfnamefont {T.}~\bibnamefont
  {Pruschke}},\ and\ \bibinfo {author} {\bibfnamefont {M.}~\bibnamefont
  {Jarrell}},\ }\bibfield  {title} {\bibinfo {title} {{Parquet approximation
  for the $4\ifmmode\times\else\texttimes\fi{}4$ Hubbard cluster}},\ }\href
  {https://doi.org/10.1103/PhysRevE.80.046706} {\bibfield  {journal} {\bibinfo
  {journal} {Phys. Rev. E}\ }\textbf {\bibinfo {volume} {80}},\ \bibinfo
  {pages} {046706} (\bibinfo {year} {2009})}\BibitemShut {NoStop}%
\bibitem [{\citenamefont {Tam}\ \emph {et~al.}(2013)\citenamefont {Tam},
  \citenamefont {Fotso}, \citenamefont {Yang}, \citenamefont {Lee},
  \citenamefont {Moreno}, \citenamefont {Ramanujam},\ and\ \citenamefont
  {Jarrell}}]{Tam2013}%
  \BibitemOpen
  \bibfield  {author} {\bibinfo {author} {\bibfnamefont {K.-M.}\ \bibnamefont
  {Tam}}, \bibinfo {author} {\bibfnamefont {H.}~\bibnamefont {Fotso}}, \bibinfo
  {author} {\bibfnamefont {S.-X.}\ \bibnamefont {Yang}}, \bibinfo {author}
  {\bibfnamefont {T.-W.}\ \bibnamefont {Lee}}, \bibinfo {author} {\bibfnamefont
  {J.}~\bibnamefont {Moreno}}, \bibinfo {author} {\bibfnamefont
  {J.}~\bibnamefont {Ramanujam}},\ and\ \bibinfo {author} {\bibfnamefont
  {M.}~\bibnamefont {Jarrell}},\ }\bibfield  {title} {\bibinfo {title}
  {{Solving the parquet equations for the Hubbard model beyond weak
  coupling}},\ }\href {https://doi.org/10.1103/PhysRevE.87.013311} {\bibfield
  {journal} {\bibinfo  {journal} {Phys. Rev. E}\ }\textbf {\bibinfo {volume}
  {87}},\ \bibinfo {pages} {013311} (\bibinfo {year} {2013})}\BibitemShut
  {NoStop}%
\bibitem [{\citenamefont {Valli}\ \emph {et~al.}(2015)\citenamefont {Valli},
  \citenamefont {Sch\"afer}, \citenamefont {Thunstr\"om}, \citenamefont
  {Rohringer}, \citenamefont {Andergassen}, \citenamefont {Sangiovanni},
  \citenamefont {Held},\ and\ \citenamefont {Toschi}}]{Valli2015}%
  \BibitemOpen
  \bibfield  {author} {\bibinfo {author} {\bibfnamefont {A.}~\bibnamefont
  {Valli}}, \bibinfo {author} {\bibfnamefont {T.}~\bibnamefont {Sch\"afer}},
  \bibinfo {author} {\bibfnamefont {P.}~\bibnamefont {Thunstr\"om}}, \bibinfo
  {author} {\bibfnamefont {G.}~\bibnamefont {Rohringer}}, \bibinfo {author}
  {\bibfnamefont {S.}~\bibnamefont {Andergassen}}, \bibinfo {author}
  {\bibfnamefont {G.}~\bibnamefont {Sangiovanni}}, \bibinfo {author}
  {\bibfnamefont {K.}~\bibnamefont {Held}},\ and\ \bibinfo {author}
  {\bibfnamefont {A.}~\bibnamefont {Toschi}},\ }\bibfield  {title} {\bibinfo
  {title} {{Dynamical vertex approximation in its parquet implementation:
  Application to Hubbard nanorings}},\ }\href
  {https://doi.org/10.1103/PhysRevB.91.115115} {\bibfield  {journal} {\bibinfo
  {journal} {Phys. Rev. B}\ }\textbf {\bibinfo {volume} {91}},\ \bibinfo
  {pages} {115115} (\bibinfo {year} {2015})}\BibitemShut {NoStop}%
\bibitem [{\citenamefont {Kauch}\ \emph {et~al.}(2019)\citenamefont {Kauch},
  \citenamefont {H\"{o}rbinger}, \citenamefont {Li},\ and\ \citenamefont
  {Held}}]{Kauch2019}%
  \BibitemOpen
  \bibfield  {author} {\bibinfo {author} {\bibfnamefont {A.}~\bibnamefont
  {Kauch}}, \bibinfo {author} {\bibfnamefont {F.}~\bibnamefont
  {H\"{o}rbinger}}, \bibinfo {author} {\bibfnamefont {G.}~\bibnamefont {Li}},\
  and\ \bibinfo {author} {\bibfnamefont {K.}~\bibnamefont {Held}},\ }\href@noop
  {} {\bibinfo {title} {{Interplay between magnetic and superconducting
  fluctuations in the doped 2d Hubbard model}}} (\bibinfo {year} {2019}),\
  \Eprint {https://arxiv.org/abs/1901.09743} {arXiv:1901.09743} \BibitemShut
  {NoStop}%
\bibitem [{\citenamefont {Li}\ \emph {et~al.}(2019)\citenamefont {Li},
  \citenamefont {Kauch}, \citenamefont {Pudleiner},\ and\ \citenamefont
  {Held}}]{Li2019}%
  \BibitemOpen
  \bibfield  {author} {\bibinfo {author} {\bibfnamefont {G.}~\bibnamefont
  {Li}}, \bibinfo {author} {\bibfnamefont {A.}~\bibnamefont {Kauch}}, \bibinfo
  {author} {\bibfnamefont {P.}~\bibnamefont {Pudleiner}},\ and\ \bibinfo
  {author} {\bibfnamefont {K.}~\bibnamefont {Held}},\ }\bibfield  {title}
  {\bibinfo {title} {{The victory project v1.0: An efficient parquet equations
  solver}},\ }\href {https://doi.org/https://doi.org/10.1016/j.cpc.2019.03.008}
  {\bibfield  {journal} {\bibinfo  {journal} {Comput. Phys. Commun.}\ }\textbf
  {\bibinfo {volume} {241}},\ \bibinfo {pages} {146} (\bibinfo {year}
  {2019})}\BibitemShut {NoStop}%
\bibitem [{\citenamefont {Krien}\ \emph
  {et~al.}(2020{\natexlab{a}})\citenamefont {Krien}, \citenamefont {Valli},
  \citenamefont {Chalupa}, \citenamefont {Capone}, \citenamefont
  {Lichtenstein},\ and\ \citenamefont {Toschi}}]{Krien2020}%
  \BibitemOpen
  \bibfield  {author} {\bibinfo {author} {\bibfnamefont {F.}~\bibnamefont
  {Krien}}, \bibinfo {author} {\bibfnamefont {A.}~\bibnamefont {Valli}},
  \bibinfo {author} {\bibfnamefont {P.}~\bibnamefont {Chalupa}}, \bibinfo
  {author} {\bibfnamefont {M.}~\bibnamefont {Capone}}, \bibinfo {author}
  {\bibfnamefont {A.~I.}\ \bibnamefont {Lichtenstein}},\ and\ \bibinfo {author}
  {\bibfnamefont {A.}~\bibnamefont {Toschi}},\ }\bibfield  {title} {\bibinfo
  {title} {{Boson-exchange parquet solver for dual fermions}},\ }\href
  {https://doi.org/10.1103/PhysRevB.102.195131} {\bibfield  {journal} {\bibinfo
   {journal} {Phys. Rev. B}\ }\textbf {\bibinfo {volume} {102}},\ \bibinfo
  {pages} {195131} (\bibinfo {year} {2020}{\natexlab{a}})}\BibitemShut
  {NoStop}%
\bibitem [{\citenamefont {Krien}\ \emph {et~al.}(2021)\citenamefont {Krien},
  \citenamefont {Kauch},\ and\ \citenamefont {Held}}]{Krien2021}%
  \BibitemOpen
  \bibfield  {author} {\bibinfo {author} {\bibfnamefont {F.}~\bibnamefont
  {Krien}}, \bibinfo {author} {\bibfnamefont {A.}~\bibnamefont {Kauch}},\ and\
  \bibinfo {author} {\bibfnamefont {K.}~\bibnamefont {Held}},\ }\bibfield
  {title} {\bibinfo {title} {{Tiling with triangles: parquet and $GW\gamma$
  methods unified}},\ }\href {https://doi.org/10.1103/PhysRevResearch.3.013149}
  {\bibfield  {journal} {\bibinfo  {journal} {Phys. Rev. Research}\ }\textbf
  {\bibinfo {volume} {3}},\ \bibinfo {pages} {013149} (\bibinfo {year}
  {2021})}\BibitemShut {NoStop}%
\bibitem [{\citenamefont {Tagliavini}\ \emph {et~al.}(2018)\citenamefont
  {Tagliavini}, \citenamefont {Hummel}, \citenamefont {Wentzell}, \citenamefont
  {Andergassen}, \citenamefont {Toschi},\ and\ \citenamefont
  {Rohringer}}]{Tagliavini2018}%
  \BibitemOpen
  \bibfield  {author} {\bibinfo {author} {\bibfnamefont {A.}~\bibnamefont
  {Tagliavini}}, \bibinfo {author} {\bibfnamefont {S.}~\bibnamefont {Hummel}},
  \bibinfo {author} {\bibfnamefont {N.}~\bibnamefont {Wentzell}}, \bibinfo
  {author} {\bibfnamefont {S.}~\bibnamefont {Andergassen}}, \bibinfo {author}
  {\bibfnamefont {A.}~\bibnamefont {Toschi}},\ and\ \bibinfo {author}
  {\bibfnamefont {G.}~\bibnamefont {Rohringer}},\ }\bibfield  {title} {\bibinfo
  {title} {{Efficient Bethe-Salpeter equation treatment in dynamical mean-field
  theory}},\ }\href {https://doi.org/10.1103/PhysRevB.97.235140} {\bibfield
  {journal} {\bibinfo  {journal} {Phys. Rev. B}\ }\textbf {\bibinfo {volume}
  {97}},\ \bibinfo {pages} {235140} (\bibinfo {year} {2018})}\BibitemShut
  {NoStop}%
\bibitem [{\citenamefont {Sch\"afer}\ \emph {et~al.}(2013)\citenamefont
  {Sch\"afer}, \citenamefont {Rohringer}, \citenamefont {Gunnarsson},
  \citenamefont {Ciuchi}, \citenamefont {Sangiovanni},\ and\ \citenamefont
  {Toschi}}]{Schaefer2013}%
  \BibitemOpen
  \bibfield  {author} {\bibinfo {author} {\bibfnamefont {T.}~\bibnamefont
  {Sch\"afer}}, \bibinfo {author} {\bibfnamefont {G.}~\bibnamefont
  {Rohringer}}, \bibinfo {author} {\bibfnamefont {O.}~\bibnamefont
  {Gunnarsson}}, \bibinfo {author} {\bibfnamefont {S.}~\bibnamefont {Ciuchi}},
  \bibinfo {author} {\bibfnamefont {G.}~\bibnamefont {Sangiovanni}},\ and\
  \bibinfo {author} {\bibfnamefont {A.}~\bibnamefont {Toschi}},\ }\bibfield
  {title} {\bibinfo {title} {{Divergent Precursors of the Mott-Hubbard
  Transition at the Two-Particle Level}},\ }\href
  {https://doi.org/10.1103/PhysRevLett.110.246405} {\bibfield  {journal}
  {\bibinfo  {journal} {Phys. Rev. Lett.}\ }\textbf {\bibinfo {volume} {110}},\
  \bibinfo {pages} {246405} (\bibinfo {year} {2013})}\BibitemShut {NoStop}%
\bibitem [{\citenamefont {Gunnarsson}\ \emph {et~al.}(2016)\citenamefont
  {Gunnarsson}, \citenamefont {Sch\"afer}, \citenamefont {LeBlanc},
  \citenamefont {Merino}, \citenamefont {Sangiovanni}, \citenamefont
  {Rohringer},\ and\ \citenamefont {Toschi}}]{Gunnarsson2016}%
  \BibitemOpen
  \bibfield  {author} {\bibinfo {author} {\bibfnamefont {O.}~\bibnamefont
  {Gunnarsson}}, \bibinfo {author} {\bibfnamefont {T.}~\bibnamefont
  {Sch\"afer}}, \bibinfo {author} {\bibfnamefont {J.~P.~F.}\ \bibnamefont
  {LeBlanc}}, \bibinfo {author} {\bibfnamefont {J.}~\bibnamefont {Merino}},
  \bibinfo {author} {\bibfnamefont {G.}~\bibnamefont {Sangiovanni}}, \bibinfo
  {author} {\bibfnamefont {G.}~\bibnamefont {Rohringer}},\ and\ \bibinfo
  {author} {\bibfnamefont {A.}~\bibnamefont {Toschi}},\ }\bibfield  {title}
  {\bibinfo {title} {Parquet decomposition calculations of the electronic
  self-energy},\ }\href {https://doi.org/10.1103/PhysRevB.93.245102} {\bibfield
   {journal} {\bibinfo  {journal} {Phys. Rev. B}\ }\textbf {\bibinfo {volume}
  {93}},\ \bibinfo {pages} {245102} (\bibinfo {year} {2016})}\BibitemShut
  {NoStop}%
\bibitem [{\citenamefont {Sch\"afer}\ \emph {et~al.}(2016)\citenamefont
  {Sch\"afer}, \citenamefont {Ciuchi}, \citenamefont {Wallerberger},
  \citenamefont {Thunstr\"om}, \citenamefont {Gunnarsson}, \citenamefont
  {Sangiovanni}, \citenamefont {Rohringer},\ and\ \citenamefont
  {Toschi}}]{Schaefer2016}%
  \BibitemOpen
  \bibfield  {author} {\bibinfo {author} {\bibfnamefont {T.}~\bibnamefont
  {Sch\"afer}}, \bibinfo {author} {\bibfnamefont {S.}~\bibnamefont {Ciuchi}},
  \bibinfo {author} {\bibfnamefont {M.}~\bibnamefont {Wallerberger}}, \bibinfo
  {author} {\bibfnamefont {P.}~\bibnamefont {Thunstr\"om}}, \bibinfo {author}
  {\bibfnamefont {O.}~\bibnamefont {Gunnarsson}}, \bibinfo {author}
  {\bibfnamefont {G.}~\bibnamefont {Sangiovanni}}, \bibinfo {author}
  {\bibfnamefont {G.}~\bibnamefont {Rohringer}},\ and\ \bibinfo {author}
  {\bibfnamefont {A.}~\bibnamefont {Toschi}},\ }\bibfield  {title} {\bibinfo
  {title} {Nonperturbative landscape of the mott-hubbard transition: Multiple
  divergence lines around the critical endpoint},\ }\href
  {https://doi.org/10.1103/PhysRevB.94.235108} {\bibfield  {journal} {\bibinfo
  {journal} {Phys. Rev. B}\ }\textbf {\bibinfo {volume} {94}},\ \bibinfo
  {pages} {235108} (\bibinfo {year} {2016})}\BibitemShut {NoStop}%
\bibitem [{\citenamefont {Ribic}\ \emph {et~al.}(2016)\citenamefont {Ribic},
  \citenamefont {Rohringer},\ and\ \citenamefont {Held}}]{Ribic2016}%
  \BibitemOpen
  \bibfield  {author} {\bibinfo {author} {\bibfnamefont {T.}~\bibnamefont
  {Ribic}}, \bibinfo {author} {\bibfnamefont {G.}~\bibnamefont {Rohringer}},\
  and\ \bibinfo {author} {\bibfnamefont {K.}~\bibnamefont {Held}},\ }\bibfield
  {title} {\bibinfo {title} {{Nonlocal correlations and spectral properties of
  the Falicov-Kimball model}},\ }\href
  {https://doi.org/10.1103/PhysRevB.93.195105} {\bibfield  {journal} {\bibinfo
  {journal} {Phys. Rev. B}\ }\textbf {\bibinfo {volume} {93}},\ \bibinfo
  {pages} {195105} (\bibinfo {year} {2016})}\BibitemShut {NoStop}%
\bibitem [{\citenamefont {Gunnarsson}\ \emph {et~al.}(2017)\citenamefont
  {Gunnarsson}, \citenamefont {Rohringer}, \citenamefont {Sch\"afer},
  \citenamefont {Sangiovanni},\ and\ \citenamefont {Toschi}}]{Gunnarsson2017}%
  \BibitemOpen
  \bibfield  {author} {\bibinfo {author} {\bibfnamefont {O.}~\bibnamefont
  {Gunnarsson}}, \bibinfo {author} {\bibfnamefont {G.}~\bibnamefont
  {Rohringer}}, \bibinfo {author} {\bibfnamefont {T.}~\bibnamefont
  {Sch\"afer}}, \bibinfo {author} {\bibfnamefont {G.}~\bibnamefont
  {Sangiovanni}},\ and\ \bibinfo {author} {\bibfnamefont {A.}~\bibnamefont
  {Toschi}},\ }\bibfield  {title} {\bibinfo {title} {{Breakdown of Traditional
  Many-Body Theories for Correlated Electrons}},\ }\href
  {https://doi.org/10.1103/PhysRevLett.119.056402} {\bibfield  {journal}
  {\bibinfo  {journal} {Phys. Rev. Lett.}\ }\textbf {\bibinfo {volume} {119}},\
  \bibinfo {pages} {056402} (\bibinfo {year} {2017})}\BibitemShut {NoStop}%
\bibitem [{\citenamefont {Vu\v{c}i\v{c}evi\'c}\ \emph
  {et~al.}(2018)\citenamefont {Vu\v{c}i\v{c}evi\'c}, \citenamefont {Wentzell},
  \citenamefont {Ferrero},\ and\ \citenamefont {Parcollet}}]{Vucicevic2018}%
  \BibitemOpen
  \bibfield  {author} {\bibinfo {author} {\bibfnamefont {J.}~\bibnamefont
  {Vu\v{c}i\v{c}evi\'c}}, \bibinfo {author} {\bibfnamefont {N.}~\bibnamefont
  {Wentzell}}, \bibinfo {author} {\bibfnamefont {M.}~\bibnamefont {Ferrero}},\
  and\ \bibinfo {author} {\bibfnamefont {O.}~\bibnamefont {Parcollet}},\
  }\bibfield  {title} {\bibinfo {title} {{Practical consequences of the
  Luttinger-Ward functional multivaluedness for cluster DMFT methods}},\ }\href
  {https://doi.org/10.1103/PhysRevB.97.125141} {\bibfield  {journal} {\bibinfo
  {journal} {Phys. Rev. B}\ }\textbf {\bibinfo {volume} {97}},\ \bibinfo
  {pages} {125141} (\bibinfo {year} {2018})}\BibitemShut {NoStop}%
\bibitem [{\citenamefont {Chalupa}\ \emph {et~al.}(2018)\citenamefont
  {Chalupa}, \citenamefont {Gunacker}, \citenamefont {Sch\"afer}, \citenamefont
  {Held},\ and\ \citenamefont {Toschi}}]{Chalupa2018}%
  \BibitemOpen
  \bibfield  {author} {\bibinfo {author} {\bibfnamefont {P.}~\bibnamefont
  {Chalupa}}, \bibinfo {author} {\bibfnamefont {P.}~\bibnamefont {Gunacker}},
  \bibinfo {author} {\bibfnamefont {T.}~\bibnamefont {Sch\"afer}}, \bibinfo
  {author} {\bibfnamefont {K.}~\bibnamefont {Held}},\ and\ \bibinfo {author}
  {\bibfnamefont {A.}~\bibnamefont {Toschi}},\ }\bibfield  {title} {\bibinfo
  {title} {{Divergences of the irreducible vertex functions in correlated
  metallic systems: Insights from the Anderson impurity model}},\ }\href
  {https://doi.org/10.1103/PhysRevB.97.245136} {\bibfield  {journal} {\bibinfo
  {journal} {Phys. Rev. B}\ }\textbf {\bibinfo {volume} {97}},\ \bibinfo
  {pages} {245136} (\bibinfo {year} {2018})}\BibitemShut {NoStop}%
\bibitem [{\citenamefont {Springer}\ \emph {et~al.}(2020)\citenamefont
  {Springer}, \citenamefont {Chalupa}, \citenamefont {Ciuchi}, \citenamefont
  {Sangiovanni},\ and\ \citenamefont {Toschi}}]{Springer2020}%
  \BibitemOpen
  \bibfield  {author} {\bibinfo {author} {\bibfnamefont {D.}~\bibnamefont
  {Springer}}, \bibinfo {author} {\bibfnamefont {P.}~\bibnamefont {Chalupa}},
  \bibinfo {author} {\bibfnamefont {S.}~\bibnamefont {Ciuchi}}, \bibinfo
  {author} {\bibfnamefont {G.}~\bibnamefont {Sangiovanni}},\ and\ \bibinfo
  {author} {\bibfnamefont {A.}~\bibnamefont {Toschi}},\ }\bibfield  {title}
  {\bibinfo {title} {Interplay between local response and vertex divergences in
  many-fermion systems with on-site attraction},\ }\href
  {https://doi.org/10.1103/PhysRevB.101.155148} {\bibfield  {journal} {\bibinfo
   {journal} {Phys. Rev. B}\ }\textbf {\bibinfo {volume} {101}},\ \bibinfo
  {pages} {155148} (\bibinfo {year} {2020})}\BibitemShut {NoStop}%
\bibitem [{\citenamefont {Krahl}\ and\ \citenamefont
  {Wetterich}(2007)}]{Krahl2007}%
  \BibitemOpen
  \bibfield  {author} {\bibinfo {author} {\bibfnamefont {H.}~\bibnamefont
  {Krahl}}\ and\ \bibinfo {author} {\bibfnamefont {C.}~\bibnamefont
  {Wetterich}},\ }\bibfield  {title} {\bibinfo {title} {Functional
  renormalization group for d-wave superconductivity},\ }\href
  {https://doi.org/10.1016/j.physleta.2007.03.028} {\bibfield  {journal}
  {\bibinfo  {journal} {Phys. Lett. A}\ }\textbf {\bibinfo {volume} {367}},\
  \bibinfo {pages} {263–267} (\bibinfo {year} {2007})}\BibitemShut {NoStop}%
\bibitem [{\citenamefont {Friederich}\ \emph {et~al.}(2010)\citenamefont
  {Friederich}, \citenamefont {Krahl},\ and\ \citenamefont
  {Wetterich}}]{Friederich2010}%
  \BibitemOpen
  \bibfield  {author} {\bibinfo {author} {\bibfnamefont {S.}~\bibnamefont
  {Friederich}}, \bibinfo {author} {\bibfnamefont {H.~C.}\ \bibnamefont
  {Krahl}},\ and\ \bibinfo {author} {\bibfnamefont {C.}~\bibnamefont
  {Wetterich}},\ }\bibfield  {title} {\bibinfo {title} {{Four-point vertex in
  the Hubbard model and partial bosonization}},\ }\href
  {https://doi.org/10.1103/PhysRevB.81.235108} {\bibfield  {journal} {\bibinfo
  {journal} {Phys. Rev. B}\ }\textbf {\bibinfo {volume} {81}},\ \bibinfo
  {pages} {235108} (\bibinfo {year} {2010})}\BibitemShut {NoStop}%
\bibitem [{\citenamefont {Denz}\ \emph {et~al.}(2020)\citenamefont {Denz},
  \citenamefont {Mitter}, \citenamefont {Pawlowski}, \citenamefont
  {Wetterich},\ and\ \citenamefont {Yamada}}]{Denz2020}%
  \BibitemOpen
  \bibfield  {author} {\bibinfo {author} {\bibfnamefont {T.}~\bibnamefont
  {Denz}}, \bibinfo {author} {\bibfnamefont {M.}~\bibnamefont {Mitter}},
  \bibinfo {author} {\bibfnamefont {J.~M.}\ \bibnamefont {Pawlowski}}, \bibinfo
  {author} {\bibfnamefont {C.}~\bibnamefont {Wetterich}},\ and\ \bibinfo
  {author} {\bibfnamefont {M.}~\bibnamefont {Yamada}},\ }\bibfield  {title}
  {\bibinfo {title} {{Partial bosonization for the two-dimensional Hubbard
  model}},\ }\href {https://doi.org/10.1103/PhysRevB.101.155115} {\bibfield
  {journal} {\bibinfo  {journal} {Phys. Rev. B}\ }\textbf {\bibinfo {volume}
  {101}},\ \bibinfo {pages} {155115} (\bibinfo {year} {2020})}\BibitemShut
  {NoStop}%
\bibitem [{\citenamefont {Karrasch}\ \emph {et~al.}(2008)\citenamefont
  {Karrasch}, \citenamefont {Hedden}, \citenamefont {Peters}, \citenamefont
  {Pruschke}, \citenamefont {Schönhammer},\ and\ \citenamefont
  {Meden}}]{Karrasch2008}%
  \BibitemOpen
  \bibfield  {author} {\bibinfo {author} {\bibfnamefont {C.}~\bibnamefont
  {Karrasch}}, \bibinfo {author} {\bibfnamefont {R.}~\bibnamefont {Hedden}},
  \bibinfo {author} {\bibfnamefont {R.}~\bibnamefont {Peters}}, \bibinfo
  {author} {\bibfnamefont {T.}~\bibnamefont {Pruschke}}, \bibinfo {author}
  {\bibfnamefont {K.}~\bibnamefont {Schönhammer}},\ and\ \bibinfo {author}
  {\bibfnamefont {V.}~\bibnamefont {Meden}},\ }\bibfield  {title} {\bibinfo
  {title} {{A finite-frequency functional renormalization group approach to the
  single impurity Anderson model}},\ }\href
  {https://doi.org/10.1088/0953-8984/20/34/345205} {\bibfield  {journal}
  {\bibinfo  {journal} {J. Phys. Condens. Matter}\ }\textbf {\bibinfo {volume}
  {20}},\ \bibinfo {pages} {345205} (\bibinfo {year} {2008})}\BibitemShut
  {NoStop}%
\bibitem [{\citenamefont {Husemann}\ and\ \citenamefont
  {Salmhofer}(2009)}]{Husemann2009}%
  \BibitemOpen
  \bibfield  {author} {\bibinfo {author} {\bibfnamefont {C.}~\bibnamefont
  {Husemann}}\ and\ \bibinfo {author} {\bibfnamefont {M.}~\bibnamefont
  {Salmhofer}},\ }\bibfield  {title} {\bibinfo {title} {{Efficient
  parametrization of the vertex function, $\ensuremath{\Omega}$ scheme, and the
  $t,{t}^{\ensuremath{'}}$ Hubbard model at van Hove filling}},\ }\href
  {https://doi.org/10.1103/PhysRevB.79.195125} {\bibfield  {journal} {\bibinfo
  {journal} {Phys. Rev. B}\ }\textbf {\bibinfo {volume} {79}},\ \bibinfo
  {pages} {195125} (\bibinfo {year} {2009})}\BibitemShut {NoStop}%
\bibitem [{\citenamefont {Wang}\ \emph {et~al.}(2012)\citenamefont {Wang},
  \citenamefont {Xiang}, \citenamefont {Wang}, \citenamefont {Wang},
  \citenamefont {Yang},\ and\ \citenamefont {Lee}}]{Wang2012}%
  \BibitemOpen
  \bibfield  {author} {\bibinfo {author} {\bibfnamefont {W.-S.}\ \bibnamefont
  {Wang}}, \bibinfo {author} {\bibfnamefont {Y.-Y.}\ \bibnamefont {Xiang}},
  \bibinfo {author} {\bibfnamefont {Q.-H.}\ \bibnamefont {Wang}}, \bibinfo
  {author} {\bibfnamefont {F.}~\bibnamefont {Wang}}, \bibinfo {author}
  {\bibfnamefont {F.}~\bibnamefont {Yang}},\ and\ \bibinfo {author}
  {\bibfnamefont {D.-H.}\ \bibnamefont {Lee}},\ }\bibfield  {title} {\bibinfo
  {title} {{Functional renormalization group and variational Monte Carlo
  studies of the electronic instabilities in graphene near $\frac{1}{4}$
  doping}},\ }\href {https://doi.org/10.1103/PhysRevB.85.035414} {\bibfield
  {journal} {\bibinfo  {journal} {Phys. Rev. B}\ }\textbf {\bibinfo {volume}
  {85}},\ \bibinfo {pages} {035414} (\bibinfo {year} {2012})}\BibitemShut
  {NoStop}%
\bibitem [{\citenamefont {Vilardi}\ \emph {et~al.}(2017)\citenamefont
  {Vilardi}, \citenamefont {Taranto},\ and\ \citenamefont
  {Metzner}}]{Vilardi2017}%
  \BibitemOpen
  \bibfield  {author} {\bibinfo {author} {\bibfnamefont {D.}~\bibnamefont
  {Vilardi}}, \bibinfo {author} {\bibfnamefont {C.}~\bibnamefont {Taranto}},\
  and\ \bibinfo {author} {\bibfnamefont {W.}~\bibnamefont {Metzner}},\
  }\bibfield  {title} {\bibinfo {title} {Nonseparable frequency dependence of
  the two-particle vertex in interacting fermion systems},\ }\href
  {https://doi.org/10.1103/PhysRevB.96.235110} {\bibfield  {journal} {\bibinfo
  {journal} {Phys. Rev. B}\ }\textbf {\bibinfo {volume} {96}},\ \bibinfo
  {pages} {235110} (\bibinfo {year} {2017})}\BibitemShut {NoStop}%
\bibitem [{\citenamefont {Tagliavini}\ \emph {et~al.}(2019)\citenamefont
  {Tagliavini}, \citenamefont {Hille}, \citenamefont {Kugler}, \citenamefont
  {Andergassen}, \citenamefont {Toschi},\ and\ \citenamefont
  {Honerkamp}}]{Tagliavini2019}%
  \BibitemOpen
  \bibfield  {author} {\bibinfo {author} {\bibfnamefont {A.}~\bibnamefont
  {Tagliavini}}, \bibinfo {author} {\bibfnamefont {C.}~\bibnamefont {Hille}},
  \bibinfo {author} {\bibfnamefont {F.~B.}\ \bibnamefont {Kugler}}, \bibinfo
  {author} {\bibfnamefont {S.}~\bibnamefont {Andergassen}}, \bibinfo {author}
  {\bibfnamefont {A.}~\bibnamefont {Toschi}},\ and\ \bibinfo {author}
  {\bibfnamefont {C.}~\bibnamefont {Honerkamp}},\ }\bibfield  {title} {\bibinfo
  {title} {{Multiloop functional renormalization group for the two-dimensional
  Hubbard model: Loop convergence of the response functions}},\ }\href
  {https://doi.org/10.21468/SciPostPhys.6.1.009} {\bibfield  {journal}
  {\bibinfo  {journal} {SciPost Phys.}\ }\textbf {\bibinfo {volume} {6}},\
  \bibinfo {pages} {9} (\bibinfo {year} {2019})}\BibitemShut {NoStop}%
\bibitem [{\citenamefont {Eckhardt}\ \emph {et~al.}(2018)\citenamefont
  {Eckhardt}, \citenamefont {Schober}, \citenamefont {Ehrlich},\ and\
  \citenamefont {Honerkamp}}]{Eckhardt2020}%
  \BibitemOpen
  \bibfield  {author} {\bibinfo {author} {\bibfnamefont {C.~J.}\ \bibnamefont
  {Eckhardt}}, \bibinfo {author} {\bibfnamefont {G.~A.~H.}\ \bibnamefont
  {Schober}}, \bibinfo {author} {\bibfnamefont {J.}~\bibnamefont {Ehrlich}},\
  and\ \bibinfo {author} {\bibfnamefont {C.}~\bibnamefont {Honerkamp}},\
  }\bibfield  {title} {\bibinfo {title} {{Truncated-unity parquet equations:
  Application to the repulsive Hubbard model}},\ }\href
  {https://doi.org/10.1103/PhysRevB.98.075143} {\bibfield  {journal} {\bibinfo
  {journal} {Phys. Rev. B}\ }\textbf {\bibinfo {volume} {98}},\ \bibinfo
  {pages} {075143} (\bibinfo {year} {2018})}\BibitemShut {NoStop}%
\bibitem [{\citenamefont {Hille}\ \emph {et~al.}(2020)\citenamefont {Hille},
  \citenamefont {Kugler}, \citenamefont {Eckhardt}, \citenamefont {He},
  \citenamefont {Kauch}, \citenamefont {Honerkamp}, \citenamefont {Toschi},\
  and\ \citenamefont {Andergassen}}]{Hille2020}%
  \BibitemOpen
  \bibfield  {author} {\bibinfo {author} {\bibfnamefont {C.}~\bibnamefont
  {Hille}}, \bibinfo {author} {\bibfnamefont {F.~B.}\ \bibnamefont {Kugler}},
  \bibinfo {author} {\bibfnamefont {C.~J.}\ \bibnamefont {Eckhardt}}, \bibinfo
  {author} {\bibfnamefont {Y.-Y.}\ \bibnamefont {He}}, \bibinfo {author}
  {\bibfnamefont {A.}~\bibnamefont {Kauch}}, \bibinfo {author} {\bibfnamefont
  {C.}~\bibnamefont {Honerkamp}}, \bibinfo {author} {\bibfnamefont
  {A.}~\bibnamefont {Toschi}},\ and\ \bibinfo {author} {\bibfnamefont
  {S.}~\bibnamefont {Andergassen}},\ }\bibfield  {title} {\bibinfo {title}
  {{Quantitative functional renormalization group description of the
  two-dimensional Hubbard model}},\ }\href
  {https://doi.org/10.1103/PhysRevResearch.2.033372} {\bibfield  {journal}
  {\bibinfo  {journal} {Phys. Rev. Research}\ }\textbf {\bibinfo {volume}
  {2}},\ \bibinfo {pages} {033372} (\bibinfo {year} {2020})}\BibitemShut
  {NoStop}%
\bibitem [{\citenamefont {Astretsov}\ \emph {et~al.}(2020)\citenamefont
  {Astretsov}, \citenamefont {Rohringer},\ and\ \citenamefont
  {Rubtsov}}]{Astretsov2020}%
  \BibitemOpen
  \bibfield  {author} {\bibinfo {author} {\bibfnamefont {G.~V.}\ \bibnamefont
  {Astretsov}}, \bibinfo {author} {\bibfnamefont {G.}~\bibnamefont
  {Rohringer}},\ and\ \bibinfo {author} {\bibfnamefont {A.~N.}\ \bibnamefont
  {Rubtsov}},\ }\bibfield  {title} {\bibinfo {title} {{Dual parquet scheme for
  the two-dimensional Hubbard model: Modeling low-energy physics of
  high-${T}_{c}$ cuprates with high momentum resolution}},\ }\href
  {https://doi.org/10.1103/PhysRevB.101.075109} {\bibfield  {journal} {\bibinfo
   {journal} {Phys. Rev. B}\ }\textbf {\bibinfo {volume} {101}},\ \bibinfo
  {pages} {075109} (\bibinfo {year} {2020})}\BibitemShut {NoStop}%
\bibitem [{\citenamefont {Astleithner}\ \emph {et~al.}(2020)\citenamefont
  {Astleithner}, \citenamefont {Kauch}, \citenamefont {Ribic},\ and\
  \citenamefont {Held}}]{Astleitner2020}%
  \BibitemOpen
  \bibfield  {author} {\bibinfo {author} {\bibfnamefont {K.}~\bibnamefont
  {Astleithner}}, \bibinfo {author} {\bibfnamefont {A.}~\bibnamefont {Kauch}},
  \bibinfo {author} {\bibfnamefont {T.}~\bibnamefont {Ribic}},\ and\ \bibinfo
  {author} {\bibfnamefont {K.}~\bibnamefont {Held}},\ }\bibfield  {title}
  {\bibinfo {title} {{Parquet dual fermion approach for the Falicov-Kimball
  model}},\ }\href {https://doi.org/10.1103/PhysRevB.101.165101} {\bibfield
  {journal} {\bibinfo  {journal} {Phys. Rev. B}\ }\textbf {\bibinfo {volume}
  {101}},\ \bibinfo {pages} {165101} (\bibinfo {year} {2020})}\BibitemShut
  {NoStop}%
\bibitem [{\citenamefont {Harkov}\ \emph
  {et~al.}(2021{\natexlab{a}})\citenamefont {Harkov}, \citenamefont {Vandelli},
  \citenamefont {Brener}, \citenamefont {Lichtenstein},\ and\ \citenamefont
  {Stepanov}}]{Harkov2021}%
  \BibitemOpen
  \bibfield  {author} {\bibinfo {author} {\bibfnamefont {V.}~\bibnamefont
  {Harkov}}, \bibinfo {author} {\bibfnamefont {M.}~\bibnamefont {Vandelli}},
  \bibinfo {author} {\bibfnamefont {S.}~\bibnamefont {Brener}}, \bibinfo
  {author} {\bibfnamefont {A.~I.}\ \bibnamefont {Lichtenstein}},\ and\ \bibinfo
  {author} {\bibfnamefont {E.~A.}\ \bibnamefont {Stepanov}},\ }\bibfield
  {title} {\bibinfo {title} {Impact of partially bosonized collective
  fluctuations on electronic degrees of freedom},\ }\href
  {https://doi.org/10.1103/PhysRevB.103.245123} {\bibfield  {journal} {\bibinfo
   {journal} {Phys. Rev. B}\ }\textbf {\bibinfo {volume} {103}},\ \bibinfo
  {pages} {245123} (\bibinfo {year} {2021}{\natexlab{a}})}\BibitemShut
  {NoStop}%
\bibitem [{\citenamefont {Schmalian}\ \emph {et~al.}(1999)\citenamefont
  {Schmalian}, \citenamefont {Pines},\ and\ \citenamefont
  {Stojkovi\ifmmode~\acute{c}\else \'{c}\fi{}}}]{Schmalian1999}%
  \BibitemOpen
  \bibfield  {author} {\bibinfo {author} {\bibfnamefont {J.}~\bibnamefont
  {Schmalian}}, \bibinfo {author} {\bibfnamefont {D.}~\bibnamefont {Pines}},\
  and\ \bibinfo {author} {\bibfnamefont {B.}~\bibnamefont
  {Stojkovi\ifmmode~\acute{c}\else \'{c}\fi{}}},\ }\bibfield  {title} {\bibinfo
  {title} {{Microscopic theory of weak pseudogap behavior in the underdoped
  cuprate superconductors: General theory and quasiparticle properties}},\
  }\href {https://doi.org/10.1103/PhysRevB.60.667} {\bibfield  {journal}
  {\bibinfo  {journal} {Phys. Rev. B}\ }\textbf {\bibinfo {volume} {60}},\
  \bibinfo {pages} {667} (\bibinfo {year} {1999})}\BibitemShut {NoStop}%
\bibitem [{\citenamefont {Sadovskii}(2019)}]{Sadovskii2019}%
  \BibitemOpen
  \bibfield  {author} {\bibinfo {author} {\bibfnamefont {M.~V.}\ \bibnamefont
  {Sadovskii}},\ }\bibfield  {title} {\bibinfo {title} {Diagrammatics}\
  }(\bibinfo  {publisher} {World Scientific},\ \bibinfo {year}
  {2019})\BibitemShut {NoStop}%
\bibitem [{\citenamefont {Hedin}(1965)}]{Hedin1965}%
  \BibitemOpen
  \bibfield  {author} {\bibinfo {author} {\bibfnamefont {L.}~\bibnamefont
  {Hedin}},\ }\bibfield  {title} {\bibinfo {title} {{New Method for Calculating
  the One-Particle Green's Function with Application to the Electron-Gas
  Problem}},\ }\href {https://doi.org/10.1103/PhysRev.139.A796} {\bibfield
  {journal} {\bibinfo  {journal} {Phys. Rev.}\ }\textbf {\bibinfo {volume}
  {139}},\ \bibinfo {pages} {A796} (\bibinfo {year} {1965})}\BibitemShut
  {NoStop}%
\bibitem [{\citenamefont {Husemann}\ \emph {et~al.}(2012)\citenamefont
  {Husemann}, \citenamefont {Giering},\ and\ \citenamefont
  {Salmhofer}}]{Husemann2012}%
  \BibitemOpen
  \bibfield  {author} {\bibinfo {author} {\bibfnamefont {C.}~\bibnamefont
  {Husemann}}, \bibinfo {author} {\bibfnamefont {K.-U.}\ \bibnamefont
  {Giering}},\ and\ \bibinfo {author} {\bibfnamefont {M.}~\bibnamefont
  {Salmhofer}},\ }\bibfield  {title} {\bibinfo {title} {{Frequency-dependent
  vertex functions of the ($t,{t}^{\ensuremath{'}}$) Hubbard model at weak
  coupling}},\ }\href {https://doi.org/10.1103/PhysRevB.85.075121} {\bibfield
  {journal} {\bibinfo  {journal} {Phys. Rev. B}\ }\textbf {\bibinfo {volume}
  {85}},\ \bibinfo {pages} {075121} (\bibinfo {year} {2012})}\BibitemShut
  {NoStop}%
\bibitem [{\citenamefont {Diehl}\ \emph
  {et~al.}(2007{\natexlab{a}})\citenamefont {Diehl}, \citenamefont {Gies},
  \citenamefont {Pawlowski},\ and\ \citenamefont {Wetterich}}]{Diehl2007}%
  \BibitemOpen
  \bibfield  {author} {\bibinfo {author} {\bibfnamefont {S.}~\bibnamefont
  {Diehl}}, \bibinfo {author} {\bibfnamefont {H.}~\bibnamefont {Gies}},
  \bibinfo {author} {\bibfnamefont {J.~M.}\ \bibnamefont {Pawlowski}},\ and\
  \bibinfo {author} {\bibfnamefont {C.}~\bibnamefont {Wetterich}},\ }\bibfield
  {title} {\bibinfo {title} {{Flow equations for the BCS-BEC crossover}},\
  }\href {https://doi.org/10.1103/PhysRevA.76.021602} {\bibfield  {journal}
  {\bibinfo  {journal} {Phys. Rev. A}\ }\textbf {\bibinfo {volume} {76}},\
  \bibinfo {pages} {021602} (\bibinfo {year} {2007}{\natexlab{a}})}\BibitemShut
  {NoStop}%
\bibitem [{\citenamefont {Diehl}\ \emph
  {et~al.}(2007{\natexlab{b}})\citenamefont {Diehl}, \citenamefont {Gies},
  \citenamefont {Pawlowski},\ and\ \citenamefont {Wetterich}}]{Diehl2007_II}%
  \BibitemOpen
  \bibfield  {author} {\bibinfo {author} {\bibfnamefont {S.}~\bibnamefont
  {Diehl}}, \bibinfo {author} {\bibfnamefont {H.}~\bibnamefont {Gies}},
  \bibinfo {author} {\bibfnamefont {J.~M.}\ \bibnamefont {Pawlowski}},\ and\
  \bibinfo {author} {\bibfnamefont {C.}~\bibnamefont {Wetterich}},\ }\bibfield
  {title} {\bibinfo {title} {{Renormalization flow and universality for
  ultracold fermionic atoms}},\ }\href
  {https://doi.org/10.1103/PhysRevA.76.053627} {\bibfield  {journal} {\bibinfo
  {journal} {Phys. Rev. A}\ }\textbf {\bibinfo {volume} {76}},\ \bibinfo
  {pages} {053627} (\bibinfo {year} {2007}{\natexlab{b}})}\BibitemShut
  {NoStop}%
\bibitem [{\citenamefont {Strack}\ \emph {et~al.}(2008)\citenamefont {Strack},
  \citenamefont {Gersch},\ and\ \citenamefont {Metzner}}]{Strack2008}%
  \BibitemOpen
  \bibfield  {author} {\bibinfo {author} {\bibfnamefont {P.}~\bibnamefont
  {Strack}}, \bibinfo {author} {\bibfnamefont {R.}~\bibnamefont {Gersch}},\
  and\ \bibinfo {author} {\bibfnamefont {W.}~\bibnamefont {Metzner}},\
  }\bibfield  {title} {\bibinfo {title} {Renormalization group flow for
  fermionic superfluids at zero temperature},\ }\href
  {https://doi.org/10.1103/PhysRevB.78.014522} {\bibfield  {journal} {\bibinfo
  {journal} {Phys. Rev. B}\ }\textbf {\bibinfo {volume} {78}},\ \bibinfo
  {pages} {014522} (\bibinfo {year} {2008})}\BibitemShut {NoStop}%
\bibitem [{\citenamefont {Bartosch}\ \emph {et~al.}(2009)\citenamefont
  {Bartosch}, \citenamefont {Kopietz},\ and\ \citenamefont
  {Ferraz}}]{Bartosch2009}%
  \BibitemOpen
  \bibfield  {author} {\bibinfo {author} {\bibfnamefont {L.}~\bibnamefont
  {Bartosch}}, \bibinfo {author} {\bibfnamefont {P.}~\bibnamefont {Kopietz}},\
  and\ \bibinfo {author} {\bibfnamefont {A.}~\bibnamefont {Ferraz}},\
  }\bibfield  {title} {\bibinfo {title} {{Renormalization of the BCS-BEC
  crossover by order-parameter fluctuations}},\ }\href
  {https://doi.org/10.1103/PhysRevB.80.104514} {\bibfield  {journal} {\bibinfo
  {journal} {Phys. Rev. B}\ }\textbf {\bibinfo {volume} {80}},\ \bibinfo
  {pages} {104514} (\bibinfo {year} {2009})}\BibitemShut {NoStop}%
\bibitem [{\citenamefont {Obert}\ \emph {et~al.}(2013)\citenamefont {Obert},
  \citenamefont {Husemann},\ and\ \citenamefont {Metzner}}]{Obert2013}%
  \BibitemOpen
  \bibfield  {author} {\bibinfo {author} {\bibfnamefont {B.}~\bibnamefont
  {Obert}}, \bibinfo {author} {\bibfnamefont {C.}~\bibnamefont {Husemann}},\
  and\ \bibinfo {author} {\bibfnamefont {W.}~\bibnamefont {Metzner}},\
  }\bibfield  {title} {\bibinfo {title} {{Low-energy singularities in the
  ground state of fermionic superfluids}},\ }\href
  {https://doi.org/10.1103/PhysRevB.88.144508} {\bibfield  {journal} {\bibinfo
  {journal} {Phys. Rev. B}\ }\textbf {\bibinfo {volume} {88}},\ \bibinfo
  {pages} {144508} (\bibinfo {year} {2013})}\BibitemShut {NoStop}%
\bibitem [{\citenamefont {Baier}\ \emph {et~al.}(2004)\citenamefont {Baier},
  \citenamefont {Bick},\ and\ \citenamefont {Wetterich}}]{Baier2004}%
  \BibitemOpen
  \bibfield  {author} {\bibinfo {author} {\bibfnamefont {T.}~\bibnamefont
  {Baier}}, \bibinfo {author} {\bibfnamefont {E.}~\bibnamefont {Bick}},\ and\
  \bibinfo {author} {\bibfnamefont {C.}~\bibnamefont {Wetterich}},\ }\bibfield
  {title} {\bibinfo {title} {{Temperature dependence of antiferromagnetic order
  in the Hubbard model}},\ }\href {https://doi.org/10.1103/PhysRevB.70.125111}
  {\bibfield  {journal} {\bibinfo  {journal} {Phys. Rev. B}\ }\textbf {\bibinfo
  {volume} {70}},\ \bibinfo {pages} {125111} (\bibinfo {year}
  {2004})}\BibitemShut {NoStop}%
\bibitem [{\citenamefont {Friederich}\ \emph {et~al.}(2011)\citenamefont
  {Friederich}, \citenamefont {Krahl},\ and\ \citenamefont
  {Wetterich}}]{Friederich2011}%
  \BibitemOpen
  \bibfield  {author} {\bibinfo {author} {\bibfnamefont {S.}~\bibnamefont
  {Friederich}}, \bibinfo {author} {\bibfnamefont {H.~C.}\ \bibnamefont
  {Krahl}},\ and\ \bibinfo {author} {\bibfnamefont {C.}~\bibnamefont
  {Wetterich}},\ }\bibfield  {title} {\bibinfo {title} {{Functional
  renormalization for spontaneous symmetry breaking in the Hubbard model}},\
  }\href {https://doi.org/10.1103/PhysRevB.83.155125} {\bibfield  {journal}
  {\bibinfo  {journal} {Phys. Rev. B}\ }\textbf {\bibinfo {volume} {83}},\
  \bibinfo {pages} {155125} (\bibinfo {year} {2011})}\BibitemShut {NoStop}%
\bibitem [{\citenamefont {Stepanov}\ \emph {et~al.}(2018)\citenamefont
  {Stepanov}, \citenamefont {Brener}, \citenamefont {Krien}, \citenamefont
  {Harland}, \citenamefont {Lichtenstein},\ and\ \citenamefont
  {Katsnelson}}]{Stepanov2018}%
  \BibitemOpen
  \bibfield  {author} {\bibinfo {author} {\bibfnamefont {E.~A.}\ \bibnamefont
  {Stepanov}}, \bibinfo {author} {\bibfnamefont {S.}~\bibnamefont {Brener}},
  \bibinfo {author} {\bibfnamefont {F.}~\bibnamefont {Krien}}, \bibinfo
  {author} {\bibfnamefont {M.}~\bibnamefont {Harland}}, \bibinfo {author}
  {\bibfnamefont {A.~I.}\ \bibnamefont {Lichtenstein}},\ and\ \bibinfo {author}
  {\bibfnamefont {M.~I.}\ \bibnamefont {Katsnelson}},\ }\bibfield  {title}
  {\bibinfo {title} {Effective heisenberg model and exchange interaction for
  strongly correlated systems},\ }\href
  {https://doi.org/10.1103/PhysRevLett.121.037204} {\bibfield  {journal}
  {\bibinfo  {journal} {Phys. Rev. Lett.}\ }\textbf {\bibinfo {volume} {121}},\
  \bibinfo {pages} {037204} (\bibinfo {year} {2018})}\BibitemShut {NoStop}%
\bibitem [{\citenamefont {Stepanov}\ \emph
  {et~al.}(2019{\natexlab{a}})\citenamefont {Stepanov}, \citenamefont {Huber},
  \citenamefont {Lichtenstein},\ and\ \citenamefont
  {Katsnelson}}]{Stepanov2019A}%
  \BibitemOpen
  \bibfield  {author} {\bibinfo {author} {\bibfnamefont {E.~A.}\ \bibnamefont
  {Stepanov}}, \bibinfo {author} {\bibfnamefont {A.}~\bibnamefont {Huber}},
  \bibinfo {author} {\bibfnamefont {A.~I.}\ \bibnamefont {Lichtenstein}},\ and\
  \bibinfo {author} {\bibfnamefont {M.~I.}\ \bibnamefont {Katsnelson}},\
  }\bibfield  {title} {\bibinfo {title} {Effective ising model for correlated
  systems with charge ordering},\ }\href
  {https://doi.org/10.1103/PhysRevB.99.115124} {\bibfield  {journal} {\bibinfo
  {journal} {Phys. Rev. B}\ }\textbf {\bibinfo {volume} {99}},\ \bibinfo
  {pages} {115124} (\bibinfo {year} {2019}{\natexlab{a}})}\BibitemShut
  {NoStop}%
\bibitem [{\citenamefont {Stepanov}\ \emph
  {et~al.}(2019{\natexlab{b}})\citenamefont {Stepanov}, \citenamefont
  {Harkov},\ and\ \citenamefont {Lichtenstein}}]{Stepanov2019}%
  \BibitemOpen
  \bibfield  {author} {\bibinfo {author} {\bibfnamefont {E.~A.}\ \bibnamefont
  {Stepanov}}, \bibinfo {author} {\bibfnamefont {V.}~\bibnamefont {Harkov}},\
  and\ \bibinfo {author} {\bibfnamefont {A.~I.}\ \bibnamefont {Lichtenstein}},\
  }\bibfield  {title} {\bibinfo {title} {Consistent partial bosonization of the
  extended hubbard model},\ }\href
  {https://doi.org/10.1103/PhysRevB.100.205115} {\bibfield  {journal} {\bibinfo
   {journal} {Phys. Rev. B}\ }\textbf {\bibinfo {volume} {100}},\ \bibinfo
  {pages} {205115} (\bibinfo {year} {2019}{\natexlab{b}})}\BibitemShut
  {NoStop}%
\bibitem [{Note1()}]{Note1}%
  \BibitemOpen
  \bibinfo {note} {See Ref.~\cite {Qin21} for a recent overview of
  computational results for the 2D Hubbard model.}\BibitemShut {Stop}%
\bibitem [{\citenamefont {Gunnarsson}\ \emph {et~al.}(2015)\citenamefont
  {Gunnarsson}, \citenamefont {Sch\"afer}, \citenamefont {LeBlanc},
  \citenamefont {Gull}, \citenamefont {Merino}, \citenamefont {Sangiovanni},
  \citenamefont {Rohringer},\ and\ \citenamefont {Toschi}}]{Gunnarsson2015}%
  \BibitemOpen
  \bibfield  {author} {\bibinfo {author} {\bibfnamefont {O.}~\bibnamefont
  {Gunnarsson}}, \bibinfo {author} {\bibfnamefont {T.}~\bibnamefont
  {Sch\"afer}}, \bibinfo {author} {\bibfnamefont {J.~P.~F.}\ \bibnamefont
  {LeBlanc}}, \bibinfo {author} {\bibfnamefont {E.}~\bibnamefont {Gull}},
  \bibinfo {author} {\bibfnamefont {J.}~\bibnamefont {Merino}}, \bibinfo
  {author} {\bibfnamefont {G.}~\bibnamefont {Sangiovanni}}, \bibinfo {author}
  {\bibfnamefont {G.}~\bibnamefont {Rohringer}},\ and\ \bibinfo {author}
  {\bibfnamefont {A.}~\bibnamefont {Toschi}},\ }\bibfield  {title} {\bibinfo
  {title} {{Fluctuation Diagnostics of the Electron Self-Energy: Origin of the
  Pseudogap Physics}},\ }\href {https://doi.org/10.1103/PhysRevLett.114.236402}
  {\bibfield  {journal} {\bibinfo  {journal} {Phys. Rev. Lett.}\ }\textbf
  {\bibinfo {volume} {114}},\ \bibinfo {pages} {236402} (\bibinfo {year}
  {2015})}\BibitemShut {NoStop}%
\bibitem [{\citenamefont {Krien}\ \emph
  {et~al.}(2020{\natexlab{b}})\citenamefont {Krien}, \citenamefont
  {Lichtenstein},\ and\ \citenamefont {Rohringer}}]{Rohringer2020}%
  \BibitemOpen
  \bibfield  {author} {\bibinfo {author} {\bibfnamefont {F.}~\bibnamefont
  {Krien}}, \bibinfo {author} {\bibfnamefont {A.~I.}\ \bibnamefont
  {Lichtenstein}},\ and\ \bibinfo {author} {\bibfnamefont {G.}~\bibnamefont
  {Rohringer}},\ }\bibfield  {title} {\bibinfo {title} {{Fluctuation diagnostic
  of the nodal/antinodal dichotomy in the Hubbard model at weak coupling: A
  parquet dual fermion approach}},\ }\href
  {https://doi.org/10.1103/PhysRevB.102.235133} {\bibfield  {journal} {\bibinfo
   {journal} {Phys. Rev. B}\ }\textbf {\bibinfo {volume} {102}},\ \bibinfo
  {pages} {235133} (\bibinfo {year} {2020}{\natexlab{b}})}\BibitemShut
  {NoStop}%
\bibitem [{\citenamefont {Schäfer}\ and\ \citenamefont
  {Toschi}(2021)}]{Schaefer2021}%
  \BibitemOpen
  \bibfield  {author} {\bibinfo {author} {\bibfnamefont {T.}~\bibnamefont
  {Schäfer}}\ and\ \bibinfo {author} {\bibfnamefont {A.}~\bibnamefont
  {Toschi}},\ }\bibfield  {title} {\bibinfo {title} {How to read between the
  lines of electronic spectra: the diagnostics of fluctuations in strongly
  correlated electron systems},\ }\href
  {http://iopscience.iop.org/article/10.1088/1361-648X/abeb44} {\bibfield
  {journal} {\bibinfo  {journal} {Journal of Physics: Condensed Matter}\ }
  (\bibinfo {year} {2021})}\BibitemShut {NoStop}%
\bibitem [{\citenamefont {Re}\ and\ \citenamefont
  {Rohringer}(2021)}]{Delre2021}%
  \BibitemOpen
  \bibfield  {author} {\bibinfo {author} {\bibfnamefont {L.~D.}\ \bibnamefont
  {Re}}\ and\ \bibinfo {author} {\bibfnamefont {G.}~\bibnamefont {Rohringer}},\
  }\href@noop {} {\bibinfo {title} {Fluctuations diagnostic of the spin
  susceptibility: Neel ordering revisited in dmft}} (\bibinfo {year} {2021}),\
  \Eprint {https://arxiv.org/abs/2104.11737} {arXiv:2104.11737} \BibitemShut
  {NoStop}%
\bibitem [{\citenamefont {Krien}\ and\ \citenamefont
  {Valli}(2019)}]{Krien2019_II}%
  \BibitemOpen
  \bibfield  {author} {\bibinfo {author} {\bibfnamefont {F.}~\bibnamefont
  {Krien}}\ and\ \bibinfo {author} {\bibfnamefont {A.}~\bibnamefont {Valli}},\
  }\bibfield  {title} {\bibinfo {title} {{Parquetlike equations for the Hedin
  three-leg vertex}},\ }\href {https://doi.org/10.1103/PhysRevB.100.245147}
  {\bibfield  {journal} {\bibinfo  {journal} {Phys. Rev. B}\ }\textbf {\bibinfo
  {volume} {100}},\ \bibinfo {pages} {245147} (\bibinfo {year}
  {2019})}\BibitemShut {NoStop}%
\bibitem [{\citenamefont {Kugler}\ and\ \citenamefont {von
  Delft}(2018{\natexlab{a}})}]{Kugler2018_I}%
  \BibitemOpen
  \bibfield  {author} {\bibinfo {author} {\bibfnamefont {F.~B.}\ \bibnamefont
  {Kugler}}\ and\ \bibinfo {author} {\bibfnamefont {J.}~\bibnamefont {von
  Delft}},\ }\bibfield  {title} {\bibinfo {title} {{Multiloop Functional
  Renormalization Group That Sums Up All Parquet Diagrams}},\ }\href
  {https://doi.org/10.1103/PhysRevLett.120.057403} {\bibfield  {journal}
  {\bibinfo  {journal} {Phys. Rev. Lett.}\ }\textbf {\bibinfo {volume} {120}},\
  \bibinfo {pages} {057403} (\bibinfo {year} {2018}{\natexlab{a}})}\BibitemShut
  {NoStop}%
\bibitem [{\citenamefont {Kugler}\ and\ \citenamefont {von
  Delft}(2018{\natexlab{b}})}]{Kugler2018_II}%
  \BibitemOpen
  \bibfield  {author} {\bibinfo {author} {\bibfnamefont {F.~B.}\ \bibnamefont
  {Kugler}}\ and\ \bibinfo {author} {\bibfnamefont {J.}~\bibnamefont {von
  Delft}},\ }\bibfield  {title} {\bibinfo {title} {{Multiloop functional
  renormalization group for general models}},\ }\href
  {https://doi.org/10.1103/PhysRevB.97.035162} {\bibfield  {journal} {\bibinfo
  {journal} {Phys. Rev. B}\ }\textbf {\bibinfo {volume} {97}},\ \bibinfo
  {pages} {035162} (\bibinfo {year} {2018}{\natexlab{b}})}\BibitemShut
  {NoStop}%
\bibitem [{Note2()}]{Note2}%
  \BibitemOpen
  \bibinfo {note} {Evidently, this simplification does not cause significant
  deviations from the conventional fRG in the weak-coupling regime. We also
  note that the latter corresponds to the SBE formalism with the inclusion of
  the rest functions, see App.~\ref {app: fermionic weak
  coupling})}\BibitemShut {NoStop}%
\bibitem [{\citenamefont {Sch\"afer}\ \emph {et~al.}(2021)\citenamefont
  {Sch\"afer}, \citenamefont {Wentzell}, \citenamefont {\ifmmode~\check{S}\else
  \v{S}\fi{}imkovic}, \citenamefont {He}, \citenamefont {Hille}, \citenamefont
  {Klett}, \citenamefont {Eckhardt}, \citenamefont {Arzhang}, \citenamefont
  {Harkov}, \citenamefont {Le~R\'egent}, \citenamefont {Kirsch}, \citenamefont
  {Wang}, \citenamefont {Kim}, \citenamefont {Kozik}, \citenamefont {Stepanov},
  \citenamefont {Kauch}, \citenamefont {Andergassen}, \citenamefont {Hansmann},
  \citenamefont {Rohe}, \citenamefont {Vilk}, \citenamefont {LeBlanc},
  \citenamefont {Zhang}, \citenamefont {Tremblay}, \citenamefont {Ferrero},
  \citenamefont {Parcollet},\ and\ \citenamefont {Georges}}]{Review2021}%
  \BibitemOpen
  \bibfield  {author} {\bibinfo {author} {\bibfnamefont {T.}~\bibnamefont
  {Sch\"afer}}, \bibinfo {author} {\bibfnamefont {N.}~\bibnamefont {Wentzell}},
  \bibinfo {author} {\bibfnamefont {F.}~\bibnamefont {\ifmmode~\check{S}\else
  \v{S}\fi{}imkovic}}, \bibinfo {author} {\bibfnamefont {Y.-Y.}\ \bibnamefont
  {He}}, \bibinfo {author} {\bibfnamefont {C.}~\bibnamefont {Hille}}, \bibinfo
  {author} {\bibfnamefont {M.}~\bibnamefont {Klett}}, \bibinfo {author}
  {\bibfnamefont {C.~J.}\ \bibnamefont {Eckhardt}}, \bibinfo {author}
  {\bibfnamefont {B.}~\bibnamefont {Arzhang}}, \bibinfo {author} {\bibfnamefont
  {V.}~\bibnamefont {Harkov}}, \bibinfo {author} {\bibfnamefont {F.-M.}\
  \bibnamefont {Le~R\'egent}}, \bibinfo {author} {\bibfnamefont
  {A.}~\bibnamefont {Kirsch}}, \bibinfo {author} {\bibfnamefont
  {Y.}~\bibnamefont {Wang}}, \bibinfo {author} {\bibfnamefont {A.~J.}\
  \bibnamefont {Kim}}, \bibinfo {author} {\bibfnamefont {E.}~\bibnamefont
  {Kozik}}, \bibinfo {author} {\bibfnamefont {E.~A.}\ \bibnamefont {Stepanov}},
  \bibinfo {author} {\bibfnamefont {A.}~\bibnamefont {Kauch}}, \bibinfo
  {author} {\bibfnamefont {S.}~\bibnamefont {Andergassen}}, \bibinfo {author}
  {\bibfnamefont {P.}~\bibnamefont {Hansmann}}, \bibinfo {author}
  {\bibfnamefont {D.}~\bibnamefont {Rohe}}, \bibinfo {author} {\bibfnamefont
  {Y.~M.}\ \bibnamefont {Vilk}}, \bibinfo {author} {\bibfnamefont {J.~P.~F.}\
  \bibnamefont {LeBlanc}}, \bibinfo {author} {\bibfnamefont {S.}~\bibnamefont
  {Zhang}}, \bibinfo {author} {\bibfnamefont {A.-M.~S.}\ \bibnamefont
  {Tremblay}}, \bibinfo {author} {\bibfnamefont {M.}~\bibnamefont {Ferrero}},
  \bibinfo {author} {\bibfnamefont {O.}~\bibnamefont {Parcollet}},\ and\
  \bibinfo {author} {\bibfnamefont {A.}~\bibnamefont {Georges}},\ }\bibfield
  {title} {\bibinfo {title} {Tracking the footprints of spin fluctuations: A
  multimethod, multimessenger study of the two-dimensional hubbard model},\
  }\href {https://doi.org/10.1103/PhysRevX.11.011058} {\bibfield  {journal}
  {\bibinfo  {journal} {Phys. Rev. X}\ }\textbf {\bibinfo {volume} {11}},\
  \bibinfo {pages} {011058} (\bibinfo {year} {2021})}\BibitemShut {NoStop}%
\bibitem [{\citenamefont {Rohringer}\ \emph {et~al.}(2011)\citenamefont
  {Rohringer}, \citenamefont {Toschi}, \citenamefont {Katanin},\ and\
  \citenamefont {Held}}]{Rohringer2011}%
  \BibitemOpen
  \bibfield  {author} {\bibinfo {author} {\bibfnamefont {G.}~\bibnamefont
  {Rohringer}}, \bibinfo {author} {\bibfnamefont {A.}~\bibnamefont {Toschi}},
  \bibinfo {author} {\bibfnamefont {A.}~\bibnamefont {Katanin}},\ and\ \bibinfo
  {author} {\bibfnamefont {K.}~\bibnamefont {Held}},\ }\bibfield  {title}
  {\bibinfo {title} {{Critical Properties of the Half-Filled Hubbard Model in
  Three Dimensions}},\ }\href {https://doi.org/10.1103/PhysRevLett.107.256402}
  {\bibfield  {journal} {\bibinfo  {journal} {Phys. Rev. Lett.}\ }\textbf
  {\bibinfo {volume} {107}},\ \bibinfo {pages} {256402} (\bibinfo {year}
  {2011})}\BibitemShut {NoStop}%
\bibitem [{\citenamefont {Rohringer}\ and\ \citenamefont
  {Toschi}(2016)}]{Rohringer2016}%
  \BibitemOpen
  \bibfield  {author} {\bibinfo {author} {\bibfnamefont {G.}~\bibnamefont
  {Rohringer}}\ and\ \bibinfo {author} {\bibfnamefont {A.}~\bibnamefont
  {Toschi}},\ }\bibfield  {title} {\bibinfo {title} {{Impact of nonlocal
  correlations over different energy scales: A dynamical vertex approximation
  study}},\ }\href {https://doi.org/10.1103/PhysRevB.94.125144} {\bibfield
  {journal} {\bibinfo  {journal} {Phys. Rev. B}\ }\textbf {\bibinfo {volume}
  {94}},\ \bibinfo {pages} {125144} (\bibinfo {year} {2016})}\BibitemShut
  {NoStop}%
\bibitem [{Note3()}]{Note3}%
  \BibitemOpen
  \bibinfo {note} {This rigorously holds for the exact solutions of the problem
  and/or for approximations based on a definite subset of diagrams, such as
  RPA, PA, DMFT. At the level of a truncated (1$\ell $) fRG/DMF\protect
  \textsuperscript 2RG, deviations between the susceptibilities computed via
  the flow and those computed by summing the internal frequencies
  (post-processing) may occur \cite {Kugler2018_I,Tagliavini2019}.}\BibitemShut
  {Stop}%
\bibitem [{Note4()}]{Note4}%
  \BibitemOpen
  \bibinfo {note} {Precisely, these are generalized susceptibilities of the
  auxiliary AIM associated to the corresponding self-consistent DMFT solution.
  We recall that in the limit of $d\rightarrow \infty $ the physical
  susceptibilities computed via the double Matsubara summation exactly coincide
  with the local (momentum-summed) susceptibilities of the
  lattice.}\BibitemShut {Stop}%
\bibitem [{Note5()}]{Note5}%
  \BibitemOpen
  \bibinfo {note} {For $U=16t$, the negative values in the main diagonal are
  counterbalanced by the positive contributions at high frequencies, giving
  rise to an overall positive charge response.}\BibitemShut {Stop}%
\bibitem [{\citenamefont {Jani\v{s}}\ and\ \citenamefont
  {Pokorn\'y}(2014)}]{Janis2014}%
  \BibitemOpen
  \bibfield  {author} {\bibinfo {author} {\bibfnamefont {V.}~\bibnamefont
  {Jani\v{s}}}\ and\ \bibinfo {author} {\bibfnamefont {V.}~\bibnamefont
  {Pokorn\'y}},\ }\bibfield  {title} {\bibinfo {title} {Critical
  metal-insulator transition and divergence in a two-particle irreducible
  vertex in disordered and interacting electron systems},\ }\href
  {https://doi.org/10.1103/PhysRevB.90.045143} {\bibfield  {journal} {\bibinfo
  {journal} {Phys. Rev. B}\ }\textbf {\bibinfo {volume} {90}},\ \bibinfo
  {pages} {045143} (\bibinfo {year} {2014})}\BibitemShut {NoStop}%
\bibitem [{\citenamefont {Kozik}\ \emph {et~al.}(2015)\citenamefont {Kozik},
  \citenamefont {Ferrero},\ and\ \citenamefont {Georges}}]{Kozik2015}%
  \BibitemOpen
  \bibfield  {author} {\bibinfo {author} {\bibfnamefont {E.}~\bibnamefont
  {Kozik}}, \bibinfo {author} {\bibfnamefont {M.}~\bibnamefont {Ferrero}},\
  and\ \bibinfo {author} {\bibfnamefont {A.}~\bibnamefont {Georges}},\
  }\bibfield  {title} {\bibinfo {title} {{Nonexistence of the Luttinger-Ward
  Functional and Misleading Convergence of Skeleton Diagrammatic Series for
  Hubbard-Like Models}},\ }\href
  {https://doi.org/10.1103/PhysRevLett.114.156402} {\bibfield  {journal}
  {\bibinfo  {journal} {Phys. Rev. Lett.}\ }\textbf {\bibinfo {volume} {114}},\
  \bibinfo {pages} {156402} (\bibinfo {year} {2015})}\BibitemShut {NoStop}%
\bibitem [{\citenamefont {Stan}\ \emph {et~al.}(2015)\citenamefont {Stan},
  \citenamefont {Romaniello}, \citenamefont {Rigamonti}, \citenamefont
  {Reining},\ and\ \citenamefont {Berger}}]{Stan2015}%
  \BibitemOpen
  \bibfield  {author} {\bibinfo {author} {\bibfnamefont {A.}~\bibnamefont
  {Stan}}, \bibinfo {author} {\bibfnamefont {P.}~\bibnamefont {Romaniello}},
  \bibinfo {author} {\bibfnamefont {S.}~\bibnamefont {Rigamonti}}, \bibinfo
  {author} {\bibfnamefont {L.}~\bibnamefont {Reining}},\ and\ \bibinfo {author}
  {\bibfnamefont {J.~A.}\ \bibnamefont {Berger}},\ }\bibfield  {title}
  {\bibinfo {title} {Unphysical and physical solutions in many-body theories:
  from weak to strong correlation},\ }\href
  {http://stacks.iop.org/1367-2630/17/i=9/a=093045} {\bibfield  {journal}
  {\bibinfo  {journal} {New J. Phys.}\ }\textbf {\bibinfo {volume} {17}},\
  \bibinfo {pages} {093045} (\bibinfo {year} {2015})}\BibitemShut {NoStop}%
\bibitem [{\citenamefont {Rossi}\ and\ \citenamefont
  {Werner}(2015)}]{Rossi2015}%
  \BibitemOpen
  \bibfield  {author} {\bibinfo {author} {\bibfnamefont {R.}~\bibnamefont
  {Rossi}}\ and\ \bibinfo {author} {\bibfnamefont {F.}~\bibnamefont {Werner}},\
  }\bibfield  {title} {\bibinfo {title} {{Skeleton series and multivaluedness
  of the self-energy functional in zero space-time dimensions}},\ }\href
  {https://doi.org/10.1088/1751-8113/48/48/485202} {\bibfield  {journal}
  {\bibinfo  {journal} {J. Phys. A}\ }\textbf {\bibinfo {volume} {48}},\
  \bibinfo {pages} {485202} (\bibinfo {year} {2015})}\BibitemShut {NoStop}%
\bibitem [{\citenamefont {Tarantino}\ \emph {et~al.}(2017)\citenamefont
  {Tarantino}, \citenamefont {Romaniello}, \citenamefont {Berger},\ and\
  \citenamefont {Reining}}]{Tarantino2017}%
  \BibitemOpen
  \bibfield  {author} {\bibinfo {author} {\bibfnamefont {W.}~\bibnamefont
  {Tarantino}}, \bibinfo {author} {\bibfnamefont {P.}~\bibnamefont
  {Romaniello}}, \bibinfo {author} {\bibfnamefont {J.~A.}\ \bibnamefont
  {Berger}},\ and\ \bibinfo {author} {\bibfnamefont {L.}~\bibnamefont
  {Reining}},\ }\bibfield  {title} {\bibinfo {title} {{Self-consistent Dyson
  equation and self-energy functionals: An analysis and illustration on the
  example of the Hubbard atom}},\ }\href
  {https://doi.org/10.1103/PhysRevB.96.045124} {\bibfield  {journal} {\bibinfo
  {journal} {Phys. Rev. B}\ }\textbf {\bibinfo {volume} {96}},\ \bibinfo
  {pages} {045124} (\bibinfo {year} {2017})}\BibitemShut {NoStop}%
\bibitem [{\citenamefont {Kim}\ \emph {et~al.}(2020)\citenamefont {Kim},
  \citenamefont {Werner},\ and\ \citenamefont {Kozik}}]{Kozik2020}%
  \BibitemOpen
  \bibfield  {author} {\bibinfo {author} {\bibfnamefont {A.~J.}\ \bibnamefont
  {Kim}}, \bibinfo {author} {\bibfnamefont {P.}~\bibnamefont {Werner}},\ and\
  \bibinfo {author} {\bibfnamefont {E.}~\bibnamefont {Kozik}},\ }\href@noop {}
  {\bibinfo {title} {{Strange Metal Solution in the Diagrammatic Theory for the
  $2d$ Hubbard Model}}} (\bibinfo {year} {2020}),\ \Eprint
  {https://arxiv.org/abs/2012.06159} {arXiv:2012.06159} \BibitemShut {NoStop}%
\bibitem [{\citenamefont {van Loon}\ \emph {et~al.}(2018)\citenamefont {van
  Loon}, \citenamefont {Krien}, \citenamefont {Hafermann}, \citenamefont
  {Lichtenstein},\ and\ \citenamefont {Katsnelson}}]{VanLoon2018}%
  \BibitemOpen
  \bibfield  {author} {\bibinfo {author} {\bibfnamefont {E.~G. C.~P.}\
  \bibnamefont {van Loon}}, \bibinfo {author} {\bibfnamefont {F.}~\bibnamefont
  {Krien}}, \bibinfo {author} {\bibfnamefont {H.}~\bibnamefont {Hafermann}},
  \bibinfo {author} {\bibfnamefont {A.~I.}\ \bibnamefont {Lichtenstein}},\ and\
  \bibinfo {author} {\bibfnamefont {M.~I.}\ \bibnamefont {Katsnelson}},\
  }\bibfield  {title} {\bibinfo {title} {{Fermion-boson vertex within dynamical
  mean-field theory}},\ }\href {https://doi.org/10.1103/PhysRevB.98.205148}
  {\bibfield  {journal} {\bibinfo  {journal} {Phys. Rev. B}\ }\textbf {\bibinfo
  {volume} {98}},\ \bibinfo {pages} {205148} (\bibinfo {year}
  {2018})}\BibitemShut {NoStop}%
\bibitem [{\citenamefont {Harkov}\ \emph
  {et~al.}(2021{\natexlab{b}})\citenamefont {Harkov}, \citenamefont
  {Lichtenstein},\ and\ \citenamefont {Krien}}]{Harkov2021a}%
  \BibitemOpen
  \bibfield  {author} {\bibinfo {author} {\bibfnamefont {V.}~\bibnamefont
  {Harkov}}, \bibinfo {author} {\bibfnamefont {A.~I.}\ \bibnamefont
  {Lichtenstein}},\ and\ \bibinfo {author} {\bibfnamefont {F.}~\bibnamefont
  {Krien}},\ }\bibfield  {title} {\bibinfo {title} {{Parametrizations of local
  vertex corrections from weak to strong coupling: Importance of the Hedin
  three-leg vertex}},\ }\href {https://doi.org/10.1103/PhysRevB.104.125141}
  {\bibfield  {journal} {\bibinfo  {journal} {Phys. Rev. B}\ }\textbf {\bibinfo
  {volume} {104}},\ \bibinfo {pages} {125141} (\bibinfo {year}
  {2021}{\natexlab{b}})}\BibitemShut {NoStop}%
\bibitem [{\citenamefont {Toschi}\ \emph {et~al.}(2007)\citenamefont {Toschi},
  \citenamefont {Katanin},\ and\ \citenamefont {Held}}]{Toschi2007}%
  \BibitemOpen
  \bibfield  {author} {\bibinfo {author} {\bibfnamefont {A.}~\bibnamefont
  {Toschi}}, \bibinfo {author} {\bibfnamefont {A.~A.}\ \bibnamefont
  {Katanin}},\ and\ \bibinfo {author} {\bibfnamefont {K.}~\bibnamefont
  {Held}},\ }\bibfield  {title} {\bibinfo {title} {{Dynamical vertex
  approximation: A step beyond dynamical mean-field theory}},\ }\href
  {https://doi.org/10.1103/PhysRevB.75.045118} {\bibfield  {journal} {\bibinfo
  {journal} {Phys. Rev. B}\ }\textbf {\bibinfo {volume} {75}},\ \bibinfo
  {pages} {045118} (\bibinfo {year} {2007})}\BibitemShut {NoStop}%
\bibitem [{\citenamefont {Katanin}(2009)}]{Katanin2009}%
  \BibitemOpen
  \bibfield  {author} {\bibinfo {author} {\bibfnamefont {A.~A.}\ \bibnamefont
  {Katanin}},\ }\bibfield  {title} {\bibinfo {title} {{Two-loop functional
  renormalization group approach to the one- and two-dimensional Hubbard
  model}},\ }\href {https://doi.org/10.1103/PhysRevB.79.235119} {\bibfield
  {journal} {\bibinfo  {journal} {Phys. Rev. B}\ }\textbf {\bibinfo {volume}
  {79}},\ \bibinfo {pages} {235119} (\bibinfo {year} {2009})}\BibitemShut
  {NoStop}%
\bibitem [{\citenamefont {Reitner}\ \emph {et~al.}(2020)\citenamefont
  {Reitner}, \citenamefont {Chalupa}, \citenamefont {Del~Re}, \citenamefont
  {Springer}, \citenamefont {Ciuchi}, \citenamefont {Sangiovanni},\ and\
  \citenamefont {Toschi}}]{Reitner2020}%
  \BibitemOpen
  \bibfield  {author} {\bibinfo {author} {\bibfnamefont {M.}~\bibnamefont
  {Reitner}}, \bibinfo {author} {\bibfnamefont {P.}~\bibnamefont {Chalupa}},
  \bibinfo {author} {\bibfnamefont {L.}~\bibnamefont {Del~Re}}, \bibinfo
  {author} {\bibfnamefont {D.}~\bibnamefont {Springer}}, \bibinfo {author}
  {\bibfnamefont {S.}~\bibnamefont {Ciuchi}}, \bibinfo {author} {\bibfnamefont
  {G.}~\bibnamefont {Sangiovanni}},\ and\ \bibinfo {author} {\bibfnamefont
  {A.}~\bibnamefont {Toschi}},\ }\bibfield  {title} {\bibinfo {title}
  {{Attractive Effect of a Strong Electronic Repulsion: The Physics of Vertex
  Divergences}},\ }\href {https://doi.org/10.1103/PhysRevLett.125.196403}
  {\bibfield  {journal} {\bibinfo  {journal} {Phys. Rev. Lett.}\ }\textbf
  {\bibinfo {volume} {125}},\ \bibinfo {pages} {196403} (\bibinfo {year}
  {2020})}\BibitemShut {NoStop}%
\bibitem [{\citenamefont {van Loon}\ \emph {et~al.}(2020)\citenamefont {van
  Loon}, \citenamefont {Krien},\ and\ \citenamefont {Katanin}}]{VanLoon2020}%
  \BibitemOpen
  \bibfield  {author} {\bibinfo {author} {\bibfnamefont {E.~G. C.~P.}\
  \bibnamefont {van Loon}}, \bibinfo {author} {\bibfnamefont {F.}~\bibnamefont
  {Krien}},\ and\ \bibinfo {author} {\bibfnamefont {A.~A.}\ \bibnamefont
  {Katanin}},\ }\bibfield  {title} {\bibinfo {title} {{Bethe-Salpeter Equation
  at the Critical End Point of the Mott Transition}},\ }\href
  {https://doi.org/10.1103/PhysRevLett.125.136402} {\bibfield  {journal}
  {\bibinfo  {journal} {Phys. Rev. Lett.}\ }\textbf {\bibinfo {volume} {125}},\
  \bibinfo {pages} {136402} (\bibinfo {year} {2020})}\BibitemShut {NoStop}%
\bibitem [{Note6()}]{Note6}%
  \BibitemOpen
  \bibinfo {note} {S. Heinzelmann, private communication.}\BibitemShut {Stop}%
\bibitem [{\citenamefont {Halboth}\ and\ \citenamefont
  {Metzner}(2000{\natexlab{a}})}]{Halboth00A}%
  \BibitemOpen
  \bibfield  {author} {\bibinfo {author} {\bibfnamefont {C.~J.}\ \bibnamefont
  {Halboth}}\ and\ \bibinfo {author} {\bibfnamefont {W.}~\bibnamefont
  {Metzner}},\ }\bibfield  {title} {\bibinfo {title} {{Renormalization-group
  analysis of the two-dimensional Hubbard model}},\ }\href
  {https://doi.org/10.1103/PhysRevB.61.7364} {\bibfield  {journal} {\bibinfo
  {journal} {Phys. Rev. B}\ }\textbf {\bibinfo {volume} {61}},\ \bibinfo
  {pages} {7364} (\bibinfo {year} {2000}{\natexlab{a}})}\BibitemShut {NoStop}%
\bibitem [{\citenamefont {Halboth}\ and\ \citenamefont
  {Metzner}(2000{\natexlab{b}})}]{Halboth00B}%
  \BibitemOpen
  \bibfield  {author} {\bibinfo {author} {\bibfnamefont {C.~J.}\ \bibnamefont
  {Halboth}}\ and\ \bibinfo {author} {\bibfnamefont {W.}~\bibnamefont
  {Metzner}},\ }\bibfield  {title} {\bibinfo {title} {{$\mathit{d}$-Wave
  Superconductivity and Pomeranchuk Instability in the Two-Dimensional Hubbard
  Model}},\ }\href {https://doi.org/10.1103/PhysRevLett.85.5162} {\bibfield
  {journal} {\bibinfo  {journal} {Phys. Rev. Lett.}\ }\textbf {\bibinfo
  {volume} {85}},\ \bibinfo {pages} {5162} (\bibinfo {year}
  {2000}{\natexlab{b}})}\BibitemShut {NoStop}%
\bibitem [{\citenamefont {Yamase}\ \emph {et~al.}(2016)\citenamefont {Yamase},
  \citenamefont {Eberlein},\ and\ \citenamefont {Metzner}}]{Yamase16}%
  \BibitemOpen
  \bibfield  {author} {\bibinfo {author} {\bibfnamefont {H.}~\bibnamefont
  {Yamase}}, \bibinfo {author} {\bibfnamefont {A.}~\bibnamefont {Eberlein}},\
  and\ \bibinfo {author} {\bibfnamefont {W.}~\bibnamefont {Metzner}},\
  }\bibfield  {title} {\bibinfo {title} {{Coexistence of Incommensurate
  Magnetism and Superconductivity in the Two-Dimensional Hubbard Model}},\
  }\href {https://doi.org/10.1103/PhysRevLett.116.096402} {\bibfield  {journal}
  {\bibinfo  {journal} {Phys. Rev. Lett.}\ }\textbf {\bibinfo {volume} {116}},\
  \bibinfo {pages} {096402} (\bibinfo {year} {2016})}\BibitemShut {NoStop}%
\bibitem [{\citenamefont {Schulz}(1990)}]{Schulz90}%
  \BibitemOpen
  \bibfield  {author} {\bibinfo {author} {\bibfnamefont {H.~J.}\ \bibnamefont
  {Schulz}},\ }\bibfield  {title} {\bibinfo {title} {{Incommensurate
  antiferromagnetism in the two-dimensional Hubbard model}},\ }\href
  {https://doi.org/10.1103/PhysRevLett.64.1445} {\bibfield  {journal} {\bibinfo
   {journal} {Phys. Rev. Lett.}\ }\textbf {\bibinfo {volume} {64}},\ \bibinfo
  {pages} {1445} (\bibinfo {year} {1990})}\BibitemShut {NoStop}%
\bibitem [{\citenamefont {{T. Dombre}}(1990)}]{Dombre90}%
  \BibitemOpen
  \bibfield  {author} {\bibinfo {author} {\bibnamefont {{T. Dombre}}},\
  }\bibfield  {title} {\bibinfo {title} {{Modulated spiral phases in doped
  quantum antiferromagnets}},\ }\href
  {https://doi.org/10.1051/jphys:01990005109084700} {\bibfield  {journal}
  {\bibinfo  {journal} {J. Phys. France}\ }\textbf {\bibinfo {volume} {51}},\
  \bibinfo {pages} {847} (\bibinfo {year} {1990})}\BibitemShut {NoStop}%
\bibitem [{\citenamefont {Fr{\'{e}}sard}\ \emph {et~al.}(1991)\citenamefont
  {Fr{\'{e}}sard}, \citenamefont {Dzierzawa},\ and\ \citenamefont
  {W\"{o}lfle}}]{Fresard91}%
  \BibitemOpen
  \bibfield  {author} {\bibinfo {author} {\bibfnamefont {R.}~\bibnamefont
  {Fr{\'{e}}sard}}, \bibinfo {author} {\bibfnamefont {M.}~\bibnamefont
  {Dzierzawa}},\ and\ \bibinfo {author} {\bibfnamefont {P.}~\bibnamefont
  {W\"{o}lfle}},\ }\bibfield  {title} {\bibinfo {title} {{Slave-Boson Approach
  to Spiral Magnetic Order in the Hubbard Model}},\ }\href
  {https://doi.org/10.1209/0295-5075/15/3/016} {\bibfield  {journal} {\bibinfo
  {journal} {Europhys. Lett.}\ }\textbf {\bibinfo {volume} {15}},\ \bibinfo
  {pages} {325} (\bibinfo {year} {1991})}\BibitemShut {NoStop}%
\bibitem [{\citenamefont {Igoshev}\ \emph {et~al.}(2010)\citenamefont
  {Igoshev}, \citenamefont {Timirgazin}, \citenamefont {Katanin}, \citenamefont
  {Arzhnikov},\ and\ \citenamefont {Irkhin}}]{Igoshev10}%
  \BibitemOpen
  \bibfield  {author} {\bibinfo {author} {\bibfnamefont {P.~A.}\ \bibnamefont
  {Igoshev}}, \bibinfo {author} {\bibfnamefont {M.~A.}\ \bibnamefont
  {Timirgazin}}, \bibinfo {author} {\bibfnamefont {A.~A.}\ \bibnamefont
  {Katanin}}, \bibinfo {author} {\bibfnamefont {A.~K.}\ \bibnamefont
  {Arzhnikov}},\ and\ \bibinfo {author} {\bibfnamefont {V.~Y.}\ \bibnamefont
  {Irkhin}},\ }\bibfield  {title} {\bibinfo {title} {{Incommensurate magnetic
  order and phase separation in the two-dimensional Hubbard model with nearest-
  and next-nearest-neighbor hopping}},\ }\href
  {https://doi.org/10.1103/PhysRevB.81.094407} {\bibfield  {journal} {\bibinfo
  {journal} {Phys. Rev. B}\ }\textbf {\bibinfo {volume} {81}},\ \bibinfo
  {pages} {094407} (\bibinfo {year} {2010})}\BibitemShut {NoStop}%
\bibitem [{\citenamefont {Shraiman}\ and\ \citenamefont
  {Siggia}(1989)}]{Shraiman89}%
  \BibitemOpen
  \bibfield  {author} {\bibinfo {author} {\bibfnamefont {B.~I.}\ \bibnamefont
  {Shraiman}}\ and\ \bibinfo {author} {\bibfnamefont {E.~D.}\ \bibnamefont
  {Siggia}},\ }\bibfield  {title} {\bibinfo {title} {Spiral phase of a doped
  quantum antiferromagnet},\ }\href
  {https://doi.org/10.1103/PhysRevLett.62.1564} {\bibfield  {journal} {\bibinfo
   {journal} {Phys. Rev. Lett.}\ }\textbf {\bibinfo {volume} {62}},\ \bibinfo
  {pages} {1564} (\bibinfo {year} {1989})}\BibitemShut {NoStop}%
\bibitem [{\citenamefont {Chubukov}\ and\ \citenamefont
  {Frenkel}(1992)}]{Chubukov92}%
  \BibitemOpen
  \bibfield  {author} {\bibinfo {author} {\bibfnamefont {A.~V.}\ \bibnamefont
  {Chubukov}}\ and\ \bibinfo {author} {\bibfnamefont {D.~M.}\ \bibnamefont
  {Frenkel}},\ }\bibfield  {title} {\bibinfo {title} {{Renormalized
  perturbation theory of magnetic instabilities in the two-dimensional Hubbard
  model at small doping}},\ }\href {https://doi.org/10.1103/PhysRevB.46.11884}
  {\bibfield  {journal} {\bibinfo  {journal} {Phys. Rev. B}\ }\textbf {\bibinfo
  {volume} {46}},\ \bibinfo {pages} {11884} (\bibinfo {year}
  {1992})}\BibitemShut {NoStop}%
\bibitem [{\citenamefont {Chubukov}\ and\ \citenamefont
  {Musaelian}(1995)}]{Chubukov95}%
  \BibitemOpen
  \bibfield  {author} {\bibinfo {author} {\bibfnamefont {A.~V.}\ \bibnamefont
  {Chubukov}}\ and\ \bibinfo {author} {\bibfnamefont {K.~A.}\ \bibnamefont
  {Musaelian}},\ }\bibfield  {title} {\bibinfo {title} {{Magnetic phases of the
  two-dimensional Hubbard model at low doping}},\ }\href
  {https://doi.org/10.1103/PhysRevB.51.12605} {\bibfield  {journal} {\bibinfo
  {journal} {Phys. Rev. B}\ }\textbf {\bibinfo {volume} {51}},\ \bibinfo
  {pages} {12605} (\bibinfo {year} {1995})}\BibitemShut {NoStop}%
\bibitem [{\citenamefont {Kotov}\ and\ \citenamefont
  {Sushkov}(2004)}]{Kotov04}%
  \BibitemOpen
  \bibfield  {author} {\bibinfo {author} {\bibfnamefont {V.~N.}\ \bibnamefont
  {Kotov}}\ and\ \bibinfo {author} {\bibfnamefont {O.~P.}\ \bibnamefont
  {Sushkov}},\ }\bibfield  {title} {\bibinfo {title} {{Stability of the spiral
  phase in the two-dimensional extended $t\text{\ensuremath{-}}J$ model}},\
  }\href {https://doi.org/10.1103/PhysRevB.70.195105} {\bibfield  {journal}
  {\bibinfo  {journal} {Phys. Rev. B}\ }\textbf {\bibinfo {volume} {70}},\
  \bibinfo {pages} {195105} (\bibinfo {year} {2004})}\BibitemShut {NoStop}%
\bibitem [{\citenamefont {Fleck}\ \emph {et~al.}(1999)\citenamefont {Fleck},
  \citenamefont {Lichtenstein}, \citenamefont {Ole\ifmmode~\acute{s}\else
  \'{s}\fi{}},\ and\ \citenamefont {Hedin}}]{Fleck99}%
  \BibitemOpen
  \bibfield  {author} {\bibinfo {author} {\bibfnamefont {M.}~\bibnamefont
  {Fleck}}, \bibinfo {author} {\bibfnamefont {A.~I.}\ \bibnamefont
  {Lichtenstein}}, \bibinfo {author} {\bibfnamefont {A.~M.}\ \bibnamefont
  {Ole\ifmmode~\acute{s}\else \'{s}\fi{}}},\ and\ \bibinfo {author}
  {\bibfnamefont {L.}~\bibnamefont {Hedin}},\ }\bibfield  {title} {\bibinfo
  {title} {{Spectral and transport properties of doped Mott-Hubbard systems
  with incommensurate magnetic order}},\ }\href
  {https://doi.org/10.1103/PhysRevB.60.5224} {\bibfield  {journal} {\bibinfo
  {journal} {Phys. Rev. B}\ }\textbf {\bibinfo {volume} {60}},\ \bibinfo
  {pages} {5224} (\bibinfo {year} {1999})}\BibitemShut {NoStop}%
\bibitem [{\citenamefont {Sch\"afer}\ \emph {et~al.}(2017)\citenamefont
  {Sch\"afer}, \citenamefont {Katanin}, \citenamefont {Held},\ and\
  \citenamefont {Toschi}}]{Schaefer17}%
  \BibitemOpen
  \bibfield  {author} {\bibinfo {author} {\bibfnamefont {T.}~\bibnamefont
  {Sch\"afer}}, \bibinfo {author} {\bibfnamefont {A.~A.}\ \bibnamefont
  {Katanin}}, \bibinfo {author} {\bibfnamefont {K.}~\bibnamefont {Held}},\ and\
  \bibinfo {author} {\bibfnamefont {A.}~\bibnamefont {Toschi}},\ }\bibfield
  {title} {\bibinfo {title} {{Interplay of Correlations and Kohn Anomalies in
  Three Dimensions: Quantum Criticality with a Twist}},\ }\href
  {https://doi.org/10.1103/PhysRevLett.119.046402} {\bibfield  {journal}
  {\bibinfo  {journal} {Phys. Rev. Lett.}\ }\textbf {\bibinfo {volume} {119}},\
  \bibinfo {pages} {046402} (\bibinfo {year} {2017})}\BibitemShut {NoStop}%
\bibitem [{\citenamefont {Vilardi}\ \emph {et~al.}(2018)\citenamefont
  {Vilardi}, \citenamefont {Taranto},\ and\ \citenamefont
  {Metzner}}]{Vilardi18}%
  \BibitemOpen
  \bibfield  {author} {\bibinfo {author} {\bibfnamefont {D.}~\bibnamefont
  {Vilardi}}, \bibinfo {author} {\bibfnamefont {C.}~\bibnamefont {Taranto}},\
  and\ \bibinfo {author} {\bibfnamefont {W.}~\bibnamefont {Metzner}},\
  }\bibfield  {title} {\bibinfo {title} {{Dynamically enhanced magnetic
  incommensurability: Effects of local dynamics on nonlocal spin correlations
  in a strongly correlated metal}},\ }\href
  {https://doi.org/10.1103/PhysRevB.97.235110} {\bibfield  {journal} {\bibinfo
  {journal} {Phys. Rev. B}\ }\textbf {\bibinfo {volume} {97}},\ \bibinfo
  {pages} {235110} (\bibinfo {year} {2018})}\BibitemShut {NoStop}%
\bibitem [{\citenamefont {Kauch}\ \emph {et~al.}(2020)\citenamefont {Kauch},
  \citenamefont {Pudleiner}, \citenamefont {Astleithner}, \citenamefont
  {Thunstr\"om}, \citenamefont {Ribic},\ and\ \citenamefont
  {Held}}]{Kauch2020}%
  \BibitemOpen
  \bibfield  {author} {\bibinfo {author} {\bibfnamefont {A.}~\bibnamefont
  {Kauch}}, \bibinfo {author} {\bibfnamefont {P.}~\bibnamefont {Pudleiner}},
  \bibinfo {author} {\bibfnamefont {K.}~\bibnamefont {Astleithner}}, \bibinfo
  {author} {\bibfnamefont {P.}~\bibnamefont {Thunstr\"om}}, \bibinfo {author}
  {\bibfnamefont {T.}~\bibnamefont {Ribic}},\ and\ \bibinfo {author}
  {\bibfnamefont {K.}~\bibnamefont {Held}},\ }\bibfield  {title} {\bibinfo
  {title} {Generic optical excitations of correlated systems:
  $\ensuremath{\pi}$-tons},\ }\href
  {https://doi.org/10.1103/PhysRevLett.124.047401} {\bibfield  {journal}
  {\bibinfo  {journal} {Phys. Rev. Lett.}\ }\textbf {\bibinfo {volume} {124}},\
  \bibinfo {pages} {047401} (\bibinfo {year} {2020})}\BibitemShut {NoStop}%
\bibitem [{\citenamefont {Eberlein}\ and\ \citenamefont
  {Metzner}(2014)}]{Eberlein14}%
  \BibitemOpen
  \bibfield  {author} {\bibinfo {author} {\bibfnamefont {A.}~\bibnamefont
  {Eberlein}}\ and\ \bibinfo {author} {\bibfnamefont {W.}~\bibnamefont
  {Metzner}},\ }\bibfield  {title} {\bibinfo {title} {{Superconductivity in the
  two-dimensional $t$-${t}^{\ensuremath{'}}$-Hubbard model}},\ }\href
  {https://doi.org/10.1103/PhysRevB.89.035126} {\bibfield  {journal} {\bibinfo
  {journal} {Phys. Rev. B}\ }\textbf {\bibinfo {volume} {89}},\ \bibinfo
  {pages} {035126} (\bibinfo {year} {2014})}\BibitemShut {NoStop}%
\bibitem [{\citenamefont {Wu}\ \emph {et~al.}(2017)\citenamefont {Wu},
  \citenamefont {Ferrero}, \citenamefont {Georges},\ and\ \citenamefont
  {Kozik}}]{Wu2017}%
  \BibitemOpen
  \bibfield  {author} {\bibinfo {author} {\bibfnamefont {W.}~\bibnamefont
  {Wu}}, \bibinfo {author} {\bibfnamefont {M.}~\bibnamefont {Ferrero}},
  \bibinfo {author} {\bibfnamefont {A.}~\bibnamefont {Georges}},\ and\ \bibinfo
  {author} {\bibfnamefont {E.}~\bibnamefont {Kozik}},\ }\bibfield  {title}
  {\bibinfo {title} {{Controlling Feynman diagrammatic expansions: Physical
  nature of the pseudogap in the two-dimensional Hubbard model}},\ }\href
  {https://doi.org/10.1103/PhysRevB.96.041105} {\bibfield  {journal} {\bibinfo
  {journal} {Phys. Rev. B}\ }\textbf {\bibinfo {volume} {96}},\ \bibinfo
  {pages} {041105} (\bibinfo {year} {2017})}\BibitemShut {NoStop}%
\bibitem [{\citenamefont {Wang}\ \emph {et~al.}(2014)\citenamefont {Wang},
  \citenamefont {Eberlein},\ and\ \citenamefont {Metzner}}]{Wang2014}%
  \BibitemOpen
  \bibfield  {author} {\bibinfo {author} {\bibfnamefont {J.}~\bibnamefont
  {Wang}}, \bibinfo {author} {\bibfnamefont {A.}~\bibnamefont {Eberlein}},\
  and\ \bibinfo {author} {\bibfnamefont {W.}~\bibnamefont {Metzner}},\
  }\bibfield  {title} {\bibinfo {title} {{Competing order in correlated
  electron systems made simple: Consistent fusion of functional renormalization
  and mean-field theory}},\ }\href {https://doi.org/10.1103/PhysRevB.89.121116}
  {\bibfield  {journal} {\bibinfo  {journal} {Phys. Rev. B}\ }\textbf {\bibinfo
  {volume} {89}},\ \bibinfo {pages} {121116} (\bibinfo {year}
  {2014})}\BibitemShut {NoStop}%
\bibitem [{\citenamefont {Vilardi}\ \emph {et~al.}(2020)\citenamefont
  {Vilardi}, \citenamefont {Bonetti},\ and\ \citenamefont
  {Metzner}}]{Vilardi20}%
  \BibitemOpen
  \bibfield  {author} {\bibinfo {author} {\bibfnamefont {D.}~\bibnamefont
  {Vilardi}}, \bibinfo {author} {\bibfnamefont {P.~M.}\ \bibnamefont
  {Bonetti}},\ and\ \bibinfo {author} {\bibfnamefont {W.}~\bibnamefont
  {Metzner}},\ }\bibfield  {title} {\bibinfo {title} {{Dynamical functional
  renormalization group computation of order parameters and critical
  temperatures in the two-dimensional Hubbard model}},\ }\href
  {https://doi.org/10.1103/PhysRevB.102.245128} {\bibfield  {journal} {\bibinfo
   {journal} {Phys. Rev. B}\ }\textbf {\bibinfo {volume} {102}},\ \bibinfo
  {pages} {245128} (\bibinfo {year} {2020})}\BibitemShut {NoStop}%
\bibitem [{\citenamefont {Bonetti}(2020)}]{Bonetti20}%
  \BibitemOpen
  \bibfield  {author} {\bibinfo {author} {\bibfnamefont {P.~M.}\ \bibnamefont
  {Bonetti}},\ }\bibfield  {title} {\bibinfo {title} {{Accessing the ordered
  phase of correlated Fermi systems: Vertex bosonization and mean-field theory
  within the functional renormalization group}},\ }\href
  {https://doi.org/10.1103/PhysRevB.102.235160} {\bibfield  {journal} {\bibinfo
   {journal} {Phys. Rev. B}\ }\textbf {\bibinfo {volume} {102}},\ \bibinfo
  {pages} {235160} (\bibinfo {year} {2020})}\BibitemShut {NoStop}%
\bibitem [{\citenamefont {Qin}\ \emph {et~al.}(2021)\citenamefont {Qin},
  \citenamefont {Schäfer}, \citenamefont {Andergassen}, \citenamefont
  {Corboz},\ and\ \citenamefont {Gull}}]{Qin21}%
  \BibitemOpen
  \bibfield  {author} {\bibinfo {author} {\bibfnamefont {M.}~\bibnamefont
  {Qin}}, \bibinfo {author} {\bibfnamefont {T.}~\bibnamefont {Schäfer}},
  \bibinfo {author} {\bibfnamefont {S.}~\bibnamefont {Andergassen}}, \bibinfo
  {author} {\bibfnamefont {P.}~\bibnamefont {Corboz}},\ and\ \bibinfo {author}
  {\bibfnamefont {E.}~\bibnamefont {Gull}},\ }\href@noop {} {\bibinfo {title}
  {{The Hubbard model: A computational perspective}}} (\bibinfo {year}
  {2021}),\ \Eprint {https://arxiv.org/abs/2104.00064} {arXiv:2104.00064}
  \BibitemShut {NoStop}%
\end{thebibliography}%
\end{document}